\documentclass[a4paper,onecolumn,11pt,accepted=2026-05-20]{quantumarticle}
\pdfoutput=1
\pdfoutput=1
\usepackage[font=small,labelfont=bf]{caption}
\usepackage[utf8]{inputenc}
\usepackage[english]{babel}
\usepackage[T1]{fontenc}
\usepackage{amsmath}
\usepackage{parskip}
\usepackage[numbers,sort&compress]{natbib}
\usepackage[colorlinks=true, citecolor=gray]{hyperref}
\usepackage[dvipsnames,table]{xcolor}
\usepackage{tikz}
\usepackage{lipsum}

\usepackage{pdfsync}
\usepackage{amsfonts}
\usepackage{amsthm}
\usepackage{amssymb}
\usepackage{thmtools}
\usepackage{dsfont}
\usepackage{braket}
\usepackage{mathtools}
\usepackage[export]{adjustbox}
\usepackage{scalerel}
\usepackage[nameinlink]{cleveref}
\crefname{subsection}{subsection}{subsections}
\usepackage{xstring}
\usepackage{ifthen}
\usepackage{stmaryrd}
\usepackage{subcaption}
\usepackage[normalem]{ulem}
\usepackage{tabularray}
\usepackage{makecell}
\usepackage{xkeyval}
\usepackage{wasysym}

\usepackage{graphicx}
\usepackage{float}
\usepackage{svg}
\usepackage{appendix}
\usepackage{multirow}
\usepackage{longtable}
\usepackage{tabularx}
\usepackage{placeins}

\theoremstyle{plain}
\newtheorem{theorem}{Theorem}[section]

\newtheorem{keypoint}[theorem]{Key point}

\newtheorem{definition}[theorem]{Definition}

\newtheorem{difference}[theorem]{Difference}

\crefname{keypoint}{key point}{key points}
\Crefname{keypoint}{Key point}{Key points}
\crefname{difference}{difference}{differences}
\Crefname{difference}{Difference}{Differences}

\declaretheoremstyle[qed={$\blacksquare$}]{exampleStyle}

\declaretheorem[style=exampleStyle, sibling=theorem, name=Example, numbered=no]{example*}
\makeatletter
\newcounter{subsubsubsection}[subsubsection]
\renewcommand{\thesubsubsubsection}{\thesubsubsection.\arabic{subsubsubsection}}
\newcommand{\subsubsubsection}[1]{%
    \refstepcounter{subsubsubsection}%
    \vskip 2ex \@plus 1ex \@minus .2ex% Reduced vertical space before
    {\sffamily\normalsize% Sans-serif font family, normal size, no bold
        \thesubsubsubsection \quad #1\par}% Put heading on its own line
    \nobreak% Prevent page break after heading
}
\makeatother

\newcommand{\ccite}[2][]{%
    \IfSubStr{#2}{,}{refs.~}{ref.~}%
    \ifthenelse{\equal{#1}{}}{\cite{#2}}{\cite[#1]{#2}}}
\newcommand{\Ccite}[2][]{%
    \IfSubStr{#2}{,}{Refs.~}{Ref.~}%
    \ifthenelse{\equal{#1}{}}{\cite{#2}}{\cite[#1]{#2}}}
\newcommand{\LLL}{\mathcal{L}}
\newcommand{\MMM}{\mathcal{M}}
\newcommand{\NNN}{\mathcal{N}}
\newcommand{\PPP}{\mathcal{P}}
\newcommand{\SSS}{\mathcal{S}}

\newif\ifShowComments
\ShowCommentstrue
\newcommand{\defeq}{\coloneqq}

\newcommand{\groupPres}[1]{\left\langle #1 \right\rangle}

\newcommand{\isg}[1]{\SSS_{#1}}

\newcommand{\lpglong}[1]{\NNN(\isg{#1}) / \isg{#1}}

\newcommand{\dcccOp}[3]{#1_{\colourful{#2}, #3}}
\newcommand{\colourful}[1]{%
    \IfEqCase{#1}{%
            {r}{\textcolor{red}{r}}%
            {g}{\textcolor{Green}{g}}%
            {b}{\textcolor{blue}{b}}%
            {c}{\textcolor{cyan}{c}}%
            {m}{\textcolor{magenta}{m}}%
            {y}{\textcolor{Dandelion}{y}}% Yellow!90!Black? Goldenrod? Dandelion?
            {X}{\textcolor{red}{X}}% TODO: Match with ZX red
            {Y}{\textcolor{blue}{Y}}% TODO: Match with ZX blue
            {Z}{\textcolor{Green}{Z}}% TODO: Match with ZX green
    }[#1]}
\DeclareFontEncoding{LS1}{}{}
\DeclareFontSubstitution{LS1}{stix}{m}{n}
\DeclareSymbolFont{symbols4}{LS1}{stixbb}{m}{it}
\DeclareMathSymbol{\filledHexagon}{\mathord}{symbols4}{"DE}
\newcommand{\confinedColourDefault}{gray!50}

\newcommand{\magneticColourDefault}{orange!20}
\newcommand{\condensedRow}{1}
\newcommand{\condensedColumn}{1}
\newcommand{\confinedRows}{2,3}
\newcommand{\confinedColumns}{2,3}
\newcommand{\condensedRowColour}{white}
\newcommand{\condensedColumnColour}{white}
\newcommand{\confinedColour}{white}
\newcommand{\bosonTableRx}{}
\newcommand{\bosonTableGx}{}
\newcommand{\bosonTableBx}{}
\newcommand{\bosonTableRy}{}
\newcommand{\bosonTableGy}{}
\newcommand{\bosonTableBy}{}
\newcommand{\bosonTableRz}{}
\newcommand{\bosonTableGz}{}
\newcommand{\bosonTableBz}{}
\makeatletter
\define@key{bosonTable}{condensed}[]{%
    \IfEqCase{#1}{%
        {}{%
            \renewcommand{\condensedRowColour}{white}%
            \renewcommand{\condensedColumnColour}{white}%
            \renewcommand{\confinedColour}{white}%
        }{rx}{%
            \renewcommand{\condensedRow}{1}%
            \renewcommand{\condensedColumn}{1}%
            \renewcommand{\confinedRows}{2, 3}%
            \renewcommand{\confinedColumns}{2, 3}%
            \renewcommand{\confinedColour}{\confinedColourDefault}%
            \renewcommand{\condensedColumnColour}{\deconfinedColourDefault}%
            \renewcommand{\condensedRowColour}{\deconfinedColourDefault}%
        }{gx}{%
            \renewcommand{\condensedRow}{1}%
            \renewcommand{\condensedColumn}{2}%
            \renewcommand{\confinedRows}{2, 3}%
            \renewcommand{\confinedColumns}{1, 3}%
            \renewcommand{\confinedColour}{\confinedColourDefault}%
            \renewcommand{\condensedColumnColour}{\deconfinedColourDefault}%
            \renewcommand{\condensedRowColour}{\deconfinedColourDefault}%
        }{bx}{%
            \renewcommand{\condensedRow}{1}%
            \renewcommand{\condensedColumn}{3}%
            \renewcommand{\confinedRows}{2, 3}%
            \renewcommand{\confinedColumns}{1, 2}%
            \renewcommand{\confinedColour}{\confinedColourDefault}%
            \renewcommand{\condensedColumnColour}{\deconfinedColourDefault}%
            \renewcommand{\condensedRowColour}{\deconfinedColourDefault}%
        }{ry}{%
            \renewcommand{\condensedRow}{2}%
            \renewcommand{\condensedColumn}{1}%
            \renewcommand{\confinedRows}{1, 3}%
            \renewcommand{\confinedColumns}{2, 3}%
            \renewcommand{\confinedColour}{\confinedColourDefault}%
            \renewcommand{\condensedColumnColour}{\deconfinedColourDefault}%
            \renewcommand{\condensedRowColour}{\deconfinedColourDefault}%
        }{gy}{%
            \renewcommand{\condensedRow}{2}%
            \renewcommand{\condensedColumn}{2}%
            \renewcommand{\confinedRows}{1, 3}%
            \renewcommand{\confinedColumns}{1, 3}%
            \renewcommand{\confinedColour}{\confinedColourDefault}%
            \renewcommand{\condensedColumnColour}{\deconfinedColourDefault}%
            \renewcommand{\condensedRowColour}{\deconfinedColourDefault}%
        }{by}{%
            \renewcommand{\condensedRow}{2}%
            \renewcommand{\condensedColumn}{3}%
            \renewcommand{\confinedRows}{1, 3}%
            \renewcommand{\confinedColumns}{1, 2}%
            \renewcommand{\confinedColour}{\confinedColourDefault}%
            \renewcommand{\condensedColumnColour}{\deconfinedColourDefault}%
            \renewcommand{\condensedRowColour}{\deconfinedColourDefault}%
        }{rz}{%
            \renewcommand{\condensedRow}{3}%
            \renewcommand{\condensedColumn}{1}%
            \renewcommand{\confinedRows}{1, 2}%
            \renewcommand{\confinedColumns}{2, 3}%
            \renewcommand{\confinedColour}{\confinedColourDefault}%
            \renewcommand{\condensedColumnColour}{\deconfinedColourDefault}%
            \renewcommand{\condensedRowColour}{\deconfinedColourDefault}%
        }{gz}{%
            \renewcommand{\condensedRow}{3}%
            \renewcommand{\condensedColumn}{2}%
            \renewcommand{\confinedRows}{1, 2}%
            \renewcommand{\confinedColumns}{1, 3}%
            \renewcommand{\confinedColour}{\confinedColourDefault}%
            \renewcommand{\condensedColumnColour}{\deconfinedColourDefault}%
            \renewcommand{\condensedRowColour}{\deconfinedColourDefault}%
        }{bz}{%
            \renewcommand{\condensedRow}{3}%
            \renewcommand{\condensedColumn}{3}%
            \renewcommand{\confinedRows}{1, 2}%
            \renewcommand{\confinedColumns}{1, 2}%
            \renewcommand{\confinedColour}{\confinedColourDefault}%
            \renewcommand{\condensedColumnColour}{\deconfinedColourDefault}%
            \renewcommand{\condensedRowColour}{\deconfinedColourDefault}%
        }%
    }[\PackageError%
        {bosonTableError}%
        {Unknown boson label: `#1'}%
        {Accepted values: `', `rx', `gx', `bx', `ry', `gy', `by', `rz', `gz', `bz'}%
    ]%
}
\define@key{bosonTable}{rowLabel}[]{%
    \IfEqCase{#1}{%
        {}{\renewcommand{\condensedRowColour}{white}}%
        {electric}{\renewcommand{\condensedRowColour}{\electricColourDefault}}%
        {e}{\renewcommand{\condensedRowColour}{\electricColourDefault}}%
        {magnetic}{\renewcommand{\condensedRowColour}{\magneticColourDefault}}%
        {m}{\renewcommand{\condensedRowColour}{\magneticColourDefault}}%
    }[\PackageError%
        {bosonTableError}%
        {Unknown charge label: `#1'}%
        {Accepted values: `', `electric', `e', `magnetic', `m'}%
    ]%
}
\define@key{bosonTable}{columnLabel}[]{%
    \IfEqCase{#1}{%
        {}{\renewcommand{\condensedColumnColour}{white}}%
        {electric}{\renewcommand{\condensedColumnColour}{\electricColourDefault}}%
        {e}{\renewcommand{\condensedColumnColour}{\electricColourDefault}}%
        {magnetic}{\renewcommand{\condensedColumnColour}{\magneticColourDefault}}%
        {m}{\renewcommand{\condensedColumnColour}{\magneticColourDefault}}%
    }[\PackageError%
        {bosonTableError}%
        {Unknown charge label: `#1'}%
        {Accepted values: `', `electric', `e', `magnetic', `m'}%
    ]%
}
\define@key{bosonTable}{rx}[]{\renewcommand{\bosonTableRx}{#1}}
\define@key{bosonTable}{gx}[]{\renewcommand{\bosonTableGx}{#1}}
\define@key{bosonTable}{bx}[]{\renewcommand{\bosonTableBx}{#1}}
\define@key{bosonTable}{ry}[]{\renewcommand{\bosonTableRy}{#1}}
\define@key{bosonTable}{gy}[]{\renewcommand{\bosonTableGy}{#1}}
\define@key{bosonTable}{by}[]{\renewcommand{\bosonTableBy}{#1}}
\define@key{bosonTable}{rz}[]{\renewcommand{\bosonTableRz}{#1}}
\define@key{bosonTable}{gz}[]{\renewcommand{\bosonTableGz}{#1}}
\define@key{bosonTable}{bz}[]{\renewcommand{\bosonTableBz}{#1}}
\setkeys{bosonTable}{%
    condensed={},
    rowLabel={},
    columnLabel={},
    rx={},
    gx={},
    bx={},
    ry={},
    gy={},
    by={},
    rz={},
    gz={},
    bz={},
}
\newcommand{\bosonTable}[1]{%
    \begingroup%
        \setkeys{bosonTable}{#1}%
        \ExpandArgs{ne}\begin{tblr}{
            colspec = {ccc},
            cell{\condensedRow}{\confinedColumns} = {\condensedRowColour},
            cell{\confinedRows}{\condensedColumn} = {\condensedColumnColour},
            cell{\confinedRows}{\confinedColumns} = {\confinedColour},
            vlines = {1-3}{solid},
            vline{1, 4} = {0pt},
            hlines = {1-3}{solid},
            hline{1, 4} = {0pt},
        }
            \ensuremath{\bosonTableRx} & \ensuremath{\bosonTableGx} & \ensuremath{\bosonTableBx} \\
            \ensuremath{\bosonTableRy} & \ensuremath{\bosonTableGy} & \ensuremath{\bosonTableBy} \\
            \ensuremath{\bosonTableRz} & \ensuremath{\bosonTableGz} & \ensuremath{\bosonTableBz}
        \end{tblr}
    \endgroup%
}
\makeatother
\newcommand{\qqquad}{\quad\qquad}

\title{Dynamical codes for hardware with noisy readouts}

\author{Peter-Jan H.~S.~Derks}
\email{peter-janderks@hotmail.com}

\affiliation{Dahlem Center for Complex Quantum Systems, Freie Universit\"at Berlin, 14195 Berlin, Germany}

\author{Alex Townsend-Teague}
\email{alex.townsend-teague@outlook.com}
\affiliation{Dahlem Center for Complex Quantum Systems, Freie Universit\"at Berlin, 14195 Berlin, Germany}

\author{Jens Eisert}
\affiliation{Dahlem Center for Complex Quantum Systems, Freie Universit\"at Berlin, 14195 Berlin, Germany}

\author{Markus S.~Kesselring}
\affiliation{Dahlem Center for Complex Quantum Systems, Freie Universit\"at Berlin, 14195 Berlin, Germany}

\author{Oscar Higgott}
\thanks{Now at Google Quantum AI, Venice, CA 90291, USA.}
\affiliation{Department of Physics \& Astronomy, University College London, WC1E 6BT London, United Kingdom}

\author{Benjamin J.~Brown}
\affiliation{IBM Quantum, T. J. Watson Research Center, Yorktown Heights, New York 10598, USA}
\affiliation{IBM Denmark, Sundkrogsgade 11, 2100 Copenhagen, Denmark}

\begin{document}
    \maketitle
    
    \begin{abstract}
    \noindent
Dynamical stabilizer codes may offer a practical route to large-scale quantum computation. Such codes are defined by a schedule of error-detecting measurements, which allows for flexibility in their construction. In this work, we ask how best to optimise the measurement schedule of dynamically condensed colour codes in various limits of noise bias. We take a particular focus on the setting where measurements introduce more noise than unitary and idling operations -- a noise model relevant to some hardware proposals.
For measurement-biased noise models, we improve code performance by strategically repeating measurements within the schedule.
For unbiased or $Z$-biased noise models, we find repeating measurements offers little improvement -- somewhat contrary to our expectations -- and investigate why this is.
To perform this analysis,
we generalise a metric called the teraquop footprint to the teraquop volume.
This is the product of the number of qubits and number of rounds of measurements required such that the probability of a spacelike or timelike logical error occurring is less than $10^{-12}$.
In most cases, we find differences in performance are primarily due to the number of rounds of measurements required, rather than the number of qubits -- emphasising the importance of using the teraquop volume in the analysis.
Additionally, our results provide another example of the importance of making use of correlated errors when decoding, in that using belief matching rather than minimum-weight perfect matching can turn a worst-performing code under a given noise model into a best-performing code.
\end{abstract}

  \vspace{10pt}
    \noindent\textbf{Code availability:} \\
    \url{https://github.com/peter-janderks/Floquet-codes-and-measurement-bias}
    \newpage
    \tableofcontents
    \newpage

    \section{Introduction}

To be able to execute currently known useful quantum algorithms, a computer with near-perfect qubits and gates is needed.
As individual qubits and gates are far from perfect,
the plan is to employ \emph{quantum error correction} (QEC) to build a fault-tolerant quantum computer.
A quantum computing scheme is considered fault-tolerant if its output is correct with high probability, even though its individual components are faulty.
A popular strategy for implementing fault-tolerant quantum computing is to perform gates through lattice surgery on qubits encoded in the surface code \cite{dennis2002topological,Horsman_2012_lattice_surgery,Litinski2019gameofsurfacecodes, gidney2021factor}.
The basic unit of a lattice surgery computer is a single qubit encoded in the surface code for a fixed unit of time. 
We refer to this unit as a  \textit{surface code block}.
It is an exciting time for lattice surgery computing, with several experimental prototypes of surface code blocks demonstrated 
\cite{acharya2024quantum, caune2024demonstrating, wootton2022measurements, eickbusch2024demonstrating, ali2024reducing, krinner2022realizing, harper2025characterising}.

In parallel with the development of these prototypes, the recent discovery of \textit{dynamic stabilizer codes} has led to new ways of implementing surface code blocks. This started with the introduction of the Hastings-Haah honeycomb code \cite{hastings2021dynamically}.
Subsequently, a larger class of codes containing the honeycomb code was discovered, which we refer to as \textit{dynamically condensed colour codes} (DCCCs) \cite{kesselring2022anyon}.
Despite having `colour code' in their name, DCCCs are topologically equivalent to the surface code, so can be used as surface code blocks
\cite{davydova2022floquet, bombin2024unifying, kesselring2022anyon}. 

Schemes for fault-tolerant quantum computing have an enormous resource cost for implementation with modern technology.
For this reason, the main motivation for looking for new QEC codes is to reduce the resource requirements of a fault-tolerant computer. Importantly, the performance of a QEC code depends on the characteristics of the noise that affects a computer. 
Notably, for many hardware platforms, the noise can be biased towards certain types of operations.
For example, a certain Pauli error ($Z$, say) can be more likely to occur than another ($X$ or $Y$, say).
In the past, it has been shown that QEC codes can be tailored to deal with biased noise
\cite{bonilla2021xzzx, darmawan2021practical, OscarSubsystemGaugeFixing, Miguel2023cellularautomaton, Srivastava2022xyzhexagonal, NonIID, PhysRevLett.133.110601,Tuckett2018,Tuckett2019, Tuckett2020}.
These results have been so promising that some quantum computers are now designed such that the noise is
deliberately biased \cite{darmawan2021practical}.
However, fewer works look at tailoring codes 
to hardware platforms where measurement errors are dominant --
i.e.\ where measurements are more faulty than other types of quantum operations.
This type of biased noise is relevant to multiple hardware proposals based on superconducting qubits~\cite{harper2025characterising, rigetti_qpus, iqm_radiance}.

In this work, we investigate the resource requirements of a lattice surgery quantum computer that uses a DCCC.
Simulations have shown promising results for the \textit{space} cost of two specific DCCCs under an unbiased noise model \cite{gidney2021fault, gidney2022benchmarking, kesselring2022anyon}.
However, the \textit{spacetime} cost provides a more precise measure for resource estimation of a lattice surgery quantum computer. This has not previously been studied for either uniform or biased noise models.
To measure the spacetime cost, we generalise the teraquop footprint~\cite{gidney2021fault} to the \textit{teraquop volume}, defined as the product of the number of qubits and the number of rounds required to achieve a logical error rate of $10^{-12}$, the so-called \textit{teraquop regime}.
A logical error rate of $10^{-12}$ or lower would be required for a lattice surgery quantum computer to be able to solve important problems~\cite{lee2021even, gidney2021factor, haner2020improved,gidney2025factor,bourdoncle2026two}.
% We develop a numerical procedure to find the teraquop volume and carry it out for several biased noise models, including models biased towards measurement noise.

\subsection{Previous numerical work}\label{subsec:previous-numerical-work}

The footprint of what we'll refer to as the $X^1 Y^1 Z^1$ honeycomb code has been studied in \Ccite{gidney2022benchmarking, gidney2021fault, dessertaine2025, chan2025tailoring}.
The footprints of hyperbolic and semi-hyperbolic $X^1 Y^1 Z^1$ honeycomb codes have been studied in \Ccite{higgott2024constructions,sutcliffe2025distributedquantumerrorcorrection}. 
The memory threshold of what we'll refer to as the $X^1 Z^1$ honeycomb code has been studied in \Ccite{kesselring2022anyon, Hetnyi2024}. 
The memory threshold of the $X^1 Z^1$ honeycomb code tailored to biased noise has been studied in \Ccite{setiawan2024tailoring}. 
Measurements of $X^1 Y^1 Z^1$ and $X^1 Z^1$ honeycomb code plaquette stabilizers have been performed on superconducting hardware \cite{wootton2022measurements}.

\subsection{Overview of results}\label{subsec:results}

We simulate a large class of DCCCs which are defined by their \textit{measurement schedules}.
All DCCCs are implemented by repeatedly measuring sets of $XX$, $YY$ or $ZZ$ operators according to certain rules.
Additionally, one can choose to repeat a particular set of weight-two measurements any number of times~\cite{OscarSubsystemGaugeFixing}.
A glossary of the terms used in the figures within this section is provided in Table \ref{tab:glossary}.

\subsubsection{Direct parity measurement noise results}\label{subsubsec:direct-parity-measurement-noise-results}

\Cref{fig:phenomenological_noise_volume_plot_beliefmatching,fig:phenomenological_noise_volume_plot_pymatching} show the performance of DCCCs for direct parity measurement noise models decoded using MWPM and belief matching.
We have obtained volumes for 16 different noise models, which differ in measurement bias and $Z$ error bias.
From the results presented in \Cref{fig:phenomenological_noise_volume_plot_beliefmatching,fig:phenomenological_noise_volume_plot_pymatching}, we can draw the following conclusions:

\begin{keypoint}
When decoding with MWPM, it is better to only measure $XX$ and $ZZ$, while when decoding with belief matching, it is better to measure $XX$, $YY$, and $ZZ$.
\end{keypoint}

\begin{keypoint}
Repeating measurements can improve the teraquop volume under measure\-ment-biased noise.
\end{keypoint}

\subsubsection{Auxiliary qubit circuit noise results}\label{subsubsec:auxiliary-qubit-circuit-noise-results}

\begin{figure}[p]
    \centering
    \includegraphics[width=0.9\columnwidth]{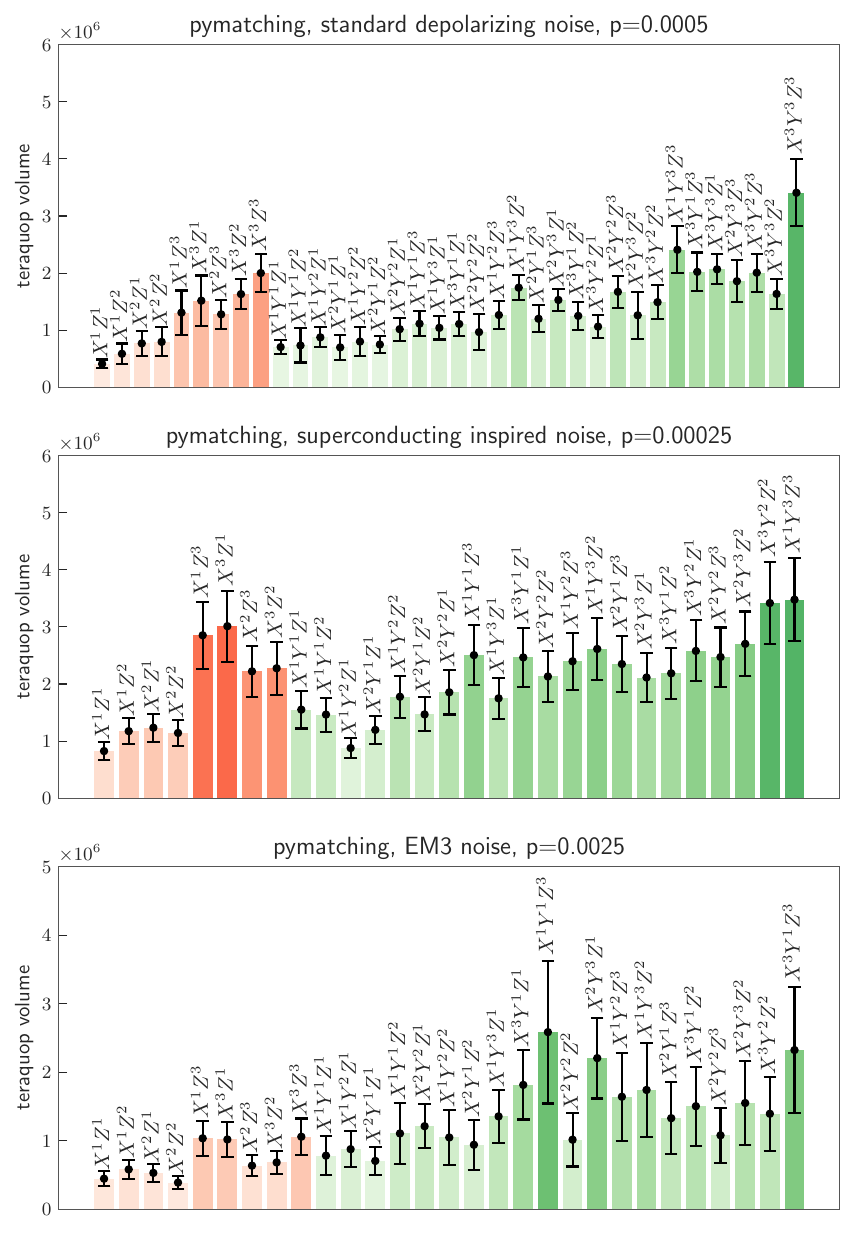}
    \caption{Teraquop volumes of $X^a Y^b Z^c$ and $X^a Z^b$ codes for three auxiliary qubit circuit noise models; standard depolarizing noise with $p=0.0005$, superconducting inspired noise with $p=0.00025$, and entangling measurement noise with $p=0.0025$. Decoding is performed using MWPM. 
    Codes are ordered first by type ($X^a Z^b$ code, then $X^a Y^b Z^c$ code), then by increasing schedule length ($a + b$ or $a+b+c$). 
    Codes with equal schedule length are ordered by balance, with more uniformly distributed repetition parameters appearing first.
    In each case, the code with the best teraquop volume is from the $X^a Z^b$ family, though the difference is small.
    Repeating measurements broadly has negligible effect, though does seem to improve the performance of the $X^1 Y^1 Z^1$ code in the superconducting inspired noise model, which features a moderate measurement bias.
    The error bars are explained in \Cref{sec:methodology}.}
    \label{fig:pymatching_circuit_level_noise_volume_plot}
\end{figure}

\begin{figure}[p]
    \centering
    \includegraphics[width=0.9\columnwidth]{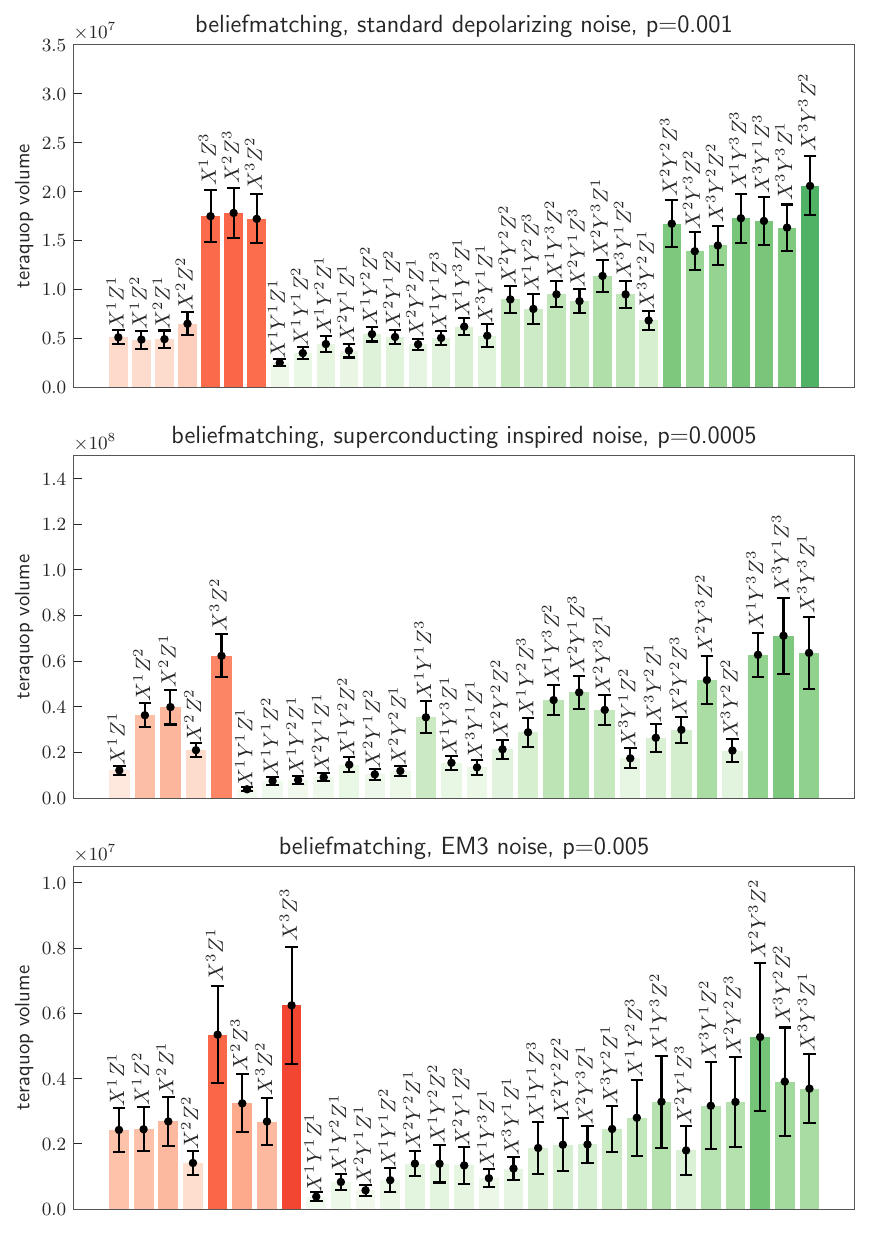}
  \caption{
    Same plot as in \Cref{fig:pymatching_circuit_level_noise_volume_plot} but with decoding performed using belief matching and higher physical error rates;
    standard depolarizing noise with $p=0.001$, superconducting inspired noise with $p=0.0005$, and entangling measurement noise with $p=0.005$. 
    As before, repeating measurements broadly has negligible effect, though does improve the performance of the $X^1 Z^1$ code in the entangling measurements noise model.}
    \label{fig:beliefmatching_circuit_level_noise_volume_plot}
\end{figure}

Three auxiliary qubit circuit noise models are simulated: \emph{superconducting-inspired} (SI) noise, \emph{standard depolarizing} (SD) noise, and \emph{entangling
measurement} (EM) noise.
\Cref{fig:pymatching_circuit_level_noise_volume_plot,fig:beliefmatching_circuit_level_noise_volume_plot} show the performance of DCCCs when decoded with MWPM and belief matching, respectively. 
We can draw the following conclusions:

\begin{keypoint}
For all three auxiliary qubit circuit noise models, when decoding with MWPM, it is better to only measure $XX$ and $ZZ$, while when decoding with belief matching, it is better to measure $XX$, $YY$, and $ZZ$.
\end{keypoint}

\begin{keypoint}
The noise bias in the SI noise model is not strong enough to make repeating measurements worthwhile.
\end{keypoint}

\begin{keypoint}
For EM noise and decoding using MWPM the best DCCC is $X^2Z^2$.
\end{keypoint}

We consider this a key point because DCCCs are especially well suited for EM noise.
Namely, for this noise model, it has been shown that the teraquop footprint of the $X^1 Y^1 Z^1$ code is smaller than the teraquop footprint of the surface code implemented with pair measurements~\cite{gidney2023pair,paetznick2023performance}.

\subsubsection{Remainder of this work}\label{subsubsec:remainder-of-this-work}

\Cref{sec:DCCCs} contains background information on DCCCs.
If you are familiar with DCCCs, detectors, logical errors, and decoding graphs, you can skip \Cref{sec:DCCCs}.
In \Cref{sec:teraquop_volume} we introduce and explain the teraquop volume, a key performance metric which we use to evaluate the performance of QEC protocols.
The remainder of this work
is dedicated to understanding the main results and explaining how these results have
been obtained.
We first focus on the results comparing the $X^1 Z^1$ and $X^1 Y^1 Z^1$ codes in \Cref{sec:choosing-a-schedule}.
Following this we discuss how and why repeating measurement influences spacetime cost in \Cref{sec:gauge_dcccs}.
We finish this work with a summary and outlook in \Cref{sec:summary_and_outlook}.
The source code used to produce the results in this paper can be found at \\
\url{https://github.com/peter-janderks/floquet_colour_codes_numerics}.

\begin{table}[htbp]
    \renewcommand{\arraystretch}{1.2}
    \scalebox{0.9}{
        \begin{tabularx}{\textwidth}{l|X}
            \hline
    %  \multicolumn{2}{c}%{\textbf{Section \ref{sec:DCCCs}}} \\
            \hline
            \makecell[lt]{$X^a Y^b Z^c$ code \\$a,b,c \in \mathbb{Z}_{>0}$} & DCCC whose measurement schedule consists of $a$ consecutive $XX$ measurements, $b$ consecutive $YY$ measurements, and $c$ consecutive $ZZ$ measurements on certain qubits. \\
            \makecell[lt]{$X^a Z^b$ code\\$a,b,c \in \mathbb{Z}_{>0}$} & DCCC whose measurement schedule consist of $a$ consecutive $XX$ measurements and $b$ consecutive $ZZ$ measurements on certain qubits.\\
            \makecell[lt]{Teraquop\\volume} & The amount of qubits times the number of measurement rounds at which the probability of any logical error occurring is $10^{-12}$.  \\
            MWPM & A decoder that uses the minimum-weight perfect-matching algorithm.\\
            \makecell[lt]{Belief\\matching} &  A decoder that uses belief propagation and the minimum-weight perfect-matching algorithm.\\
            \makecell[lt]{Direct parity\\measurement noise} & A circuit noise model in which single-qubit Pauli errors can occur before measurement rounds with probabilities $p_X$, $p_Y$ and $p_Z$, and measurement errors can occur with probability $p_m$. Assumes native two-qubit parity measurements.\\
            \makecell[lt]{Measurement\\error bias} & $\frac{p_m}{p_X + p_Y + p_Z}$, assuming $p_X, p_Y, p_Z$ and $p_m$ are chosen such that the physical error rate is kept constant at $10^{-3}$.\\
            \makecell[lt]{$Z$ error\\bias} & $\frac{2p_Z}{p_X+p_Y}$, assuming $p_X, p_Y, p_Z$ and $p_m$ are chosen such that the physical error rate is kept constant at $10^{-3}$.\\
            \makecell[lt]{Auxiliary qubit\\circuit noise} & A circuit noise model in which a $q$-qubit Pauli error can occur with probability $p_g$ after each gate $g$ on $q$ qubits, and measurement errors can occur with probability $p_m$. Multi-qubit measurements are decomposed using auxiliary qubits.\\
            \makecell[lt]{Standard\\depolarizing noise} & An auxiliary qubit circuit noise model that assumes multi-qubit measurements are implemented by single-qubit measurements on auxiliary qubits, and all error probabilities are equal.\\
            \makecell[lt]{Superconducting-\\inspired noise} & An auxiliary qubit circuit noise model that assumes multi-qubit measurements are implemented by single-qubit measurements on auxiliary qubits, and the error probabilities model those experienced by real superconducting hardware.\\
            \makecell[lt]{Entangling\\measurement noise} & An auxiliary qubit circuit noise model that assumes multi-qubit measurements are implemented directly, and all error probabilities are equal.\\
            \hline
        \end{tabularx}
    }
    \caption{
        Glossary for terms used in Figs.\  \ref{fig:phenomenological_noise_volume_plot_pymatching}, \ref{fig:phenomenological_noise_volume_plot_beliefmatching}, \ref{fig:pymatching_circuit_level_noise_volume_plot}, and \ref{fig:beliefmatching_circuit_level_noise_volume_plot}.
        More information on the noise models is given in \Cref{subsec:noise_models}.}
    \label{tab:glossary}
\end{table}

\begin{figure}[p]
    \centering
    \includegraphics[width=0.9\columnwidth]{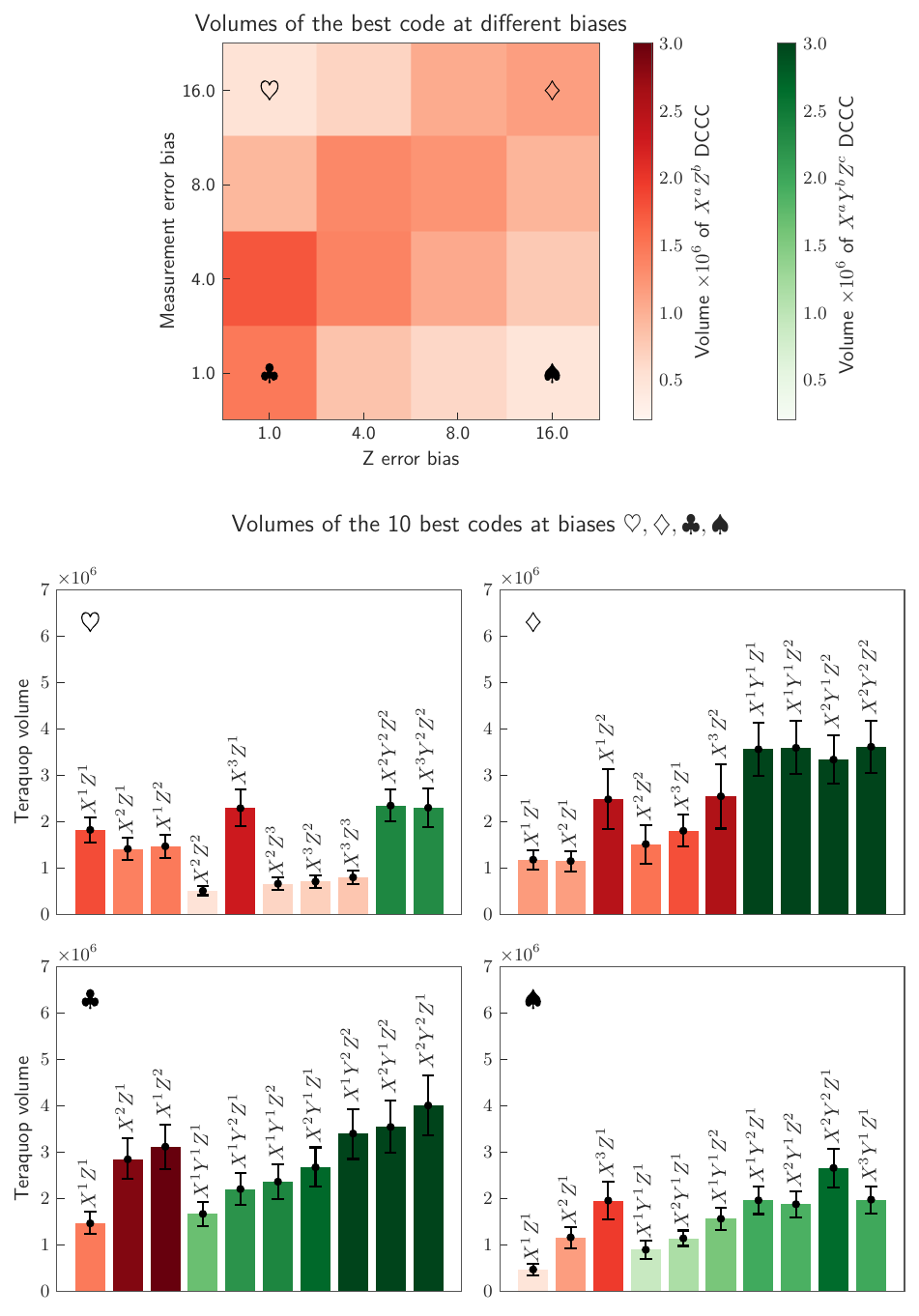}
    \caption{Squares in the top patchwork plot are red (green) if the code with the smallest teraquop volume is an $X^aZ^b$ code ($X^aZ^bY^c$ code).
        A darker shade indicates a higher volume.
        The four bar plots show the footprints of the 10 best codes at the four limits of the simulated biases.
        The error bars are explained in \Cref{sec:methodology}.
        Decoding is performed using MWPM.
        For all noise models investigated here, the $X^a Z^b$ code outperforms the $X^a Y^b Z^c$ code.
        Repeating measurements improves performance when the noise is biased strongly towards measurement errors only, and has negligible effect otherwise.
        Codes are ordered first by type ($X^a Z^b$ code, then $X^a Y^b Z^c$ code), then by increasing schedule length ($a+b$ or $a+b+c$). 
        Codes with equal schedule length are ordered by balance, with more uniformly distributed repetition parameters appearing first.
        For noise model details, definitions, and naming conventions see Table \ref{tab:glossary}.
     }
    \label{fig:phenomenological_noise_volume_plot_pymatching}
\end{figure}

\begin{figure}[p]
    \centering
    \includegraphics[width=0.9\columnwidth]{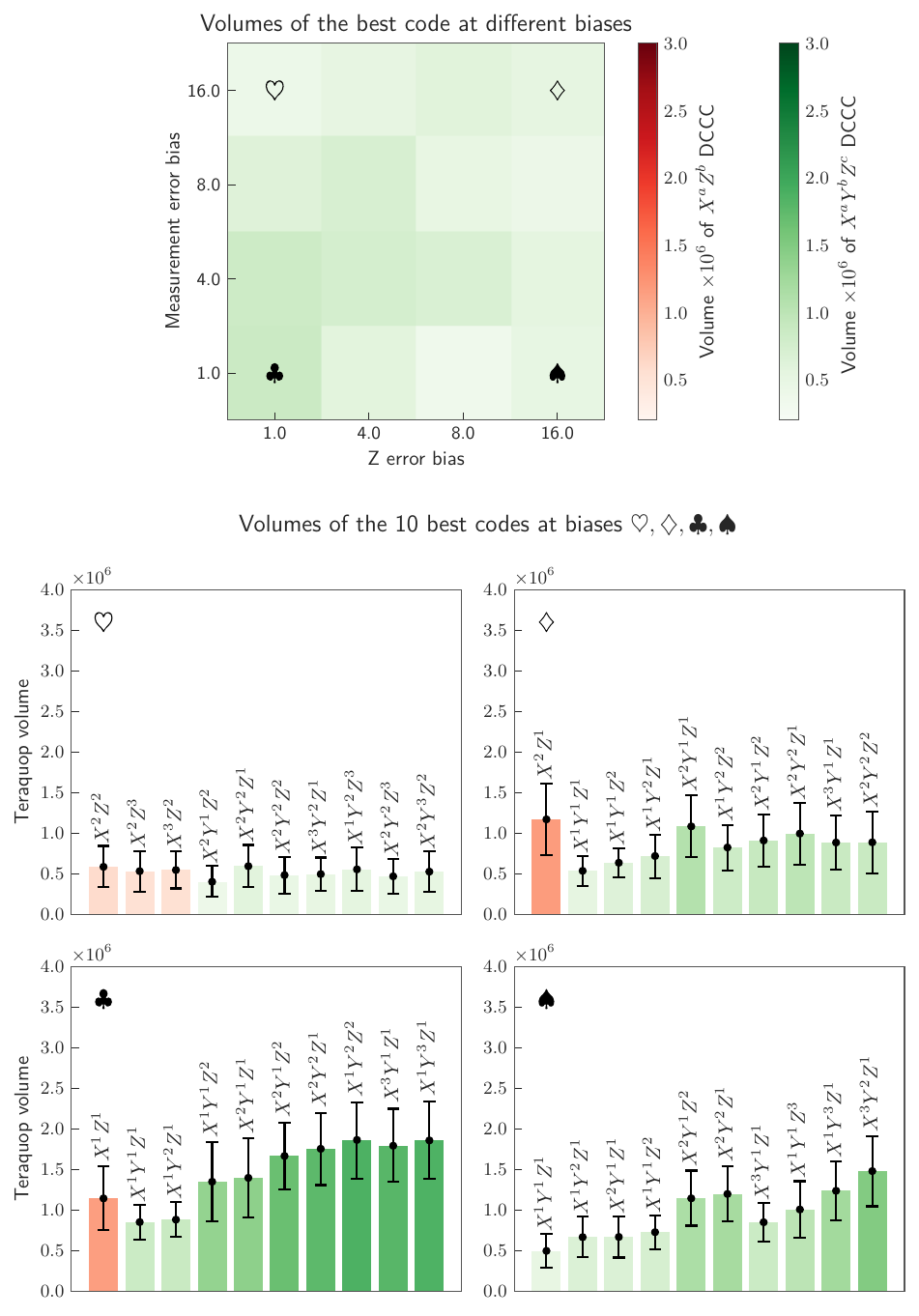}
    \caption{
        Same plot as in \Cref{fig:phenomenological_noise_volume_plot_pymatching} but with decoding performed using belief matching.
        In contrast to \Cref{fig:phenomenological_noise_volume_plot_pymatching}, for all noise models investigated here, the $X^a Y^b Z^c$ code outperforms the $X^a Z^b$ code.
        However, as before, repeating measurements improves performance when the noise is biased strongly towards measurement errors only, but has negligible effect otherwise.
    }
    \label{fig:phenomenological_noise_volume_plot_beliefmatching}
\end{figure}

\FloatBarrier

    \section{Preliminaries}\label{sec:DCCCs}

Here we recap \textit{dynamically condensed colour codes} (DCCCs) \cite{kesselring2022anyon} and how such codes can be used to detect and correct errors. DCCCs are a type of \textit{dynamic stabilizer code}. We refer those interested in a more general introduction to the latter topic to \Ccite{fu2024errorcorrectiondynamicalcodes, Townsend_Teague_2023}. We will assume the reader is comfortable with the stabilizer formalism.
Otherwise, we recommend \Ccite{NielsenChuang,GottesmanQECBook2024} for a basic introduction to this topic. 

DCCCs can be defined in terms of the colour code's \textit{boson table},
a $3 \times 3$ grid with columns labelled by colours $c, m$ and $y$ (cyan, magenta, and yellow)
and rows labelled by letters $X, Y$ and $Z$ (non-identity Pauli matrices).
Cells of the table are thus colour-Pauli pairs $(\kappa, P)$.
Often we will want to talk about generic colours $\{\kappa, \kappa', \kappa''\} = \{c, m, y\}$
and generic Pauli matrices $\{P, P', P''\} = \{X, Y, Z\}$,
for which we will use a boson table with these generic labels instead:
\begin{equation}\label{eq:cc_boson_table}
\bosonTable{
    rx={\colourful{c}, X},
    gx={\colourful{m}, X},
    bx={\colourful{y}, X},
    ry={\colourful{c}, Y},
    gy={\colourful{m}, Y},
    by={\colourful{y}, Y},
    rz={\colourful{c}, Z},
    gz={\colourful{m}, Z},
    bz={\colourful{y}, Z},
}
\qquad \text{or} \qquad
\bosonTable{
    rx={\kappa, P},
    gx={\kappa', P},
    bx={\kappa'', P},
    ry={\kappa, P'},
    gy={\kappa', P'},
    by={\kappa'', P'},
    rz={\kappa, P''},
    gz={\kappa', P''},
    bz={\kappa'', P''},
}
\end{equation}

A DCCC can be defined on any three-face-colourable, three-valent 2D lattice.
Edges of such a lattice inherit a canonical colouring;
an edge separating faces of colours $\kappa$ and $\kappa'$ is assigned the third colour $\kappa''$.
On every vertex of the lattice we place a qubit.
Throughout this work we use the honeycomb lattice.
We also assume throughout that this lattice is placed on a torus;
we comment on the limitations of this in \Cref{subsec:boundaries}.

\begin{figure}[htbp!]
    \centering
     \includegraphics[width=180pt]{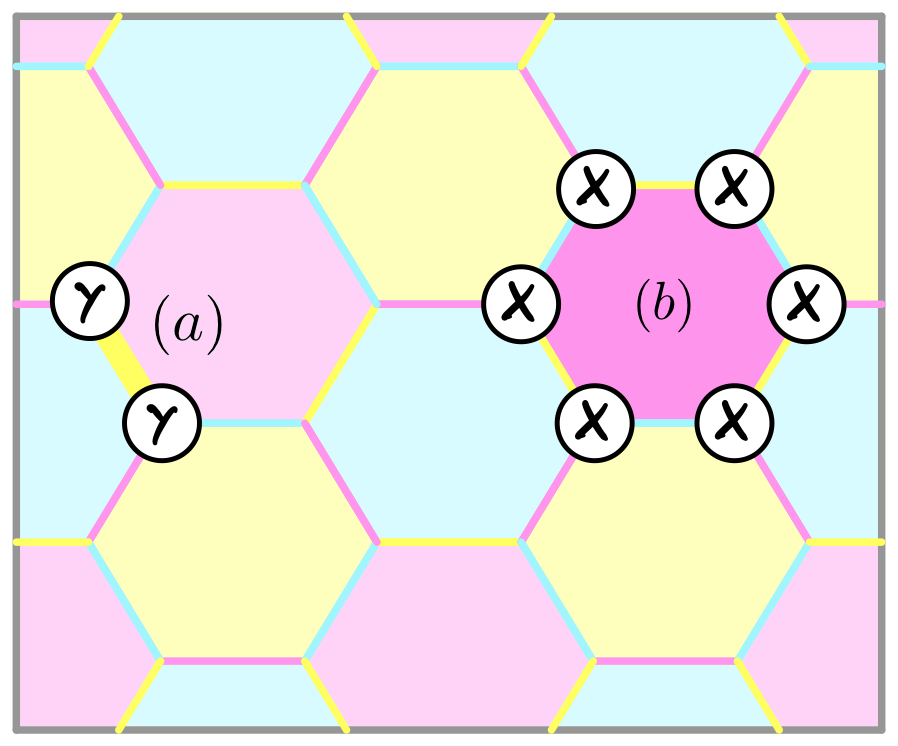}
    \caption{
        $(a)$ A $Y^{\otimes 2}$ yellow edge operator $e_{\textcolor{Dandelion}{y}, {Y}}$.
        $(b)$ An $X^{\otimes 6}$ magenta face operator $f_{\textcolor{magenta}{m}, {X}}$.
    }
    \label{fig:edge_and_face_operator}
\end{figure}

For every cell $(\kappa, P)$ of the boson table, we can define certain operators on the lattice.
Namely, for every $\kappa$-coloured edge $e_\kappa = (v_0, v_1)$,
we can define a \textit{$(\kappa, P)$ edge operator} $\dcccOp{e}{\kappa}{P} \defeq P_{v_0} P_{v_1}$,
and the set of all such edge operators as $\dcccOp{E}{\kappa}{P}$.
Similarly, for every $\kappa$-coloured face $f_\kappa$ with vertices $v_0, v_1, \ldots, v_5$
we can define a \textit{$(\kappa, P)$ face operator} $\dcccOp{f}{\kappa}{P} \defeq P_{v_0} P_{v_1} \ldots P_{v_5}$,
and the set of all such face operators as $\dcccOp{F}{\kappa}{P}$. 
Note that each face operator $\dcccOp{f}{\kappa}{P}$ can be made from edge operators in two ways; as the product of the three $(\kappa', P)$ edge operators around $f$, or the three $(\kappa'', P)$ edge operators.
Additionally, it can be written as the product of the other two face operators $\dcccOp{f}{\kappa}{P'}$ and $\dcccOp{f}{\kappa}{P''}$ on $f$, up to global phase.
Examples of such operators are shown in \Cref{fig:edge_and_face_operator}.

\subsection{Measurement schedule}\label{subsec:measurement_schedule}

A DCCC, then, is a dynamic stabilizer code defined by
repeatedly measuring sets $\dcccOp{E}{\kappa}{P}$ of all $(\kappa, P)$ edge operators.
We say it has \textit{measurement schedule} $\left(
    \langle \dcccOp{E}{\kappa_0}{P_0} \rangle,
    \langle \dcccOp{E}{\kappa_1}{P_1} \rangle,
    \ldots
    \right)$.
For DCCCs as defined in \Ccite[Section 7]{kesselring2022anyon}, there is a single restriction:
if at time $t$ the operators $\dcccOp{E}{\kappa_t}{P_{t}}$ are measured,
then at time $t+1$ we must choose a new set $\dcccOp{E}{\kappa_{t+1}}{P_{t+1}}$ of edges to measure,
such that cell $(\kappa_{t+1}, P_{t+1})$ shares neither a row nor column with $(\kappa_t, P_t)$ in the boson table.
Later, in \Cref{sec:gauge_dcccs}, we relax this restriction and investigate DCCCs with repeated measurements -- i.e. we allow $\dcccOp{E}{\kappa_{t+1}}{P_{t+1}} = \dcccOp{E}{\kappa_t}{P_t}$.
But until then, the two DCCCs we will focus on are
the \textit{$X^1Y^1Z^1$ honeycomb code} from \Ccite{hastings2021dynamically} and the \textit{$X^1Z^1$ honeycomb code} from 
\Ccite{kesselring2022anyon} \footnote{The $X^1Y^1Z^1$ honeycomb code is equivalent to the Hastings-Haah code up to conjugating certain qubits by Cliffords.}.

These are defined by the measurement schedules
\begin{equation}\label{eq:honeycomb_codes_measurement_schedules}
    \begin{aligned}
        \MMM_{X^1Y^1Z^1} &= \left(
            \langle \dcccOp{E}{c}{X} \rangle,
            \langle \dcccOp{E}{m}{Y} \rangle,
            \langle \dcccOp{E}{y}{Z} \rangle
        \right),\\
        \MMM_{X^1Z^1} &= \left(
            \langle \dcccOp{E}{c}{X} \rangle,
            \langle \dcccOp{E}{m}{Z} \rangle,
            \langle \dcccOp{E}{y}{X} \rangle,
            \langle \dcccOp{E}{c}{Z} \rangle,
            \langle \dcccOp{E}{m}{X} \rangle,
            \langle \dcccOp{E}{y}{Z} \rangle
        \right).
    \end{aligned}
\end{equation}
Much of this work is concerned with the properties of these two codes and their performances under different noise models and decoders.

\begin{figure}[h!]
    \centering
    \includegraphics[width=400pt]{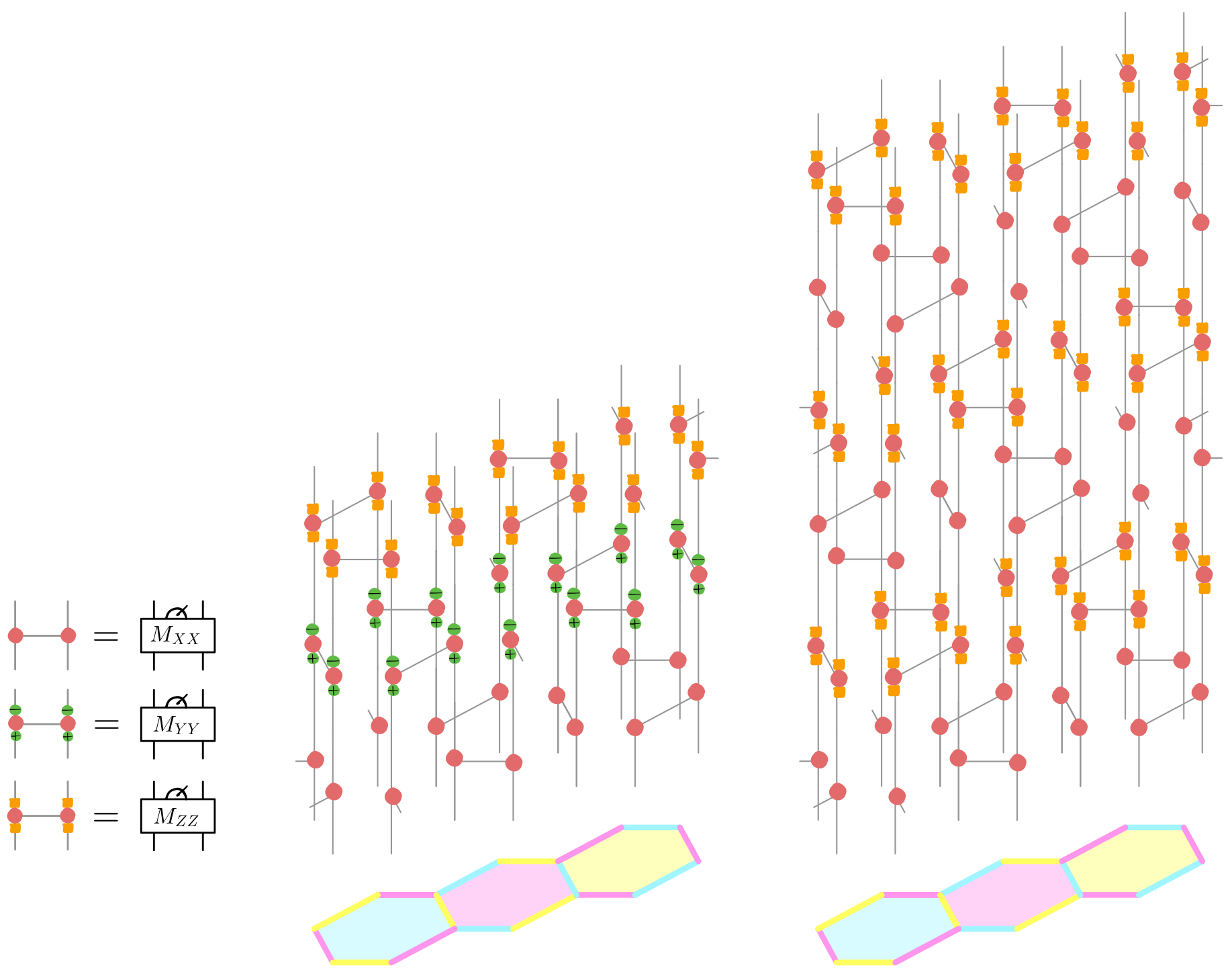}
    \caption{
        For a subset of qubits, we show the sequence of measurements defining the $X^1 Y^1 Z^1$ (middle) and $X^1 Z^1$ (right) honeycomb codes.
        This figure uses the ZX-calculus~\cite{Coecke_2011}, a rigorous graphical language for quantum mechanics.
        For those unfamiliar with the ZX-calculus, a crash course is given in \Cref{sec:zx-calculus-and-pauli-webs},
        with proper introductions available in \Ccite{vandewetering2020zxcalculusworkingquantumcomputer,KissingerWetering2024Book}.
        Alternatively, you're absolutely free to ignore these diagrams' rigorous meaning and treat them as a visual guide to the algebra in the text -- much in the same way as \Cref{fig:edge_and_face_operator} has no rigorous meaning.
        In which case, all you need to know for now is that the diagrams are read bottom-to-top and the three diagrams on the left of the figure denote measurements of $X \otimes X$, $Y \otimes Y$ and $Z \otimes Z$ respectively.
    }
    \label{fig:schedules_XYZ_and_CSS}
\end{figure}

\subsection{Stabilizers and logical operators}\label{subsec:stabilizer_and_logicals}

\begin{figure}[b]
    \centering
    \includegraphics[width=\linewidth]{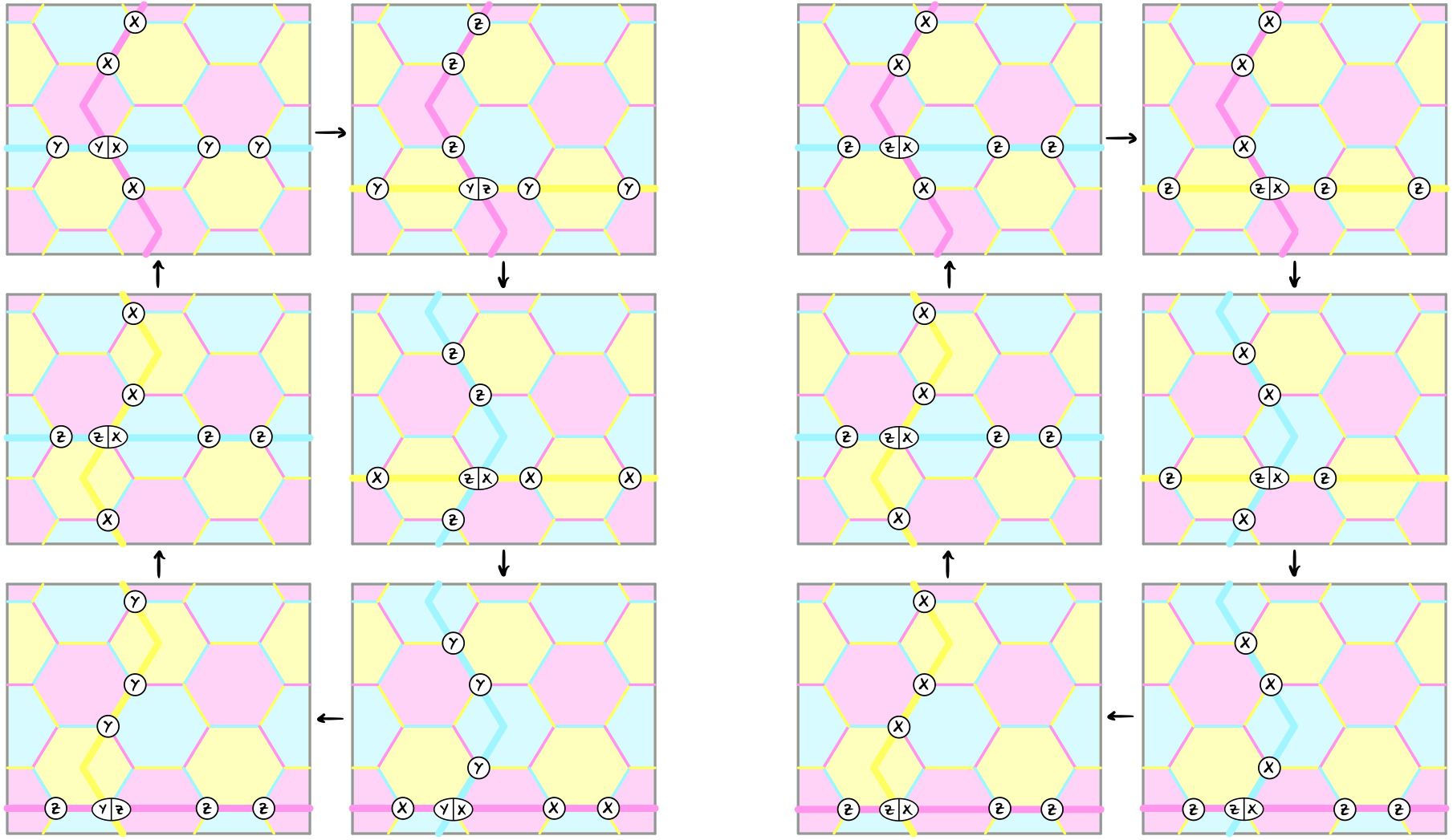}
    \caption{
        Representatives of logical $E$ and $M$ operators for the $X^1 Y^1 Z^1$ (left) and $X^1 Z^1$ (right) honeycomb codes at timesteps $t$ mod $6$.
        Timestep $0$ mod $6$ is the top left subfigure in each cycle.
        Boundaries are periodic, forming a torus.
        Throughout this work we say a DCCC encodes only one logical qubit - we just ignore the other one.
        We arbitrarily choose that the $E$ logical operator wraps vertically around the torus in this figure, with the $M$ operator wrapping around horizontally.
        Importantly, the representative of the $E$ operator in the $X^1 Z^1$ code always consists of Pauli $Z$ operators, and the $M$ operator always consists of Pauli $X$ operators.
        In the $X^1 Y^1 Z^1$ code, in contrast, the representatives of both the $E$ and $M$ operator consist one third of the time of Pauli $X$ operators, one third of the time of $Y$ operators and one third of the time of $Z$ operators.
        This will be important when discussing the effects of biased noise later on.}
    \label{fig:logicals_XYZ_and_XZ}
\end{figure}

In any dynamic stabilizer code on $n$ qubits, once a measurement schedule has been defined, two further sequences can be derived;
the \textit{instantaneous stabilizer groups} (ISGs) $\SSS = (\SSS_{-1}, \SSS_0, \SSS_1, \ldots)$
and the \textit{instantaneous logical Pauli groups} (ILPGs) $\LLL = (\LLL_{-1}, \LLL_0, \LLL_1, \ldots)$.
These respectively track -- at each timestep $t$ -- which Paulis are stabilizers of our code,
and which Paulis are representatives of logical Pauli operators.
Explicit presentations for each $\SSS_t$ and $\LLL_t$ can be found via the stabilizer formalism \cite[Section 6.2]{GottesmanQECBook2024}.
We let $r_t$ denote the rank of $\SSS_t$.
The group $\LLL_t$ is defined to be $\lpglong{t}$, the normaliser of $\SSS_t$ in the Pauli group, modulo $\SSS_t$.
Every $\LLL_t$ is isomorphic to the Pauli group $\PPP_{k_t}$ on $k_t \coloneq n - r_t$ qubits~\cite[Theorem 3.11]{GottesmanQECBook2024}.
If there is a timestep $T$ such that for all $t \geq T$ we have $r_t = r_T \eqqcolon r$, then we say the code is \textit{established}.
Correspondingly, for such $t \geq T$, we get $k_{t} = k_T \eqqcolon k$.
We thus say such a code encodes $k$ logical qubits.

Every DCCC on a torus has an establishment time $T$, after which it encodes $k=2$ logical qubits.
Alternatively, we can always just ignore one of the logical qubits and pretend it only encodes one qubit; this makes our work more straightforward to compare with the more easily implementable planar case, which encodes just one logical qubit, and allows us to tailor the code dimensions to biased noise models later on.
In the rest of this work, we will do exactly this.
In \Cref{fig:logicals_XYZ_and_XZ}, we show representatives of logical $X$ and $Z$ operators for the $X^1Y^1Z^1$ and $X^1Z^1$ honeycomb codes at timesteps $t$ mod $6$.
For reasons we discuss shortly (\Cref{subsec:E_and_M_components}), we will actually call these $E$ and $M$ logical operators rather than $X$ and $Z$.

\subsection{Noise models}\label{subsec:noise_models}

Before we can discuss how DCCCs are used to detect and correct errors, we need to meet the errors in question.

\subsubsection{Direct parity measurement noise model}
In a quantum circuit that implements a DCCC -- one that repeatedly performs the defining sequence of measurements $\MMM = (\groupPres{\dcccOp{E}{\kappa_0}{P_0}}, \groupPres{\dcccOp{E}{\kappa_1}{P_1}}, \ldots)$ --
in addition to the $n$ \textit{data qubits},
we may require additional \textit{auxiliary qubits} in order to perform the measurements.
In the \textit{direct parity measurement noise model},
we assume that two-qubit parity measurements (such as $XX$, $YY$, $ZZ$) can be performed natively.
We assume that at every timestep $t$,
before a generating set for $\MMM_t$ is measured,
any of the $n$ data qubits can be afflicted by a single-qubit Pauli error $X$, $Y$, or $Z$ with probability $p_X$, $p_Y$, or $p_Z$ respectively.
We call these \textit{data qubit errors}.
We also assume that every measurement can be erroneously reported with probability $p_m$ --
i.e.\ the actual measurement outcome is $\pm 1$, but the reported measurement outcome is $\mp 1$.
We call this a \textit{measurement error}, or \textit{measurement noise}.
A noise model is \textit{biased} unless all of these probabilities are equal.
Tailoring DCCCs to be effective against biased noise models will be a key theme of this work.
The two types of bias we will focus on are \textit{measurement bias} and \textit{$Z$ bias}.
In the direct parity measurement noise model, the measurement bias is defined as
$\eta_m = p_m/(p_X + p_Y + p_Z)$,
and the $Z$ bias is defined as
$\eta_Z = 2p_Z/(p_X+p_Y)=p_Z/p_X$, assuming $p_X=p_Y$.

In any circuit we keep the total physical error rate $p$ fixed as
\begin{equation}\label{eq:physical_error_rate_defn}
    p = \frac{p_m M + (p_X + p_Y + p_Z) Q} {M + 3Q}
\end{equation}
where $M$ is the number of measurements in the circuit and $Q$ is the number of data qubit noise locations.
To generate 
\Cref{fig:phenomenological_noise_volume_plot_beliefmatching,fig:phenomenological_noise_volume_plot_pymatching}, we have used $p = 10^{-3}$.

\begin{figure}[p]
  \centering
    \begin{subfigure}[c]{0.14\textwidth}
        \centering
        \includegraphics[width=0.8\textwidth]{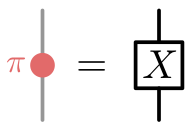}
        \caption{}
        \label{fig:ZX_X_gate}
    \end{subfigure}
    \begin{subfigure}[c]{0.14\textwidth}
        \centering
        \includegraphics[width=0.8\textwidth]{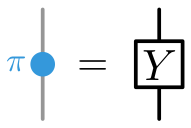}
        \caption{}
        \label{fig:ZX_Y_gate}
    \end{subfigure}
    \begin{subfigure}[c]{0.14\textwidth}
        \centering
        \includegraphics[width=0.8\textwidth]{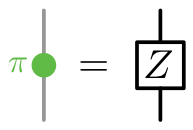}
        \caption{}
        \label{fig:ZX_Z_gate}
    \end{subfigure}
    \begin{subfigure}[c]{0.27\textwidth}
        \centering
        \includegraphics[width=0.8\textwidth]{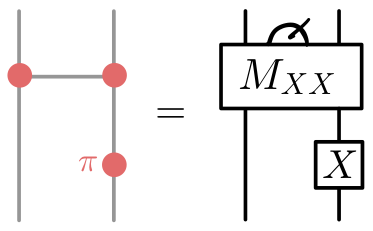}
        \caption{}
        \label{fig:ZX_XX_meas_X_error}
    \end{subfigure}
    \begin{subfigure}[c]{0.27\textwidth}
        \centering
        \includegraphics[width=0.8\textwidth]{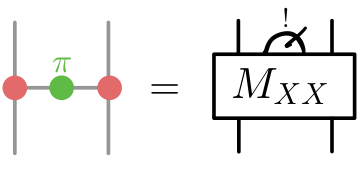}
        \caption{}
        \label{fig:ZX_XX_meas_error}
    \end{subfigure}
    \caption{
        Subfigures (a), (b) and (c) show the non-identity Pauli gates as ZX-diagrams.
        Subfigures (d) and (e) show examples of errors in a phenomenological noise model.
        Specifically, (d) shows an $X$ error occurring before an $e_{\kappa, X}$ measurement,
        while (e) shows a measurement error on an $e_{\kappa, X}$ measurement.
        Indeed, throughout this work, a measurement error on any $e_{\kappa, P}$ measurement will correspond in the ZX-calculus to a two-legged green $\pi$-spider on a `horizontal' wire.
    }
    \label{fig:phenom_error_examples}

    \vspace{20pt}
  
  \begin{subfigure}[b]{0.2\textwidth}
        \centering
        \includegraphics[width=0.6\textwidth]{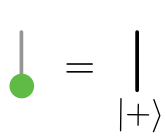}
        \caption{}
        \label{fig:ZX_plus_init}
    \end{subfigure}
    \begin{subfigure}[b]{0.2\textwidth}
        \centering
        \includegraphics[width=0.6\textwidth]{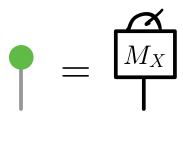}
        \caption{}
        \label{fig:ZX_X_meas}
    \end{subfigure}
    \begin{subfigure}[b]{0.3\textwidth}
        \centering
        \includegraphics[width=0.7\textwidth]{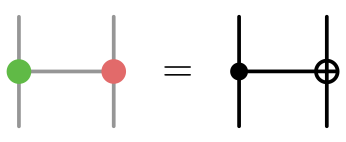}
        \caption{}
        \label{fig:ZX_CNOT}
    \end{subfigure}
    \begin{subfigure}[b]{0.45\textwidth}
        \centering
        \includegraphics[width=0.7\textwidth]{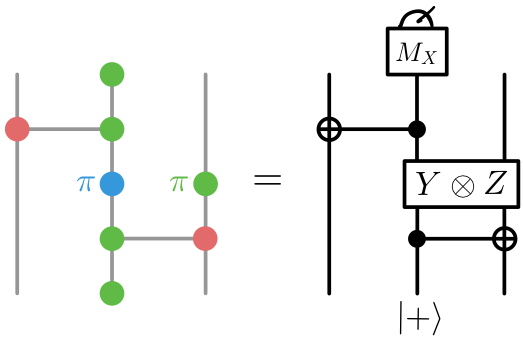}
        \caption{}
        \label{fig:ZX_XX_meas_circuit_YZ_error}
    \end{subfigure}
    \begin{subfigure}[b]{0.45\textwidth}
        \centering
        \includegraphics[width=0.7\textwidth]{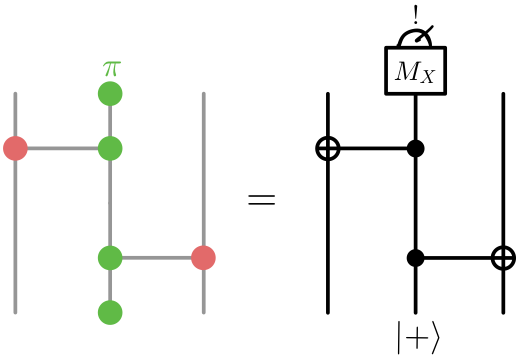}
        \caption{}
        \label{fig:ZX_XX_meas_circuit_meas_error}
    \end{subfigure}
    \caption{
        Subfigures \textbf{(a)}, \textbf{(b)} and \textbf{(c)} show the $\ket{+}$ state, $X$ measurement and $CX$ gate respectively as ZX-diagrams.
        Subfigures \textbf{(d)} and \textbf{(e)} show examples of errors in an auxiliary qubit circuit noise model,
        where two-qubit gates and a single-qubit measurement are used to measure $e_{\kappa, X}$.
        Specifically, \textbf{(d)} shows a $Y \otimes Z$ error occurring immediately after a $CX$ gate,
        while \textbf{(e)} shows a measurement error on an $X$ measurement.
    }
    \label{fig:circuit_level_error_examples}

    \vspace{20pt}
  
    \begingroup
    \renewcommand*{\arraystretch}{1.2}
        \centering
        \begin{tabular}{|p{30mm}|p{60 mm}|p{10mm}|p{10mm}|p{10mm}|}
            \hline
            Operation & Noise applied after operation & SD & SI & EM3 \\
            \hline
            \hline
             CX, CY, CZ & $\{I,X,Y,Z\}^{\otimes 2} - \{I \otimes I\}$ & $p$ & $p$ & -  \\
            \hline
            Init. $\ket{+}$ & $Z$ & $p$ & $2p$ & $p$ \\
            Init. $\ket{i}$ & $Y$ & $p$ & $2p$ & $p$ \\
            Init. $\ket{0}$ & $X$ & $p$ & $2p$ & $p$ \\
            \hline
            Meas. $X, Y, Z$ & bit-flip &  $p$ & $5p$ & $p$ \\
            \hline
            Idle & $X,Y,Z$ & $p$ & $p/10$ & $p$ \\
            \hline
            Resonator idle* & $X,Y,Z$ & 0 & $2p$ &$0$ \\
            \hline
        $M_{PP}$** & $\{I, X, Y, Z\}^{\otimes 2} \times $ \{bit-flip, no-flip\} & - & - & $p$ \\
    
        \hline
    
        \end{tabular}
        \captionof{table}{Exact details of the noise models used for numerical simulations. The leftmost column contains all types of operations that are applied in the circuits. The second column lists which errors can occur after a gate. If multiple errors are given, one is chosen uniformly at random. 
        *Resonator idle refers to a circuit location during which a qubit is not measured or reset in a time step during which other qubits are being measured or reset. **With the $M_{PP}$ operation the two-qubit Pauli errors are applied before the measurement.}
        \label{tab:noise_table}
    \endgroup
\end{figure}

\subsubsection{Auxiliary qubit circuit noise model}

An \textit{auxiliary qubit circuit noise model} decomposes multi-qubit measurements into elementary gates using auxiliary qubits, in contrast to the direct parity measurement noise model which assumes native two-qubit parity measurements.
In an auxiliary qubit circuit noise model,
we ignore the distinction between data and auxiliary qubits.
Instead, we just assume that after every gate $g$ on $q$ qubits, any weight-$q$ Pauli operator can occur on these qubits with some probability $p_g$.
We continue to assume that measurement noise can occur.
At this point, the actual circuits used to implement the DCCCs become relevant.
We consider two different families of circuits, \emph{single-qubit measurement circuits} and \emph{two-qubit measurement circuits}.
In the first, the measurements of edge operators $\dcccOp{e}{\kappa}{P}$ are implemented via auxiliary qubits, two-qubit gates, and single-qubit Pauli measurements.
In the second, the measurements of edge operators are implemented `natively' -- no auxiliary qubits are used.

The single-qubit measurement circuits are simulated with two different auxiliary qubit circuit noise models: \textit{standard depolarizing} (SD) and \textit{superconducting-inspired} (SI).
The SI model is taken from \Ccite{gidney2021fault}. 
There is a slight difference between our SI model and \textit{SI1000} in \Ccite{gidney2021fault}, namely that in the SI1000 noise model it is assumed that the only two-qubit gate used is $CZ$, while the circuits here contain $CX$, $CY$, and $CZ$ gates.
The two-qubit measurement circuits are simulated with a different auxiliary qubit circuit noise model referred to as \textit{entangling measurements} (EM) noise.
The EM noise model is exactly the EM3 model taken from \Ccite{gidney2022benchmarking}.
The different gates, errors, and probabilities of each model are shown in \Cref{tab:noise_table}.

\subsection{Detectors}\label{subsec:detectors}

We detect and correct errors using \textit{detectors}
\cite{Gidney2021stimfaststabilizer,derks2024}.
A detector is a set of measurement outcomes $\{m_1, \ldots, m_b\}$ whose product is deterministic in the absence of noise.
In a stabilizer code, detectors are particularly simple;
they consist of the outcomes of measuring the same stabilizer twice at consecutive timesteps. In dynamic stabilizer codes, we will not necessarily measure the same operators at consecutive timesteps,
so detectors can look more complicated. 
However, they are still easy to find using the stabilizer formalism;
one is created anytime we measure a Pauli operator $p$ such that
$p$ or $-p$ is a stabilizer for the code at a particular timestep.
If we ever find that the product $m_1 \ldots m_b$ differs from the expected value,
then we can conclude that some errors must have affected our system.
We say that such a detector has been \textit{violated} by these errors.
% and we say that the detectors violated by a set of errors are the \textit{syndrome} of these errors.
Often we'll write a detector $\{m_1, \ldots, m_b\}$ as a formal product $m_1 \ldots m_b$ with powers taken mod 2.\footnote{%
    By formal product, we mean we forget that symbols like $m_j$ are actually stand-ins for values $-1$ and $1$,
    and treat the symbols just as objects to be moved around algebraically.
    By taking powers mod 2, we mean -- for example -- the formal product $m^3$ is the same as $m^1 =  m$.
    It seems only ${\sim}16.7\%$ of the authors of this work think formal products are the best way to write detectors.
    If you prefer linear algebra,
    you can also think of a detector as a binary vector -- see \Ccite{derks2024}.
    Or you can continue to just think of them as sets of measurement outcomes.}

\begin{figure}[t]
    \centering
    \includegraphics[width=0.8\linewidth]{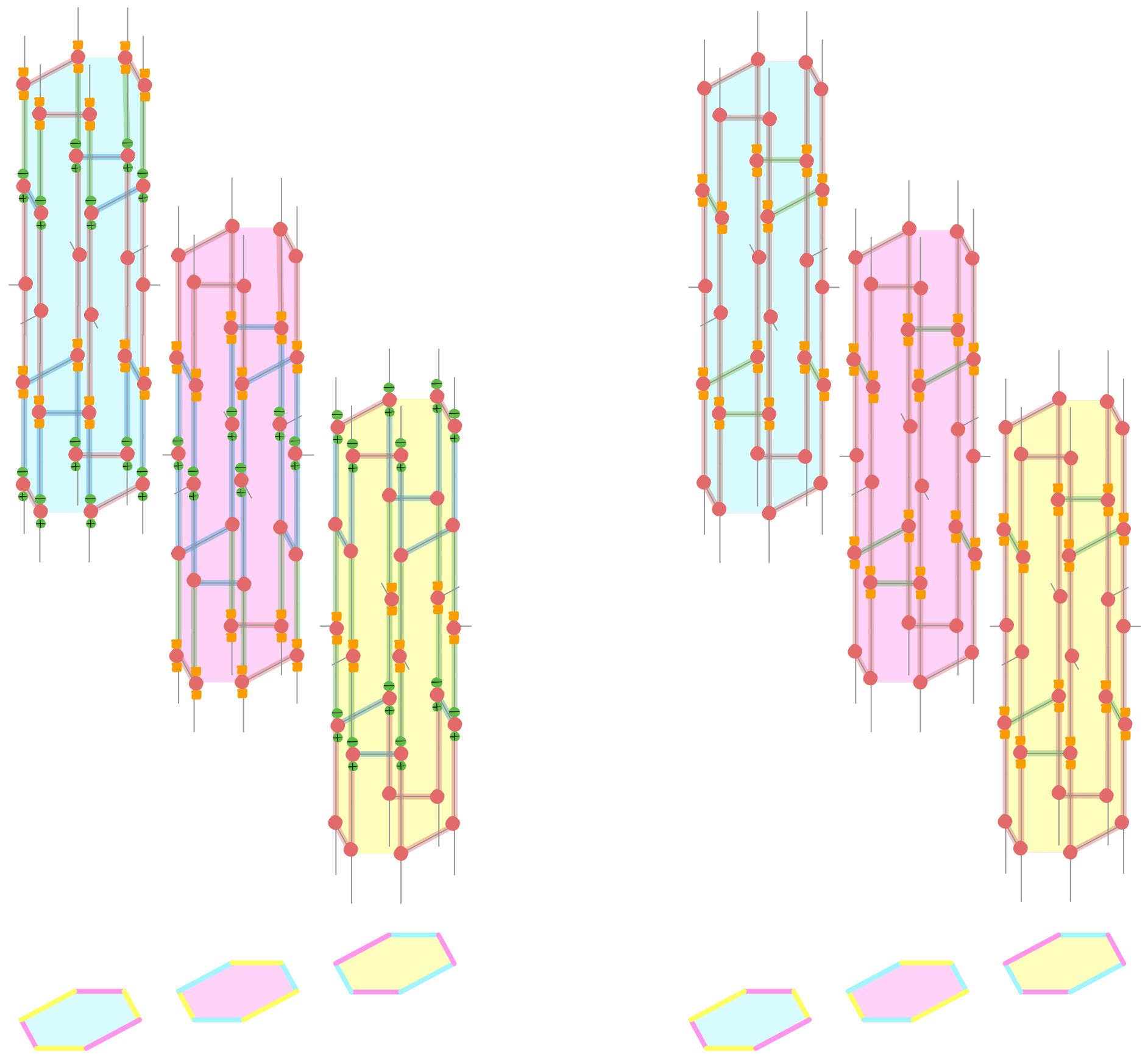}
    \caption{
        A subset of detectors of the $X^1 Y^1 Z^1$ code (left) and the $X^1 Z^1$ code (right).
        Certain wires in this ZX-diagram are highlighted red, green or blue;
        they constitute a \textit{Pauli web}, which we go into in more detail in \Cref{sec:zx-calculus-and-pauli-webs}.
        In essence, one can think of a Pauli web as telling us which types of errors violate this detector at which timesteps.
        Pauli errors and measurement errors are represented in the ZX-calculus as 2-legged $\pi$-spiders (\Cref{fig:phenom_error_examples,fig:circuit_level_error_examples}).
        A detector is violated by a set of errors if there's an odd number of them of a different colour to the highlighted wire they sit on (see \Cref{fig:XYZ_measurement_error}, for example).
        Pauli webs also tell us which measurements are actually involved in a detector and which are not; it's all mesurements corresponding to non-vertical edges highlighted red or blue.
    }
    \label{fig:detectors_XYZ_and_CSS}
\end{figure}

Let's now look at DCCC detectors.
We let $m_t(\dcccOp{e}{\kappa_t}{P_t})$ denote
the outcome of measuring the edge operator $\dcccOp{e}{\kappa_t}{P_t}$ at time $t$,
and let $e \in f$ mean edge $e$ borders face $f$.
DCCCs have detectors\footnote{%
    This statement only holds if we make a distinction between detectors and \textit{observables}.
    An observable here doesn't have the usual meaning of a self-adjoint operator.
    Instead it refers to a special detector that's used to test whether the decoder correctly decoded the errors in a circuit~\cite{Gidney2021stimfaststabilizer,derks2024}.
    There are indeed observables that are not of this form.}
of the form $d = \prod_{t \in \tau} \prod_{e_{\kappa_t} \in f_{\kappa}} m_t(\dcccOp{e}{\kappa_t}{P_t})$
for some set of timesteps $\tau$ and some $\kappa$-coloured face $f_\kappa$.
For example, on the left of \Cref{fig:detectors_XYZ_and_CSS} we show a detector in the $XYZ$ honeycomb code,
and on the right a detector in the CSS honeycomb code,
both associated to some yellow face $f_{\colourful{y}}$.
As a formal product, with timesteps taken modulo 6, the former is written as
\begin{equation}\label{eq:XYZ_detector_yellow}
    \prod_{e_{\colourful{m}} \in f_{\colourful{y}}} m_4(\dcccOp{e}{m}{Y})
    \prod_{e_{\colourful{c}} \in f_{\colourful{y}}} m_3(\dcccOp{e}{c}{X})
    \prod_{e_{\colourful{m}} \in f_{\colourful{y}}} m_1(\dcccOp{e}{m}{Y})
    \prod_{e_{\colourful{c}} \in f_{\colourful{y}}} m_0(\dcccOp{e}{c}{X}),
\end{equation}
while the latter is
\begin{equation}\label{eq:CSS_detector_yellow}
    \prod_{e_{\colourful{m}} \in f_{\colourful{y}}} m_4(\dcccOp{e}{m}{X})
    \prod_{e_{\colourful{c}} \in f_{\colourful{y}}} m_0(\dcccOp{e}{c}{X}).
\end{equation}

In \Cref{fig:detectors_XYZ_and_CSS} the detectors are shown graphically via \textit{Pauli webs}, as explained in more detail in the caption.

\subsection{Decoding graph}\label{subsec:decoding_graph}

\begin{figure}[htbp]
    \centering
    \includegraphics[width=0.8\textwidth]{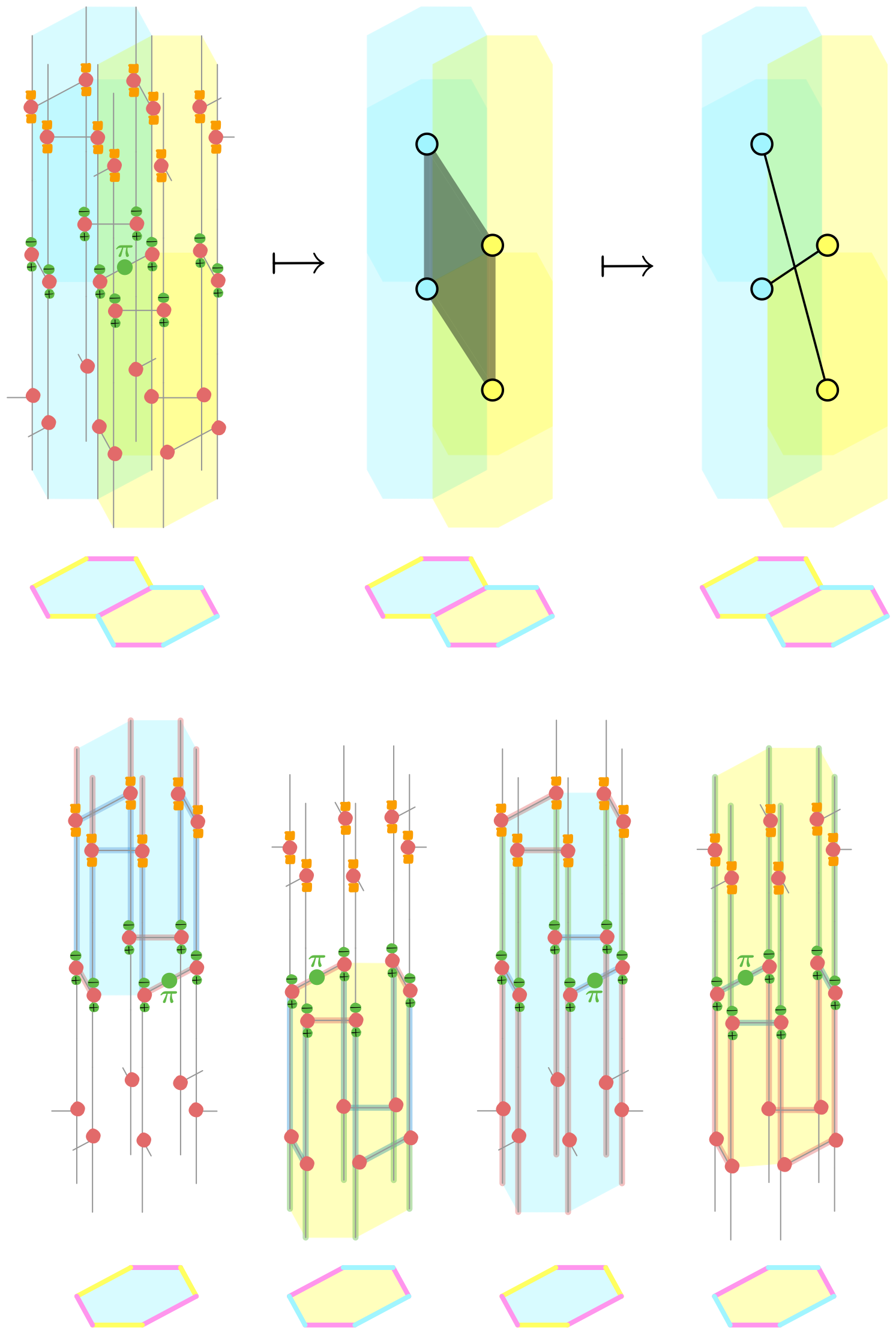}
    \caption{
        An $e_{\textcolor{magenta}{m}, Y}$ measurement error in the $X^1Y^1Z^1$ honeycomb code, which violates four detectors.
        In the top left subfigure, we show the error as a ZX-diagram.
        Since it violates four detectors, it corresponds to a hyperedge in the decoding graph (top middle).
        This hyperedge then decomposes into two ordinary edges (top right).
        The longer of these two edges in the subfigure is a remnant edge,
        because there is no actual error in the noise model that corresponds to it,
        unlike the shorter of the two edges.
        The bottom subfigures show in more detail the four detectors that this error violates.
        One can check that these really are violated by noting that, in each case, the edge on which the green $\pi$-spider representing the error sits is highlighted a colour other than green,
        as explained in \Cref{fig:detectors_XYZ_and_CSS} and in more detail in \Cref{sec:zx-calculus-and-pauli-webs}.}
    \label{fig:XYZ_measurement_error}
\end{figure}

\begin{figure}[htb]
    \centering
    \includegraphics[height=0.55\textheight]{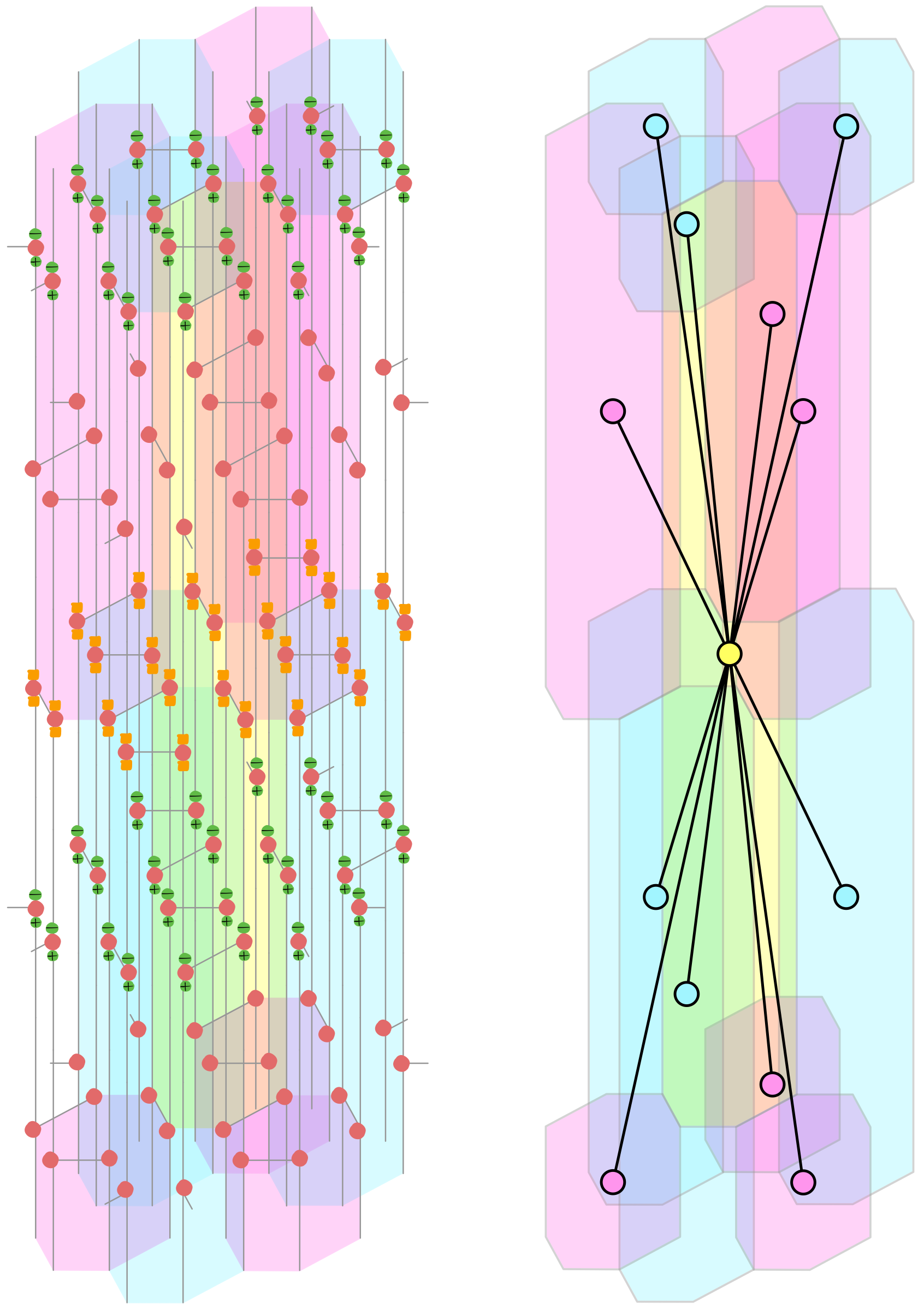}
    \caption{
        On the left we show five timesteps from the $X^1 Y^1 Z^1$ code.
        Areas coloured cyan, magenta and yellow denote detectors, though to make the diagram simpler we have not drawn the Pauli webs that define them (see \Cref{sec:detectors-cheatsheet} for a detector cheat sheet).
        On the right we show the neighbourhood of the yellow detector in the decoding graph, once all hyperedges have been decomposed into ordinary edges.
        This detector has degree 12, as does every other detector (apart from a few special cases right at the very start and end of the circuit).
        Crucially, this is also the case for the $X^1 Z^1$ code -- their decoding graphs after decomposition are isomorphic, differing only in the edge weights, i.e.\ in the probabilities of the corresponding errors occuring.}
    \label{fig:decoding_graph_face_detector_neighbourhood}
\end{figure}

The \textit{decoding graph} of a circuit implementing a quantum error-correcting code, under a given noise model, is a weighted hypergraph formed from the detectors of the code and the errors of the noise model.
The vertices of this hypergraph are exactly the detectors of the code.
For every possible independent error $e$ permitted by the noise model, letting $d_1, \ldots, d_b$ be the detectors it violates
and $p_e$ be the probability of it occurring,
a hyperedge also labelled $e$ is added to the graph, spanning vertices $\{d_1, \ldots, d_b\}$ and having weight $p_e$. 
As written, there may be multiple hyperedges spanning the same set of $\{d_1, \ldots , d_b\}$ of vertices;
if this is the case, they can be combined into a single hyperedge spanning this set of vertices,
whose weight $p$ is the probability of an odd number of the corresponding errors occurring.
We will alternatively write a hyperedge $e = \{d_1, \ldots, d_b\}$ as a formal product $e = d_1 \ldots d_b$ with powers mod 2.
As an example, in \Cref{fig:XYZ_measurement_error} we show a measurement error in the $X^1Y^1Z^1$ honeycomb code.
This violates four detectors, hence corresponds to a hyperedge spanning four vertices in the decoding graph\footnote{What we're calling a decoding graph is often called a \textit{Tanner graph}, though Tanner graphs are usually defined as bipartite graphs, with one type of vertex representing detectors, another type representing errors, and an edge between two vertices exactly if the corresponding error violates the corresponding detector. 
The two definitions are equivalent. The term \textit{detector error model} is also equivalently used \cite{gidney2022stim_error_model}.}.

Many of the best known decoding algorithms, however, can only be run on \textit{ordinary} graphs --
ones in which every hyperedge spans exactly $2$ vertices.
Decoders that require this form of graph are called \textit{matching decoders},
and include \textit{minimum-weight perfect matching} (MWPM) 
\cite{MWPM,dennis2002topological,MWPMColor}
and \textit{union-find} \cite{PhysRevResearch.2.033042,UnionFind}. 
We can transform the decoding graph into an ordinary graph, at the cost of the edge weights becoming only approximations of the corresponding errors occurring.

We start by `decomposing' errors that violate more than $2$ detectors into sets of errors that violate at most $2$ detectors.
For example, if the decoding graph contains a hyperedge $d_1 d_2 d_3 d_4$ with weight $p$,
but also a hyperedge $d_1 d_2$ with weight $p_{12}$ and a hyperedge $d_3 d_4$ with weight $p_{34}$,
then we can delete the hyperedge $d_1 d_2 d_3 d_4$ and update the weights $p_{12}$ and $p_{34}$ in some way.
In fact, even if the graph only contains hyperedges $d_1 d_2 d_3 d_4$ with weight $p$ and $d_1 d_2$ with weight $p_{12}$, 
we can delete $d_1 d_2 d_3 d_4$, update $p_{12}$ and add a new \emph{remnant edge} $d_3 d_4$ with some weight $p_{34}$.
In this work, we use the Clifford circuit simulator Stim~\cite{Gidney2021stimfaststabilizer} extensively,
and Stim's implementation of decomposing errors is described in \Ccite{stim_command_line_doc} and explained in \Ccite{gidney2022stim_error_model}.
The way in which the edge weights are updated is explained in Appendix C of \Ccite{higgott2022fragile}.
After decomposing all such hyperedges, we add a new virtual \emph{boundary vertex} $\partial$, and we replace every hyperedge spanning a single vertex $d$ with a new ordinary edge with endpoints $d$ and $\partial$ \cite{higgott2025sparse}.
At this point, every hyperedge will span exactly two vertices.

\subsection{Logical errors}\label{subsec:DCCCs_logical_errors}

\begin{figure}[htpb]
    \centering
    \includegraphics[width=\linewidth]{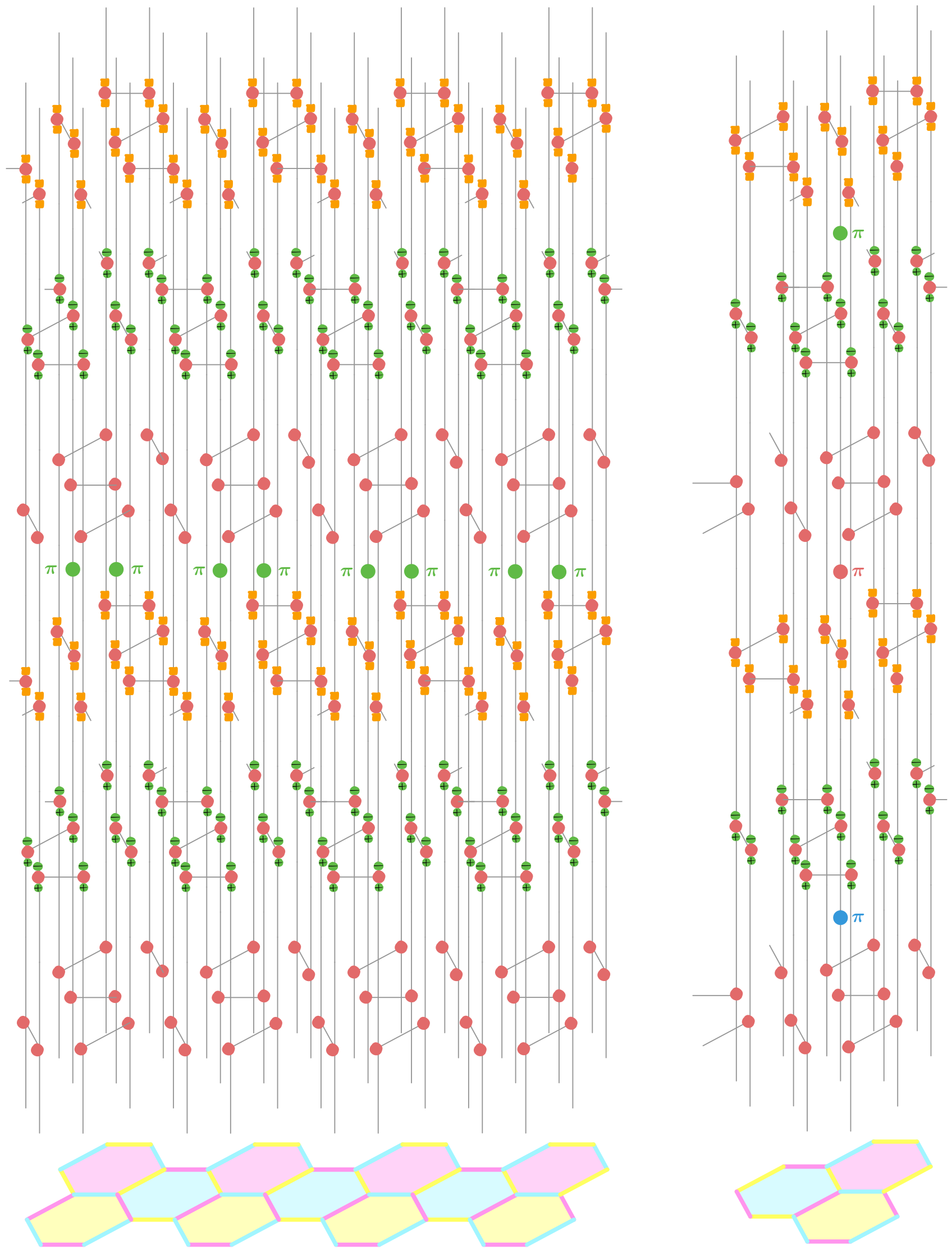}
    \caption{
        On a subset of qubits, we show a spacelike (left) and timelike (right) logical error in the XYZ honeycomb code consisting only of data qubit errors.
        \href{https://tinyurl.com/bdd7er9s}{Click here to view the spacelike} and 
        \href{https://tinyurl.com/ypwn6hca}{here to view the timelike} logical errors in
        Crumble~\cite{crumble}, a
        prototype tool for Clifford circuit exploration.}
\label{fig:XYZ_logical_data_qubit_errors}
\end{figure}

If a set of errors do not violate any detectors,
but act non-trivially on the circuit implementing our code
(i.e.\ the circuit with errors is not equal as a linear map to the circuit without errors, even up to global scalar)
then this set constitutes a \textit{logical error}.
In a DCCC, we can further label logical errors as \textit{spacelike} or \textit{timelike}.
Intuitively, spacelike logical errors are those consisting of sets of errors wrapping around the torus on which the code is defined (in either direction).
Intuitively, timelike logical errors are those consisting of sets of errors that span from the start of the circuit to the end.
For example, if after some timestep $t$ a set of qubit errors occur whose tensor product is a logical Pauli operator for the code at this timestep $t$ -- like those depicted in \Cref{fig:logicals_XYZ_and_XZ} -- then this constitutes a spacelike logical error.
In \Cref{fig:XYZ_logical_data_qubit_errors}, we show one spacelike and one timelike logical error in the $X^1Y^1Z^1$ honeycomb code.

In any particular circuit and noise model,
the minimum number of errors required to create a logical error is defined to be the \textit{fault distance} of the circuit.
One can be more specific and speak of the \textit{spacelike fault distance} or \textit{timelike fault distance} --
the minimum number of errors required to create a spacelike or timelike logical error, respectively.
% Often we will refer to the \textit{distance} rather than the fault distance --
% this is just the fault distance in the specific case of the phenomenological noise model.

\subsection{Decoding}\label{subsec:decoding}

The decoding graph contains everything we need to be able to decode a quantum error-correcting code.
In general, decoding works as follows: as we run the circuit implementing the quantum error-correcting code,
we mark the vertices $d$ that correspond to violated detectors, then pass this marked graph to a decoder.
It is the job of the decoder to find a set of hyperedges (i.e., errors) satisfying two properties:
they could have caused these detectors to be violated,
and the combination of these errors and the actual errors that occurred do not constitute a logical error.
More precisely, let $e_1, \ldots, e_c$ be the actual errors that occurred,
which necessarily have formal product $e_1 \ldots e_c$ equal to the syndrome $d_1 \ldots d_b$.
The decoder's job is to find a set $e_1', \ldots, e_{c'}'$ of hyperedges such that the formal product $e_1' \ldots e_{c'}'$ is again $d_1 \ldots d_b$,
and the set of errors $\{e_1, \ldots, e_c, e_1', \ldots, e_{c'}'\}$ is not a logical error.

In this work, we consider two decoders: \textit{minimum-weight perfect matching (MWPM)} and \textit{belief matching}~\cite{higgott2022fragile}.
These are both matching decoders, so require the decoding graph to be an ordinary graph (or decomposable into an ordinary graph).
In MWPM, we simply apply the transformation described in \Cref{subsec:decoding_graph} to the decoding graph (i.e.~we convert the hypergraph into a graph).
Then, on any given run of the circuit, we apply the MWPM decoder to the resulting marked ordinary graph~\cite{higgott2025sparse}.
In belief matching, we add a pre-processing step which first adjusts the weights of the hyperedges in the marked decoding graph.
This pre-processing step is the classical decoding algorithm \textit{belief propagation} (BP)~\cite{McKay,Leifer,PhysRevResearch.2.043423}, which is run on the full decoding hypergraph to estimate the marginal probability of each error mechanism (whether it is an edge or hyperedge).
These marginal probabilities are then used to update the edge weights when converting to a graph for MWPM~\cite{higgott2022fragile}.
The details of exactly what this does are unimportant for us,
but the key intuition is that it recovers some of the information about the probabilities of errors occurring that's otherwise lost when transforming hyperedges into ordinary edges (i.e.~it exploits knowledge of the hyperedge errors when decoding a given syndrome).
This allows for more accurate decoding for error models containing hyperedges,
at the cost of slowing down the overall decoding process.
For those interested, a longer discussion of this trade-off can be found in \Ccite{higgott2022fragile}.
After running BP, we apply the transformation described in \Cref{subsec:decoding_graph} to the marked decoding graph,
and apply MWPM to the resulting marked ordinary graph.
A key line of investigation in this work is understanding for which DCCCs and noise models using belief matching brings most benefit.

\subsection{\texorpdfstring{$E$ and $M$ components}{E and M components}}\label{subsec:E_and_M_components}

In any DCCC, after applying the transformation described in \Cref{subsec:decoding_graph} to the decoding graph,
the new ordinary graph $G$ has the property that $G - \{\partial\}$ consists of two disconnected components,
where $\partial$ is the boundary vertex we added.
We will call one of these components the $E$ (electric) component and the other the $M$ (magnetic) component.
Then many of our other objects of interest inherit a label.
Detectors corresponding to vertices in the $E$ or $M$ component are labelled $E$ or $M$ detectors respectively.
Errors whose corresponding hyperedge in the original decoding graph spans vertices entirely in the $E$ or $M$ component of $G - \{\partial\}$
are labelled $E$ or $M$ errors, respectively.
Errors whose corresponding hyperedge spans vertices in both the $E$ and $M$ components are labelled $F$ (fermion) errors\footnote{%
    This $E$, $M$ and $F$ terminology is borrowed from the condensed matter approach, as in \Ccite{kesselring2022anyon}.
    We could equally have used the more standard $X$, $Z$ and $Y$ respectively, but we felt this could ultimately cause more confusion -- e.g. in the $X^1 Y^1 Z^1$ honeycomb code, some physical Pauli $X$ errors correspond to $E$ errors, some to $M$ errors, and some to $F$ errors.
    We felt it might be confusing to say that a Pauli $X$ error was actually a $Z$ error or a $Y$ error -- indeed, we confused ourselves many times in this project with this sort of terminology.}.
Spacelike logical errors that -- intuitively speaking -- wrap around the $E$ or $M$ component are labelled spacelike logical $E$ or $M$ errors respectively.
Timelike logical errors likewise inherit an $E$ or $M$ label.
Finally, the fault distances can be further divided into a spacelike $E$ and $M$ fault distance -- the minimum number of errors needed to create a spacelike logical $E$ or $M$ error, respectively -- and a timelike $E$ and $M$ fault distance -- the minimum number of errors needed to create a timelike logical $E$ or $M$ error, respectively.

\subsection{Boundaries}\label{subsec:boundaries}

As stated throughout this section, in this work we only consider DCCCs on a torus. Primarily, this is to make comparisons between different measurement schedules fair, since there are no boundaries at which they might behave differently. Helpfully, it also makes our analysis a great deal simpler. On the other hand, if one is ultimately interested in planar codes (e.g.\ because of hardware constraints) one might very reasonably question to what extent a code's performance in the toric setting is a good proxy for its performance in the planar setting.

First, we note that a general theory of boundaries for DCCCs is given in \Ccite[Section 7.3]{kesselring2022anyon}, so every DCCC can indeed be made planar. Furthermore, for the $X^1 Y^1 Z^1$ and $X^1 Z^1$ codes that we focus on in this work, specific boundaries are defined in \Ccite{gidney2022benchmarking} and \Ccite{kesselring2022anyon} respectively. In the former work, it's shown numerically that the planar $X^1 Y^1 Z^1$ code performs essentially as well as the toric version. In the latter work, no numerical results were given, but the detectors at the boundary were fully worked out and depicted. Since these follow exactly the same pattern as the detectors in the bulk, we again anticipate that performance of this planar $X^1 Z^1$ code would essentially mirror that of its toric counterpart, but have not verified this.

    \section{Teraquop volume}\label{sec:teraquop_volume}

Before we start looking at particular DCCCs in detail, we first introduce the key 
metric we use to evaluate performance throughout this work.
It is the natural generalisation of the \emph{teraquop footprint} \cite{gidney2021fault} from a `space only' picture to a `full spacetime' picture.
We call it the \emph{teraquop volume}.

\begin{definition}
    Consider a family of circuits implementing a quantum error correcting protocol, whose $j$-th member is defined on a number of qubits $n_j$ and runs for an amount of time $T_j$.
    Given a noise model and a decoder,
    the \textbf{teraquop volume} is the smallest volume $n_j T_j$ such that the probability of any spacelike and timelike logical error in the $j$-th circuit is $10^{-12}$.
\end{definition}

Here we deliberately do not specify what exactly a quantum error correcting protocol is and what it means for a set of circuits implementing one to form a family.
For a given circuit, noise model and decoder, the \emph{overall logical error rate} is the probability of any logical error -- whether spacelike or timelike -- occurring after the decoder has been applied.
In practice, an approximation of this value will be found numerically.
If there is no $(n_j, T_j)$ such that the overall logical error rate is less than $10^{-12}$, then we say the teraquop volume is infinite.
For a given code family, noise model, and physical error rate, the teraquop volume typically corresponds to a unique point $(n_j, T_j)$ rather than a smooth curve of equivalent space-time trade-offs.
That is, one cannot generally reduce the number of qubits and compensate by increasing the number of rounds (or vice versa) while maintaining the target logical error rate.

The teraquop volume is the appropriate metric for evaluating quantum error correction performance in lattice surgery computations because it accounts for space-time tradeoffs.
In surface code architectures, circuit depth can be reduced at the cost of increased qubit overhead, or vice versa \cite{lee2021even}.
A volume-based metric captures this interchangeability, making it possible to compare codes with different spatial and temporal resource requirements on equal footing.
If the teraquop footprint were used instead, codes with high qubit overhead but low circuit depth would be unfairly penalized, and vice versa.

In the next subsection we explain how we determine the teraquop volume of a DCCC.
\Cref{fig:teraquop_volume_summary} summarizes the concept of teraquop volume and the methodology used for its determination.

\subsection{Estimating the teraquop volume}\label{subsec:estimating-teraquop-volume}

\begin{figure}[p]
  \centering
    \includegraphics[width=0.9\linewidth]{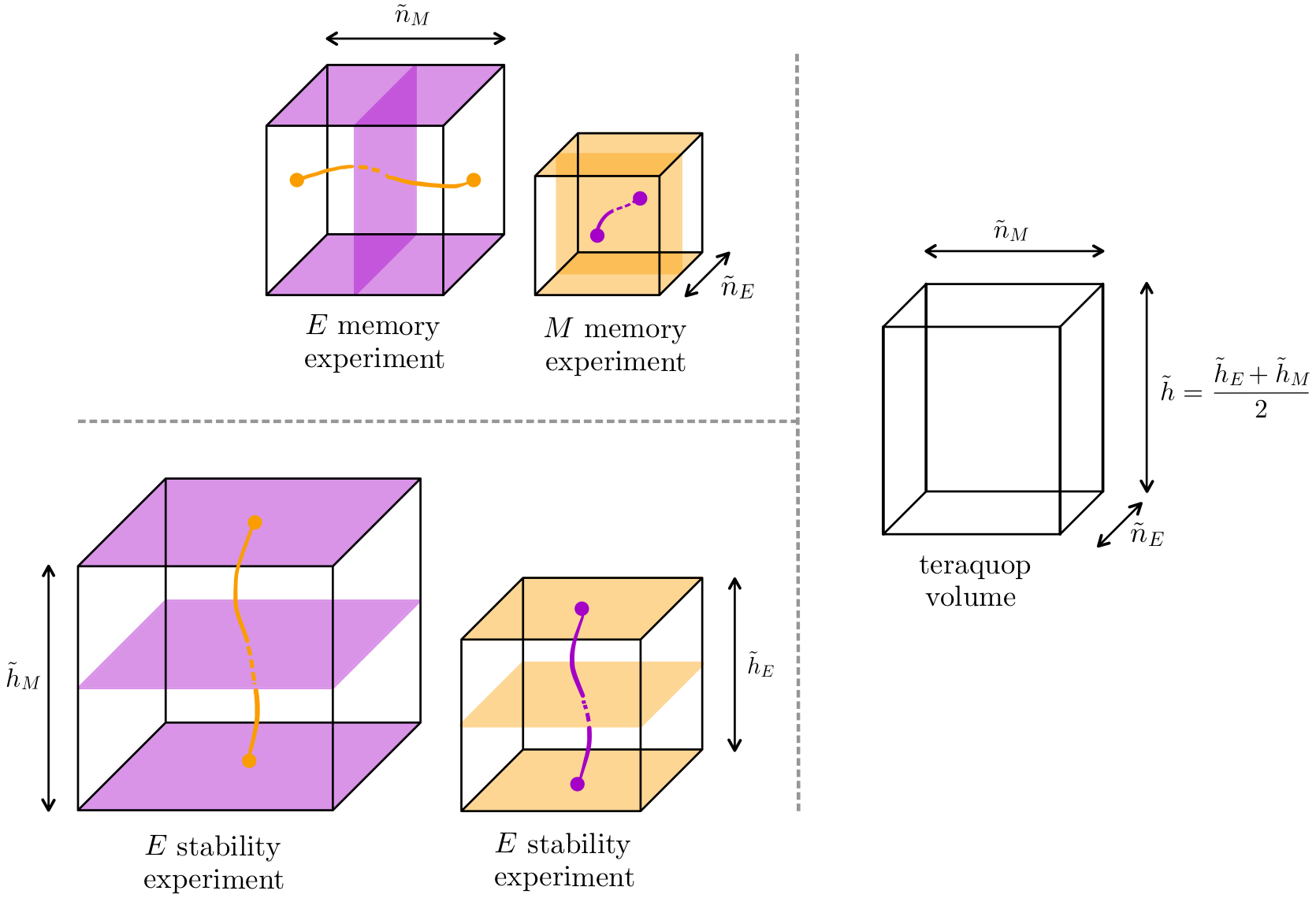}
    \caption{
        Schematic explaining how the teraquop volume of a DCCC is estimated.
        We perform $E$-memory experiments to estimate the number of columns $\tilde{n}_M$ of hexagons in any DCCC that reaches the teraquop regime. Then $M$-memory experiments estimate the number of rows $\tilde{n}_E$ required, and $E$- and $M$-stability experiments estimate the number of measurement rounds $\tilde{h}$ required.}
    \label{fig:teraquop_volume_summary}

    \vspace{10pt}
  
  \includegraphics[width=0.9\linewidth]{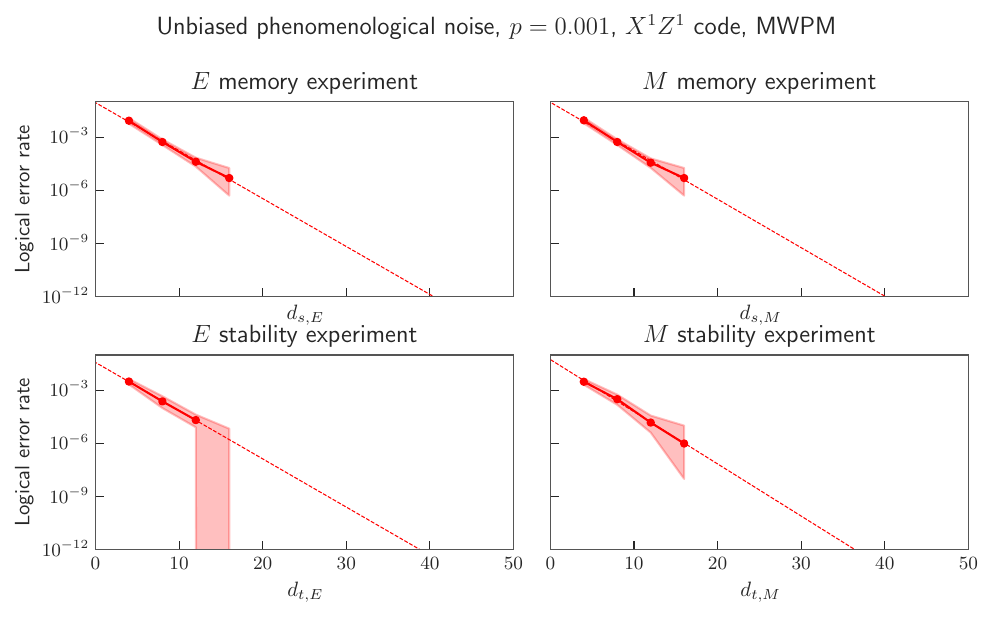}
    \caption{
        Plots showing line-fits to determine the minimal value of $n_E$, $n_M$  $h_E,$ and $h_M$ such that the probability of any logical error occurring is $10^{-12}$.
        The value of the $x$-axis wherever a dotted line would first cross the line $y = 10^{-12}$ is the value used for the corresponding teraquop dimension.
        The highlighted regions are computed with Sinter and show hypothesis probabilities within 1000× of the max likelihood hypothesis \cite{AG47_2024}.}
    \label{fig:X1Y1Z1_X1Z1_teraquop_linefits}
\end{figure}

\begin{figure}[t]
    \centering
    \includegraphics[width=\linewidth]{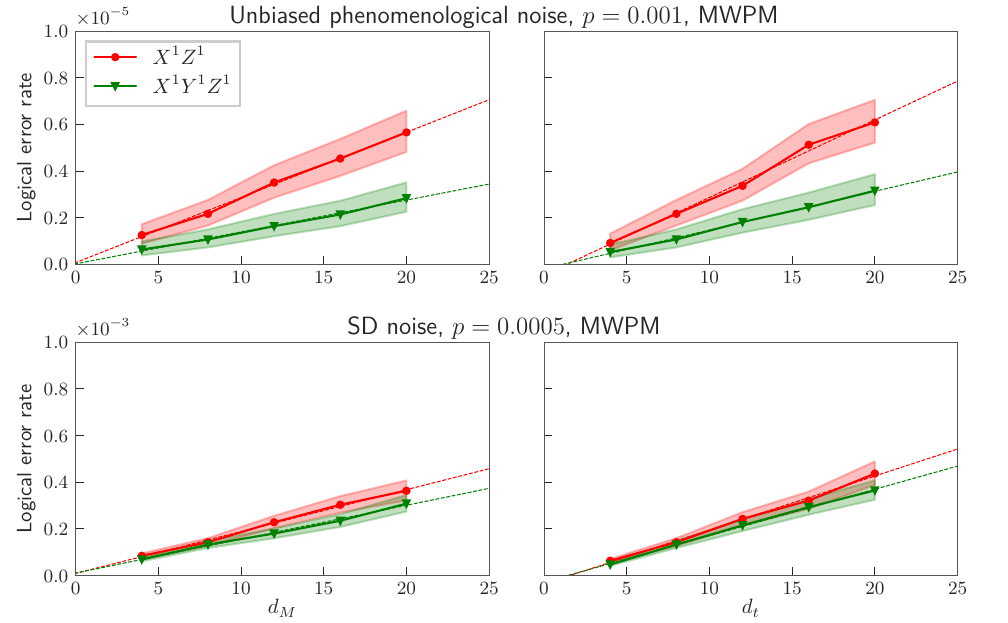}
    \caption{
        Plots showing the logical error rate of $E$ memory experiments with varying $n_M$ and $h$. Decoding is performed with MWPM.
        The plots confirm that the probability of a spacelike logical $E$ error occurring depends linearly on $n_M$ and $h$.
        The highlighted regions are computed with Sinter and show hypothesis probabilities within 1000× of the max likelihood hypothesis \cite{AG47_2024}.
    }
    \label{fig:volume_verification}
\end{figure}

Let $n_M$ and $n_E$ denote (respectively) the number of rows and columns of hexagons in a DCCC.
For example, in \Cref{fig:edge_and_face_operator,fig:logicals_XYZ_and_XZ} we showed DCCCs with $n_M = 3$ (admittedly quite wonky) rows of hexagons, and $n_E = 4$ columns.
The value of $n_M$ determines the spacelike $M$ distance $d_{s,M}$ -- namely $d_{s,M} = \frac{4}{3}n_M$.
Similarly $n_E$ determines the spacelike $E$ distance as $d_{s,E} = n_E$.

The total number of data qubits (i.e., vertices) in a DCCC with shape $(n_M, n_E)$ is $2 n_M n_E$.
The total number of auxiliary qubits varies depending on the circuits used to perform the measurements defining the code.
For example, on a heavy-hex lattice where an auxiliary qubit is placed on each edge and measurements are performed via controlled gates and single-qubit measurements, the number of auxiliary qubits is $3 n_M n_E$, for a total of $5 n_M n_E$ qubits.
If measurements are performed via two-qubit measurements directly, then no auxiliary qubits are needed, so the total number of qubits is just $2 n_M n_E$.
For a given family of circuits, we denote the total number of qubits (the \textit{footprint}) by $f(n_M n_E)$.

Given such a family, suppose we perform $h$ rounds of measurements.
In contrast to $n_M$ and $n_E$, the value of $h$ simultaneously determines both the timelike $M$ distance and the timelike $E$ distance, denoted $d_{t,M}$ and $d_{t,E}$ respectively.
Furthermore, these timelike distances depend on the specific choice of DCCC, as we will see shortly in \Cref{sec:choosing-a-schedule}.
We say the total \textit{volume} of this circuit is $f(n_M n_E) h$.
Determining the teraquop volume requires finding the minimal value of $f(n_M n_E) h$ such that the probability of any logical error occurring is $10^{-12}$.
Although this suggests a large search space, the minimal values of $n_E$, $n_M$, and $h$ can be evaluated independently to obtain a good approximation.
This is because the probability of each type of logical error occurring (spacelike $E$ or $M$, or timelike $E$ or $M$) depends exponentially on one of $n_E$, $n_M$, and $h$, and linearly on the other two.
For example, the probability of a spacelike logical $E$ error occurring depends exponentially on $n_E$ and linearly on $n_M$ and $h$.
When the minimum value of all three independently-determined dimensions are used together, each type of logical error remains well-suppressed because the exponential dependence dominates the linear scaling.
The overall logical error rate approximates $10^{-12}$, though it may be somewhat below or above this threshold by a small constant factor.

Plots verifying this linear dependence are shown in \Cref{fig:volume_verification}.
It could be the case that for different codes the probability of each type of logical error occurring depends exponentially on one parameter and non-linearly on the other two, but we have not observed this in any of the DCCCs we have studied.
If this were to occur, we expect that the approximation method we use would still give a reasonable estimate of the teraquop volume.
To verify if our approximation method is accurate, one can simulate the overall logical error rate for the estimated minimal values of $n_E$, $n_M$, and $h$ and check if it is indeed below $10^{-12}$.

To actually find these minimal values of $n_E$, $n_M$, and $h$, we perform memory and stability experiments~\cite{gidney2022stability}.
A memory experiment means initialising a logical qubit in an eigenstate of a certain logical operator,
performing the logical identity gate for some amount of time,
then measuring this logical operator.
In the absence of noise, this measurement outcome will be deterministic, but the occurrence of certain logical errors can cause the decoder to fail and return the wrong measurement outcome.
The frequency with which these failures occur is the logical error rate for this experiment.
A stability experiment instead benchmarks how reliably the product of a large region of stabilizers can be measured~\cite{gidney2022stability}.
This tests a part of a lattice surgery operation (where the product of stabilizers measured corresponds to a logical Pauli product), or a part of an operation to move a logical operator (where the region of stabilizers is multiplied into the logical operator to move it).
Whereas a memory experiment benchmarks how well a logical observable is preserved through time, a stability experiment tests how well it is preserved through space.
Both experiments test important building blocks of a lattice surgery quantum computer.

For example, in a DCCC, an $E$ memory experiment is one in which our logical qubit is initialised in the $+1$ eigenstate of the $E$ logical operator.
Its failure rate tells us how often spacelike $M$ logical errors occur,
and depends exponentially on $n_M$ but only linearly on $n_E$ and $h$.
So for variable $d$, we can perform $E$ memory experiments with $n_M$, $n_E$ and $h$ chosen such that both spacelike distances are $d$ (i.e., $d_{s,E}(n_E) = d$ and $d_{s,M}(n_M) = d$) and both timelike distances are at least $d$ (i.e., $d_{t,E}(h) \geq d$ and $d_{t,M}(h) \geq d$).
For each $d$ we obtain a logical error rate.
The minimum $n_M(d)$ at which this logical error rate goes under $10^{-12}$ is the value of $n_M$ we will use in our teraquop volume estimation.
We fit a line to the natural log of the logical error rate versus $d$ and extrapolate this line to find the required $n_M(d)$. 
\Cref{fig:X1Y1Z1_X1Z1_teraquop_linefits} contains examples of plots showing the described line fits.

We repeat the entire procedure with the roles of $E$ and $M$ swapped to find the value of $n_E$ to use in our teraquop volume estimation.
To find the value of $h$ we use, we do essentially the same thing but with stability experiments.
An $E$ stability experiment tests for protection against timelike $M$ errors, and vice versa for an $M$ stability experiment.
Plotting the logical error rates and extrapolating gives us an estimate for the numbers of rounds of measurements $h_E$ and $h_M$ needed to ensure timelike logical $E$ and $M$ errors occur with probability less than $10^{-12}$.
We then just choose the value $h$ that we use in our teraquop volume estimation to be the mean $(h_E + h_M)/2$.
By doing this we assume that places where timelike logical $E$ and $M$ errors can occur are roughly equally prevalent in a lattice surgery computation.
Additionally, we assume that logical qubits do not need to be synchronized, meaning that lattice surgery measurements in which timelike logical $E$ errors can occur can have a different duration to ones in which timelike logical $M$ errors can occur.

\subsection{Blocks}\label{teraquop-blocks}

In the rest of this work we describe a DCCC with $n_M$ rows and $n_E$ columns of hexagons implemented for $h$ rounds of measurements as an \textit{$(n_M, n_E, h)$-block} of this DCCC.
We say it has \textit{block height} $h$, \textit{block width} $n_E$ and \textit{block depth} $n_M$.
Then we'll write $\tilde{n}_M$, $\tilde{n}_E$, $\tilde{h}_M$, and $\tilde{h}_E$ for the minimum values of $n_M$, $n_E$, $h_M$, and $h_E$ such that the corresponding logical error rates are below $10^{-12}$.
An $(\tilde{n}_M, \tilde{n}_E, \tilde{h})$-block will be referred to as a \textit{teraquop block},
with \textit{teraquop height} $\tilde{h} = (\tilde{h}_E + \tilde{h}_M)/2$, and so on.

\subsection{Limitations}\label{subsec:teraquop-limitations}

The teraquop volume is a measure of the spacetime cost of the basic unit of a lattice surgery quantum computer. As with all single-number metrics, it has its drawbacks. 
One drawback is that equal cost is attributed to space and time. 
For example, we might have two (spacetime) blocks of code with equal volume but different dimensions -- one tall and narrow, one short and wide.
In a setting where qubits are limited but time is plentiful, the tall and narrow block would be the better choice.
In contrast, if qubits are plentiful but time is limited, the short and wide block should be chosen. 
This important difference between the two blocks disappears when just looking at the teraquop volumes.

A second drawback is that the teraquop volume measure does not account for optimizations achieved by altering the measurement schedule and patch size during a computation.
Consider a circuit component in which logical qubits idle for a long time. In this scenario, there are no timelike errors to worry about.
Therefore, the logical qubits should be encoded in a circuit with low teraquop \emph{footprint}, and not necessarily low teraquop \emph{volume}.

    \section{Bias-tailoring I: choosing a schedule}\label{sec:choosing-a-schedule}

In this section we discuss differences between the $X^1Z^1$ and $X^1Y^1Z^1$ codes and their impact on the teraquop volumes.
We begin by considering the simpler case of direct parity measurement noise -- the simplicity here allows us to look closely at specific error mechanisms in order to better understand the two codes -- before we move to the more realistic but messier case of auxiliary qubit circuit noise.

\subsection{Differences}\label{subsec:differences}

We first outline the major differences between the two codes, and the expected effects on the teraquop volumes.
Some of these effects are in conflict with one another -- we will see numerically in the rest of this section which features ultimately dominate performance.

\begin{difference}\label{dif:XYZ_XZ_meas_per_detector}
    Detectors in the $X^1 Y^1 Z^1$ code consist of 12 measurements, while detectors in the $X^1 Z^1$ code consist of 6 measurements (\Cref{subsec:detectors}, \Cref{fig:detectors_XYZ_and_CSS}).
\end{difference}

The intuition is that the fewer measurements involved in a detector, the better.
This is because the more measurements that are involved, the more likely the detector is to be violated,
and more violated detectors means more marked vertices in the decoding graph,
which makes the job of the decoder harder\footnote{A more accurate metric capturing this intuition is that of \textit{detector likelihood} \cite{hesner2024using}. This is the total probability of a set of errors occurring that violate a given detector. In the comparison between the $X^1 Y^1 Z^1$ and $X^1 Z^1$ codes, measurement errors are the only errors that contribute differently to the detector likelihood -- all data qubit errors affect the detector likelihood in the same way across the two codes.}.
For example, MWPM has average-case runtime that's
approximately linear in the number of violated detectors,
provided the probabilities of qubit and measurement errors occurring are not too high~\cite{higgott2025sparse}.
Indeed, this is the intuition as to why the Bacon-Shor code has no threshold~\cite{gidney2023baconthreshold}.
So this difference between the two codes would suggest the $X^1 Z^1$ code might generally perform better, and therefore have a better teraquop volume -- especially for measurement-biased noise.

\begin{difference}\label{dif:XYZ_XZ_timelike_distance}
    The $X^1 Y^1 Z^1$ code implemented for $h$ rounds has a timelike distance of $h/2$, while the $X^1 Z^1$ code implemented for $h$ rounds has a timelike distance of $h/4$ (\Cref{tab:xyz_css_summary}, \Cref{fig:repeated_measurements_timelike_distances}).
\end{difference}

In contrast to \Cref{dif:XYZ_XZ_meas_per_detector} above, this difference favours the $X^1 Y^1 Z^1$ code;
based on this alone, one would expect the $X^1 Y^1 Z^1$ code to perform better than the $X^1 Z^1$ code in stability experiments,
i.e.\ have a smaller teraquop height $\tilde{h}$.
In particular, since the weight-$h/4$ timelike logical errors in the $X^1 Z^1$ code consist entirely of measurement errors,
one would expect this to be particularly relevant for measurement-biased noise.

\begin{difference}\label{dif:XYZ_XZ_hyperedges}
    All measurement errors in the $X^1 Y^1 Z^1$ code correspond to hyperedges in the decoding graph, while no measurement errors in the $X^1 Z^1$ code correspond to hyperedges (\Cref{tab:xyz_css_summary}).
\end{difference}

Recall that hyperedges are decomposed into ordinary edges in order to use a matching decoder like MWPM,
at the cost of losing some information about the likelihood of these errors occurring (\Cref{subsec:decoding_graph}).
But belief matching recovers much of this information (\Cref{subsec:decoding}).
Therefore belief matching performs best on codes where many errors correspond to hyperedges.
Accordingly, this difference suggests the $X^1 Y^1 Z^1$ code may perform worse than the $X^1 Z^1$ code under MWPM,
but improve a lot under belief matching.

\begin{difference}\label{dif:XYZ_XZ_logicals_pauli_type}
    In the $X^1 Z^1$ code, $E$ logical errors that consist of Pauli errors can only consist of $Z$ or $Y$ errors, whereas $M$ logical errors consisting of Pauli errors can only consist of $X$ or $Y$ errors.
    In the $X^1 Y^1 Z^1$ code, both types of logical error can be formed from all types of Pauli errors.
\end{difference}

This would seem to suggest that the $X^1 Z^1$ code can be better tailored to $Z$-biased noise in memory experiments,
for the following reason.
Recall that we are considering our toric DCCCs to implement just one logical qubit, rather than two, for simplicity and for better comparison with the more widely implementable planar case (\Cref{subsec:boundaries}).
In the $X^1 Z^1$ code under a $Z$-biased noise model, spacelike $E$ logical errors become more likely to occur,
and spacelike $M$ errors become less likely to occur.
But we can in effect cancel this out by changing the code dimensions.
Since the likelihood of spacelike $E$ logical errors occurring depends exponentially on the number of columns $n_E$ of hexagons in the code, and only linearly on the number of rows $n_M$ and the number of measurement rounds $h$, we can reduce the chances of such errors occurring by increasing $n_E$.
Similarly, if we decrease $n_M$ we increase the chances of spacelike $E$ errors occurring.
Altogether one might think we should be able to define a rectangular code with $n_E > n_M$,
such that we save qubits overall while maintaining the same protection against spacelike logical errors.
In the $X^1 Y^1 Z^1$ code this is not possible, because the noise being biased towards Pauli $Z$ errors doesn't increase or decrease the probability of either type of spacelike logical error occurring.

\subsection{Direct parity measurement noise results}\label{subsec:unrepetaed-direct-parity-measurement-noise}

\begin{figure}[tbp]
    \centering
    \includegraphics[width=0.9\linewidth]{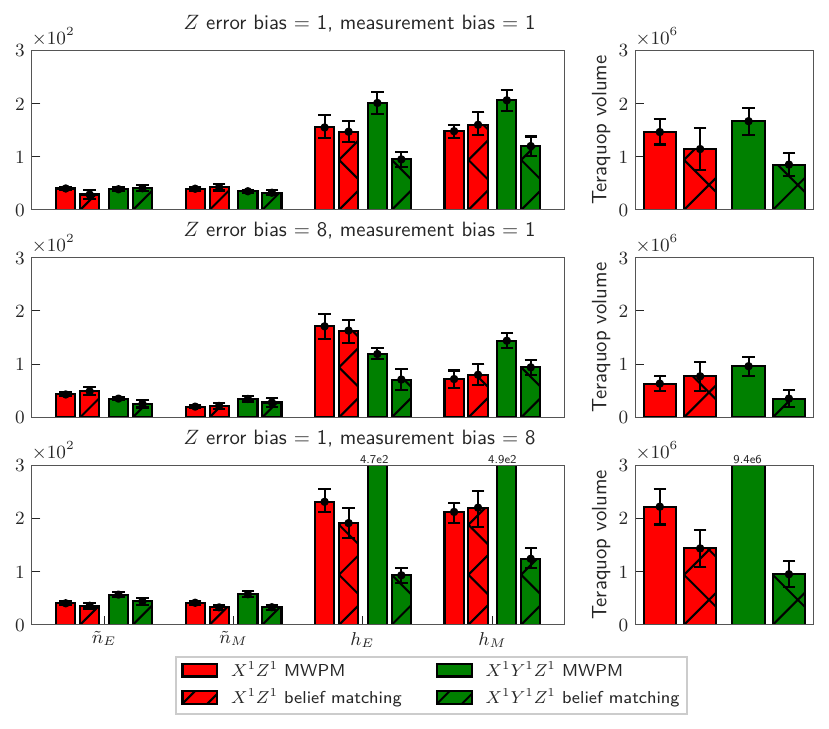}
    \caption{Comparison of $\tilde{n}_E, \tilde{n}_M, \tilde{h}_E, \tilde{h}_M,$ and teraquop volume of the $X^1Y^1Z^1$ and $X^1Z^1$ code decoded using MWPM and belief matching for unbiased, $Z$-biased, and measurement-biased phenomenological noise.
    In all cases, using belief matching rather than MWPM turns the $X^1 Y^1 Z^1$ code from the worst performer to the best performer (\Cref{key:XYZ_XZ_best_and_worst}).
    The difference in teraquop volumes comes primarily from the differences in teraquop height $\tilde{h} = (\tilde{h}_E + \tilde{h}_M)/2$ -- these dwarf the differences in teraquop footprints (\Cref{key:XYZ_XZ_footprints_similar}).
    This is thus an advert for analysing codes by looking at teraquop volumes rather than just teraquop footprints.}
    \label{fig:X1Y1Z1_X1Z1_comparison}
\end{figure}

Now we look at our simulation results.
\Cref{fig:X1Y1Z1_X1Z1_comparison} shows $\tilde{n}_E, \tilde{n}_M, \tilde{h}_E, \tilde{h}_M$ and the teraquop volume of the $X^1Y^1Z^1$ and $X^1Z^1$ codes for unbiased, $Z$-biased, and measurement-biased direct parity measurement noise.
We highlight two key points, and examine their relationship to the differences from the last subsection.

\begin{keypoint}\label{key:XYZ_XZ_best_and_worst}
    In all three noise models, the $X^1 Y^1 Z^1$ code under belief matching has the smallest teraquop volume,
    while the $X^1 Y^1 Z^1$ code under MWPM has the largest teraquop volume.
\end{keypoint}

This is exactly what was predicted by \Cref{dif:XYZ_XZ_hyperedges} -- all measurement errors in the $X^1 Y^1 Z^1$ code correspond to hyperedges in the decoding graph, which makes performance under MWPM worse.
But belief matching turns this bug into a feature.
Moreover, the effect becomes more dramatic the more the noise model is biased towards measurement noise.

This both agrees and disagrees with the intuition mentioned after \Cref{dif:XYZ_XZ_meas_per_detector} -- that the detectors in the $X^1 Y^1 Z^1$ consisting of twice as many measurements as in the $X^1 Z^1$ code would make the former perform worse, especially under measurement-biased noise.
While this is true for MWPM, using belief matching suffices to counteract this effect.
Indeed, the high number of measurements per detector and the fact all measurement errors violate four detectors are two sides of the same coin -- the latter being exactly what makes belief matching perform so well.
Similarly, the better timelike distance in the $X^1 Y^1 Z^1$ code (\Cref{dif:XYZ_XZ_timelike_distance}) doesn't seem to play a role under MWPM, but is consistent with performance under belief matching.

\begingroup
\renewcommand*{\arraystretch}{1.5}
    \begin{table}[htbp]
        \centering
        \scalebox{0.9}{\begin{tabular}{|c|w{c}{1.8cm}|w{c}{1.8cm}|w{c}{1.8cm}|w{c}{1.8cm}|w{c}{1.8cm}|w{c}{1.8cm}|}
    %    \begin{tabular}{|c|c|c|c|c|c|c|}
            \hline
            & \multicolumn{3}{c|}{$X^1 Y^1 Z^1$} & \multicolumn{3}{c|}{$X^1 Z^1$} \\
            \cline{2-7}
            & \makecell{Direct parity\\ meas.} & \makecell{Meas.\\ noise} & Z noise & \makecell{Direct parity\\ meas.} & \makecell{Meas.\\ noise} & Z noise \\
            \cline{2-7}
            \hline
            \hline
            $d_{s, E}$ & $n_E$ & $n_E$ & $n_E$ & $n_E$ & $n_E$ & $n_E$ \\
            \hline
            $d_{s, M}$ & $\frac{4}{3}n_M$ & $\frac{4}{3}n_M$ & $\frac{4}{3}n_M$ & $\frac{4}{3}n_M$ & $\frac{4}{3}n_M$ & $\infty$ \\
            \hline
            $d_{t, E}$ & $h/2$ & $h/2$ & $h/2$ & $h/4 $ & $h/4 $ & $h/2$ \\
            \hline
            $d_{t, M}$ & $h/2$ & $h/2$ & $h/2$ & $h/4$ & $h/4$ & $\infty$ \\
            \hline
            $p_h$ & $5/9$ & $1$ & $1/3$ & $2/9$ & $0$ & $0$ \\
            \hline
        \end{tabular}}
        \caption{Differences in distances between the $X^1Y^1Z^1 $ and $X^1Z^1$ codes. The first four rows show graphlike distances of spacelike and timelike logical operators. The bottom row shows the \textit{hyperedge likelihood} -- the probability that a randomly chosen error in the noise model will correspond to a hyperedge in the decoding graph.
        Of note are that the $X^1 Z^1$ code has worse timelike distance when measurement errors are possible, but infinite spacelike and timelike $M$ distances when only $Z$ errors are possible.
        The $X^1 Y^1 Z^1$ code has a higher hyperedge likelihood in all cases, explaining why it performs the worst under MWPM but the best under belief matching (\Cref{key:XYZ_XZ_best_and_worst}).}
        \label{tab:xyz_css_summary}
    \end{table}
\endgroup

\begin{figure}[htbp]
    \centering
    \includegraphics[width=0.9\linewidth]{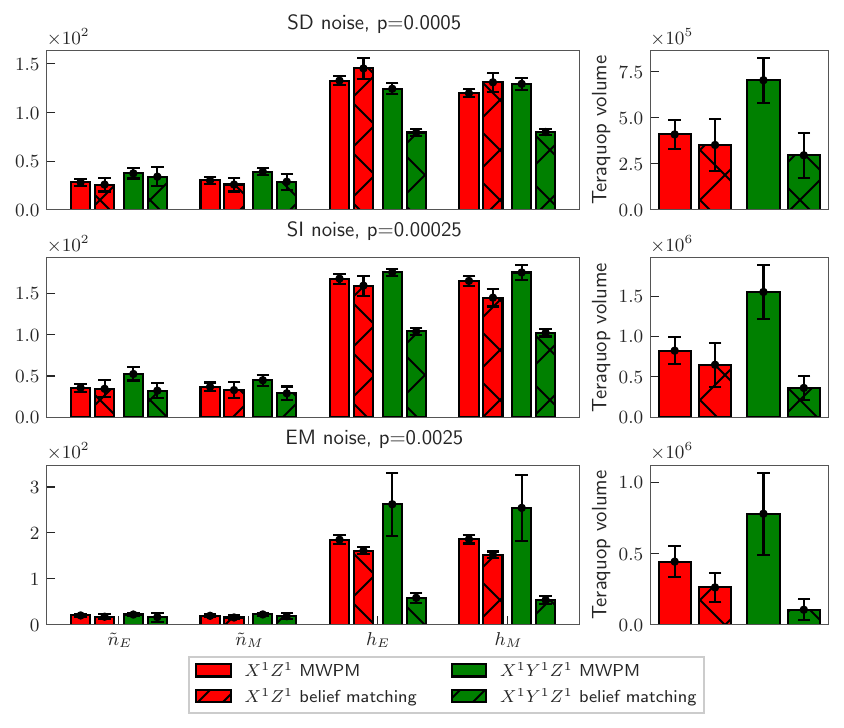}
    \caption{Comparison of $\tilde{n}_E, \tilde{n}_M, \tilde{h}_E, \tilde{h}_M,$ and teraquop volume of the $X^1Y^1Z^1$ and $X^1Z^1$ code decoded using MWPM and belief matching for three circuit level noise models; standard depolarizing noise with $p=0.0005$, superconducting inspired noise with $p=0.00025$, and entangling measurement noise with $p=0.0025$.
    As in \Cref{fig:X1Y1Z1_X1Z1_comparison}, in all cases, using belief matching rather than MWPM turns the $X^1 Y^1 Z^1$ code from the worst performer to the best performer (\Cref{key:XYZ_XZ_best_and_worst}),
        and the most dramatic differences are to be found in the teraquop heights rather than footprints (\Cref{key:XYZ_XZ_footprints_similar}) -- especially for the entangling measurements noise model.}
    \label{fig:X1Y1Z1_X1Z1_comparison_cln}
\end{figure}

\begingroup
\renewcommand*{\arraystretch}{1.35}
\begin{table}[htbp]
    \centering
    \begin{tabular}{|c|w{c}{2.4cm}|w{c}{2.4cm}|w{c}{2.4cm}|w{c}{2.4cm}|}
%    \begin{tabular}{|c|c|c|c|c|}
        \hline
        & \multicolumn{2}{c|}{$X^1 Y^1 Z^1$} & \multicolumn{2}{c|}{$X^1 Z^1$} \\
        \cline{2-5}
        & SI \& SD noise & EM noise & SI \& SD noise & EM noise \\
        \hline
        \hline
        $d_{s, E}$ & $n_E$ & $n_E/2$ & $n_E$ & $n_E/2$ \\
        \hline
        $d_{s, M}$ & $\frac{4}{3}n_M$ & $\frac{2}{3}n_M$ & $\frac{4}{3}n_M$ & $\frac{2}{3}n_M$ \\
        \hline
        $d_{t, E}$ & $h/4$ & $h/2$ & $h/4$ & $h/4$ \\
        \hline
        $d_{t, M}$ & $h/4$ & $h/2$ & $h/4$ & $h/4$ \\
        \hline
    \end{tabular}
    \caption{
        Distances of the $X^1Y^1Z^1$ and $X^1Z^1$ codes.
        Note that the entangling measurements noise model halves the spacelike distances of both codes.
        Meanwhile the SI and SD noise models halve the timelike distances of the $X^1 Y^1 Z^1$ code,
        removing an advantage they had over the $X^1 Z^1$ code from \Cref{tab:xyz_css_summary}.
        These distances were obtained using Stim's function \texttt{shortest\_graphlike\_error}.
        For distance $4$ circuits we found the exact same distances when using the function \texttt{search\_for\_undetectable\_logical\_errors}.
        This second function takes errors that correspond to hyperedges into account.
    }
    \label{tab:xyz_css_summary_cln}
\end{table}
\endgroup

\begin{keypoint}\label{key:XYZ_XZ_footprints_similar}
    The differences in teraquop volume between the $X^1 Y^1 Z^1$ and the $X^1 Z^1$ codes are primarily due to differences in the number of rounds of measurements required to reach the teraquop regime -- not in the number of physical qubits (footprint).
    
\end{keypoint}

That is, the differences in footprints $\tilde{n}_M \tilde{n}_E$ across codes and noise models are generally far smaller than the difference in the teraquop heights $\tilde{h}$.
This contradicts the intuition from \Cref{dif:XYZ_XZ_logicals_pauli_type} -- we wondered whether the teraquop width $\tilde{n}_E$ and depth $\tilde{n}_M$ of the $X^1 Z^1$ code under $Z$-biased noise would be such that the footprint $f(\tilde{n}_E \tilde{n}_M)$ would be smaller than that of the $X^1 Y^1 Z^1$ code under this noise model.
Though $\tilde{n}_M$ is indeed smaller than that of the $X^1 Y^1 Z^1$ code, $\tilde{n}_E$ is larger.
That said, it could be that for a higher $Z$ noise bias than we simulated there could be a significant difference between $\tilde{n}_E \tilde{n}_M$ for the $X^1Z^1$ and $X^1Y^1Z^1$ codes.

A meta-point to make in light of \Cref{key:XYZ_XZ_footprints_similar} above is that using the teraquop volume rather than just the teraquop footprint was important.
Had we only used the teraquop footprint, we would not have seen the dramatic differences between the two codes and the two decoders, because these differences almost entirely show up in the teraquop heights rather than the footprints.
We claim this serves as evidence in favour of using teraquop volumes in future error correction simulations. 

In \Cref{tab:xyz_css_summary}, we summarise some important metrics.
We show the four types of distances of the two codes under the three different noise models,
and additionally show the \textit{hyperedge likelihood} $p_h$ -- the likelihood of an error in that noise model corresponding to a hyperedge in the decoding graph.

\subsection{Auxiliary qubit circuit noise results}\label{subsec:unrepeated-auxiliary-qubit-circuit-noise}

\Cref{fig:X1Y1Z1_X1Z1_comparison_cln} presents $\tilde{n}_E, \tilde{n}_M, \tilde{h}_E, \tilde{h}_M$, and the teraquop volume for the $X^1Y^1Z^1$ and $X^1Z^1$ codes under three auxiliary qubit circuit noise models.
Logical operator distances are summarized in \Cref{tab:xyz_css_summary}.
The results confirm the trends identified in \Cref{key:XYZ_XZ_best_and_worst}: the $X^1Z^1$ code performs best with MWPM, whereas the $X^1Y^1Z^1$ code is superior with belief matching.

Similarly, \Cref{key:XYZ_XZ_footprints_similar} holds, as differences in teraquop volume primarily stem from variations in the number of measurement rounds.
For SI and SD noise decoded with belief matching, $\tilde{h}_E$ and $\tilde{h}_M$ are reduced by a factor of two for the $X^1Y^1Z^1$ code.
Surprisingly, the timelike distance remains identical between the two codes, suggesting that distance alone is not always a reliable performance indicator.
A similar case was made in \Ccite{gidney2023pair}.

Despite simulating entangling measurement noise with a physical error rate ten times higher, the teraquop volumes remain of the same order of magnitude across all three auxiliary qubit circuit noise models.

    \section{Bias-tailoring II: repeating measurements}\label{sec:gauge_dcccs}

There is another technique we can use to tailor our code to certain noise biases, introduced in \Ccite{OscarSubsystemGaugeFixing}, where it was referred to as \textit{schedule-induced gauge-fixing} (SIGF) and applied to \textit{subsystem} codes.
In both subsystem codes and DCCCs, the check operators measured do not all mutually commute (these check operators are known as \textit{gauge operators} in a subsystem code or edge operators in DCCCs).
As a result, using a conventional measurement schedule, the measurement of each check operator is non-deterministic when considered in isolation, and must be combined with other (different) check operators to form a detector.

The SIGF method consists of \textit{repeating} the measurements of a mutually commuting subset of the check operators.
When a subset of mutually commuting gauge operators is measured, each check operator becomes an element of the ISG.
When the measurements of these check operators are repeated, the subsequent measurements are all deterministic (in the absence of noise), and so we can construct much smaller detectors formed from consecutive measurements of individual check operators.
Intuitively, by remaining in the same ISG while we repeat the commuting measurements, we can form much smaller detectors consisting of far fewer measurements, allowing us to protect against certain types of errors more effectively.
For example, if we remain in an ISG with detectors that protect more effectively against $Z$ errors, then this can be effective for handling $Z$-biased noise models (as was done in \Ccite{OscarSubsystemGaugeFixing}).

SIGF has been applied to subsystem codes \Ccite{OscarSubsystemGaugeFixing} and to the Floquet-Bacon-Shor codes \Ccite{SohaibAlam_2025}.
However, it has not previously been studied how well it can handle a noise model biased towards \textit{measurement} errors.
Intuitively we would expect SIGF to be advantageous in this setting, since each detector is formed from far fewer of the noisy measurements.
In this work, we apply SIGF to DCCCs.
Concretely, we simply repeat the edge operator measurements made at each timestep $t$ some number of times $r_t$.
We call the resulting new dynamical code a \textit{repeated DCCC} --
or, if we want to be specific, an \emph{$\mathbf{r}$-repeated DCCC}, where $\mathbf{r} = (r_0, r_1, \ldots)$.

In the next subsection we look analytically at how this intuition manifests itself in terms of detectors, logical errors and the decoding graph, and what we expect this to mean for the teraquop volumes.
Then we present numerical results and see how well our expectations hold up -- again, first for the more tractable case of direct parity measurement noise, and then the more realistic auxiliary qubit circuit noise.
Furthermore, we explore how repeated DCCCs perform both for $Z$ error bias and measurement bias.

\subsection{Detectors}\label{subsec:repeating_measurements_detectors}

\begin{figure}[tb]
    \centering
    \includegraphics[width=0.6\linewidth]{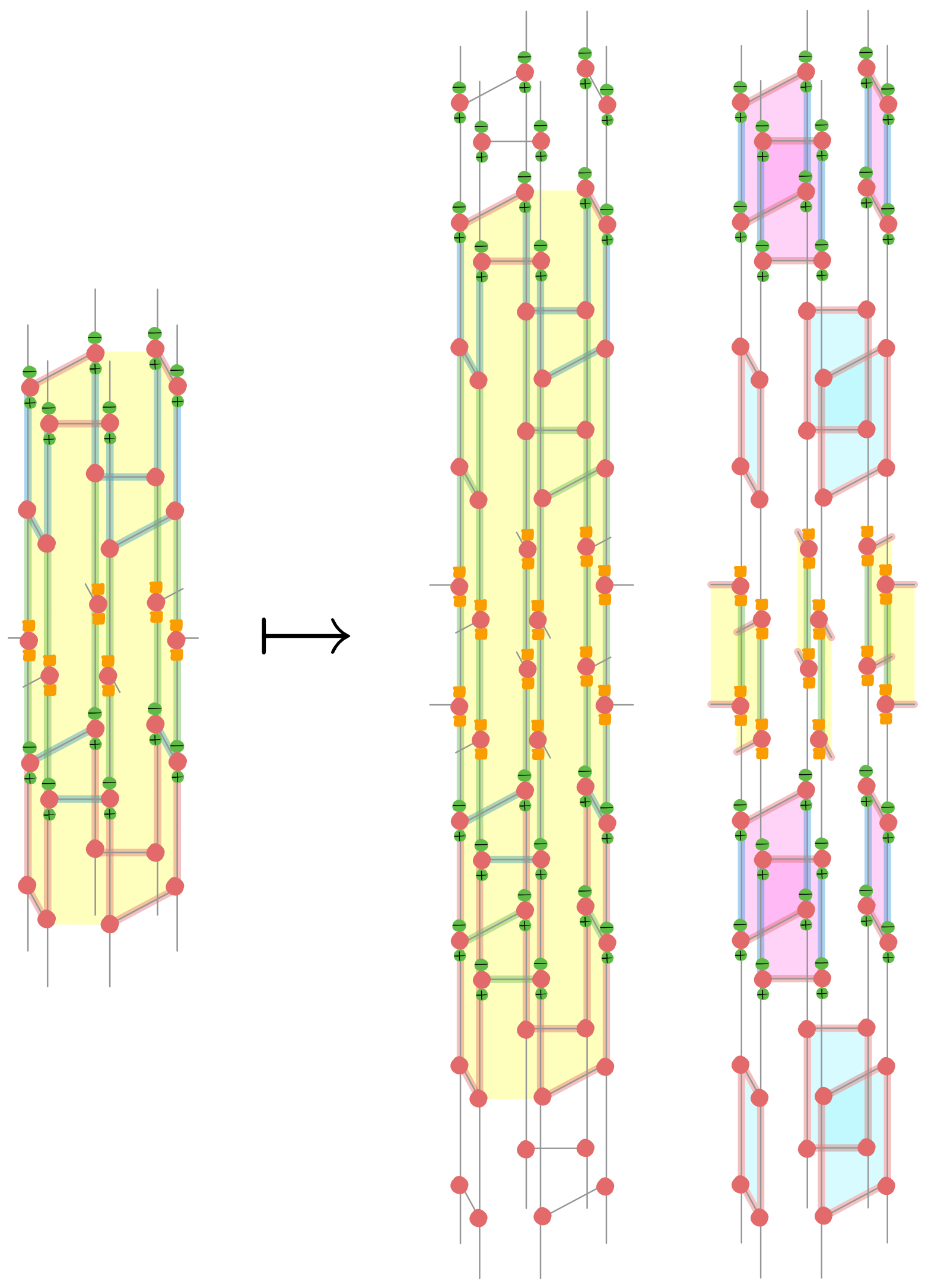}
    \caption{Left: a detector in the $X^1 Y^1 Z^1$ code with volume $6 \cdot 4 = 24$ consisting of 12 measurements (as can be read off from the Pauli web -- see \Cref{fig:detectors_XYZ_and_CSS}). Middle: the analogous detector in the repeated $X^2 Y^2 Z^2$ code, which now has volume $6 \cdot ((4-1) \cdot 2 + 1) = 42$ but still consists of 12 measurements. Right: the edge detectors on these six qubits in the $X^2 Y^2 Z^2$ code, each of which has volume 2 and consists of two consecutive edge measurements.
    As discussed more in the caption of \Cref{tab:repeated_xyz_css_averages}, repeating measurements reduces the mean detector volume and measurements per detector, while increasing the mean number of detectors per measurement -- all of which our intuition says are improvements.
    On the other hand, it increases the maximum detector volume.
    In any case, our simulations show that these improvements don't generally outweigh the negative effects of repeating measurements, except in the case of high measurement bias.}
    \label{fig:detector_if_measurements_done_twice}
\end{figure}

Suppose a (unrepeated) DCCC has a detector
\begin{equation}\label{eq:unrepeated_detector}
    \prod_{(t \in \tau)}
        \prod_{(e_{\kappa_t} \in f_{\kappa})}
            m_t(\dcccOp{e}{\kappa_t}{P_t}),
\end{equation}
where $f_\kappa$ is the $\kappa$-coloured face to which the detector is associated,
and $\tau$ is a set of timesteps.
See, for example, the leftmost diagram of \Cref{fig:detector_if_measurements_done_twice}.
Now let $r_t \geq 1$ denote the number of times each measurement at timestep $t$ in the unrepeated DCCC is repeated before moving to the next measurement type,
let $R_t = \sum_{s=0}^{t-1} r_s$ be the total number of measurement rounds up to timestep $t$ in the unrepeated DCCC,
and for all $t \in \tau$ let $u_t$ be some integer in $\{0, \ldots, r_t-1\}$.
In the corresponding repeated DCCC, for each $\dcccOp{e}{\kappa_t}{P_t}$,
one can show that using the measurement outcomes $m_{R_t + u_t}(\dcccOp{e}{\kappa_{t}}{P_{t}})$ from
any timestep $R_t + u_t$ of each set of $r_t$ repeated measurement rounds gives a detector
\begin{equation}\label{eq:repeated_detector_using_first_round_measurements}
    \prod_{(t \in \tau)}
        \prod_{(e_{\kappa_{t}} \in f_{\kappa})}
            m_{R_t + u_t}(\dcccOp{e}{\kappa_{t}}{P_{t}}).
\end{equation}

We show an example of such a detector in the middle diagram of \Cref{fig:detector_if_measurements_done_twice}, with $r_t = 2$ for all $t$.
However, some choices of $u_t$ are more sensible than others.
Firstly, we want the detector to span as few timesteps as possible;
this minimises the probability of the detector being violated,
which makes the job of the decoder easier, as discussed earlier.
Secondly, we still want to be able to use a matching decoder.

Letting $T = \max(\tau)$ be the latest timestep in $\tau$,
we can achieve these two conditions by
using measurements from the final round $R_t + r_t-1$ of $r_t$ repeated measurement rounds for all $t \neq T$,
but using measurements from the first round $R_T$ of $r_T$ repeated measurements rounds for $t = T$
\begin{equation}\label{eq:repeated_detector_sensible}
    \prod_{(t \in \tau - \{T\})}
        \prod_{(e_{\kappa_{t}} \in f_{\kappa})}
            m_{R_t + r_t-1}(\dcccOp{e}{\kappa_{t}}{P_{t}})
    \prod_{(e_{\kappa_{T}} \in f_{\kappa})}
        m_{R_T}(\dcccOp{e}{\kappa_{T}}{P_{T}}).
\end{equation}

Indeed, this is the specific choice of measurements used in the example in \Cref{fig:detector_if_measurements_done_twice}.
In this new repeated DCCC,
we also have detectors with no analogue to any of those in the original DCCC.
Suppose we have just measured all $(\kappa, P)$ edge operators at some timestep $R_t$
for the first of $r_t$ repeated measurement rounds.
Each edge operator $\dcccOp{e}{\kappa}{P}$ is therefore now a stabilizer up to some sign $m_{R_t}(\dcccOp{e}{\kappa}{P})$.
Assuming $r_t > 1$, at the next timestep $R_t + 1$ we measure all these $(\kappa, P)$ edge operators again.
The stabilizer formalism says that each such measurement will give us the deterministic outcome
$m_{R_t+1}(\dcccOp{e}{\kappa}{P}) = m_{R_t}(\dcccOp{e}{\kappa}{P})$ in the absence of any noise,
so the formal product $m_{R_t+1}(\dcccOp{e}{\kappa}{P}) m_{R_t}(\dcccOp{e}{\kappa}{P})$ is a detector,
for all $\kappa$-coloured edges $e_{\kappa}$.
Indeed, we get two-measurement detectors like these at every subsequent timestep
in the set of $r_t$ repeated measurement rounds.
That is, for every $u \in \{1, \ldots, r_t-1\}$,
we get a detector $m_{R_t + u}(\dcccOp{e}{\kappa}{P}) m_{R_t + u-1}(\dcccOp{e}{\kappa}{P})$
for all $\kappa$-coloured edges $e_{\kappa}$.
These are shown in the rightmost diagram in \Cref{fig:detector_if_measurements_done_twice}.
Since each such detector is associated to an edge, we call them \textit{edge detectors},
and will refer to the other type of detectors (those associated to a face) as \textit{face detectors}.\subsubsection{Volume, measurements per detector, detectors per measurement}\label{subsubsec:repeating_measurements_volume_measurements_detectors}

Let's now contrast these new detectors in the repeated DCCC with those of the corresponding unrepeated DCCC.
First note the number of timesteps spanned by a face detector in the repeated DCCC is at least as large as in the unrepeated DCCC.
Indeed, a face detector in an unrepeated DCCC whose first measurements were at timestep $t$ and final measurements were at timestep $T$, and thus spanned $T - t$ timesteps, will now span $R_T - (R_t + r_t - 1)$ timesteps.
If any $r_j > 1$ for $j \in \{t+1, \ldots, T-1\}$ then this is larger than $T-t$.
We'll define the \textit{volume} of a detector in a DCCC\footnote{Not to be confused with the teraquop volume of the DCCC as a whole! Also, as noted in a previous footnote, a more accurate metric capturing the same intuition is detector likelihood \cite{hesner2024using}.} as the product of the number of timesteps it spans and the number of data qubits involved in the measurements that make up this detector.
Intuitively, the smaller the volume, the better, because a smaller volume means there are fewer opportunities for this detector to be violated.
For example, the detectors in the unrepeated $X^1 Y^1 Z^1$ and $X^1 Z^1$ codes all have volume $6 \cdot 4 = 24$ (\Cref{fig:detectors_XYZ_and_CSS}).
In a repeated version with $r_t$ set to some $r$ for all $t$, these detectors would have volume $6 \cdot ((4-1)r + 1)$ (\Cref{fig:detector_if_measurements_done_twice}).

On the other hand, the number of measurements per face detector remains exactly the same -- e.g.\ $12$ for any $\mathbf{r}$-repeated $X^a Y^b Z^c$ code, and 6 for any $\mathbf{r}$-repeated $X^a Z^b$ code.
Thanks to the new edge detectors, however, which always have volume two and consist of two measurements, the \emph{average} detector volume and \emph{average} number of measurements per detector decreases.
Indeed, in the limit where $r_t \to \infty$ for all $t$, the edge detectors dominate, and these averages both tend to two.
For our two favourite DCCCs, we compute these averages exactly in \Cref{tab:repeated_xyz_css_averages}.

\begin{difference}\label{dif:repeating_detectors}
    Repeating measurements increases the maximum detector volume of a DCCC,
    while decreasing its minimum detector volume and minimum measurements per detector.
    Overall, it decreases a DCCC's average detector volume and average number of measurements per detector.
\end{difference}

Intuitively, smaller volumes and fewer measurements per detector are better, so the fact these averages decrease might suggest a general performance improvement.
However, there is an important trade-off: while the new edge detectors catch measurement errors quickly (with volume 2), data qubit errors can accumulate for longer before being detected.

\begingroup
\renewcommand*{\arraystretch}{1.5}
\begin{table}[tb]
    \centering
    \begin{tabular}{|c|c|c|c|}
        \hline
        & & $X^r Y^r Z^r$ & $X^r Z^r$ \\
        \hline
        \multirow{3}{*}{Detector volume} & Max & $6 \cdot (3r + 1) \to \infty$ & $6 \cdot (3r + 1) \to \infty$ \\
        \cline{2-4}
        & Min & $2$ & $2$ \\
        \cline{2-4}
        & Average & $2 + \frac{22}{3r - 2} \to 2$ & $2 + \frac{22}{3r - 2} \to 2$ \\
        \hline
        \multirow{3}{*}{Meas.\ per detector} & Max & $12$ & $6$ \\
        \cline{2-4}
        & Min & $2$ & $2$ \\
        \cline{2-4}
        & Average & $2 + \frac{10}{3r - 2} \to 2$ & $2 + \frac{4}{3r - 2} \to 2$ \\
        \hline
        \multirow{3}{*}{Detector degree} & Max & $12 + 6(r-1) \to \infty$ & $12 + 6(r-1) \to \infty$ \\
        \cline{2-4}
        & Min & $4$ & $4$ \\
        \cline{2-4}
        & Average & $6 + \frac{6}{3r - 2} \to 6$ & $6 + \frac{6}{3r - 2} \to 6$ \\
        \hline
    \end{tabular}
    \caption{
        Comparison of metrics concerning detectors in the $X^r Y^r Z^r$ and $X^r Z^r$ codes, for $r > 1$.
        The arrows denote the limit of an expression as $r$ tends to infinity.
        As described in \Cref{dif:repeating_detectors}, repeating measurements increases a DCCC's max.\ detector volume while decreasing its min.\ detector volume and min.\ measurements per detector.
        Overall, it decreases a DCCC's average detector volume and average number of measurements per detector.
        Intuitively smaller volumes and fewer measurements per detector are better,
        so the fact these averages decrease might suggest a general performance improvement.
        On the other hand, the key metric could be the maximum or minimum over all these values.
    }
    \label{tab:repeated_xyz_css_averages}
\end{table}
\endgroup

\subsubsection{Asymmetric schedules}\label{subsubsec:repeating_measurements_asymmetric_schedules}

By using asymmetric schedules (where $r_t$ is not the same for all $t$), we can try to further tailor our code to $Z$-biased noise models.
Here the intuition is that if physical $Z$ errors are more likely to occur, we can spend more time measuring $\dcccOp{e}{\kappa}{X}$ and $\dcccOp{e}{\kappa}{Y}$ operators, which will create more edge detectors that are violated by Pauli $Z$ errors.
The more of these we have, the more accurately we can detect and correct Pauli $Z$ errors.

For example, consider the $X^a Z^b$ code, where we repeat all $\dcccOp{e}{\kappa}{X}$ operators $a$ times
and all $\dcccOp{e}{\kappa}{Z}$ operators $b$ times.
In a $Z$-biased noise model, we expect that setting $a$ to be larger than $b$ improves performance.
Similarly, consider the $X^a Y^b Z^c$ code, suppose we repeat all $\dcccOp{e}{\kappa}{X}$ operators $a$ times, all $\dcccOp{e}{\kappa}{Y}$ operators $b$ times, and all $\dcccOp{e}{\kappa}{Z}$ operators $c$ times.
Here we expect that setting $a$ and $b$ to be larger than $c$ will improve performance.

\begin{difference}\label{dif:repeating_asymmetric_schedule}
    In a repeated DCCC, we can choose to spend more time measuring operators that will form detectors capable of detecting certain types of Pauli errors.
    This is particularly relevant for $Z$-biased noise models.
\end{difference}

\subsection{Logical errors}\label{subsec:repeating_measurements_logical_errors}

\begin{figure}[htbp]
    \centering
    \includegraphics[height=0.77\textheight]{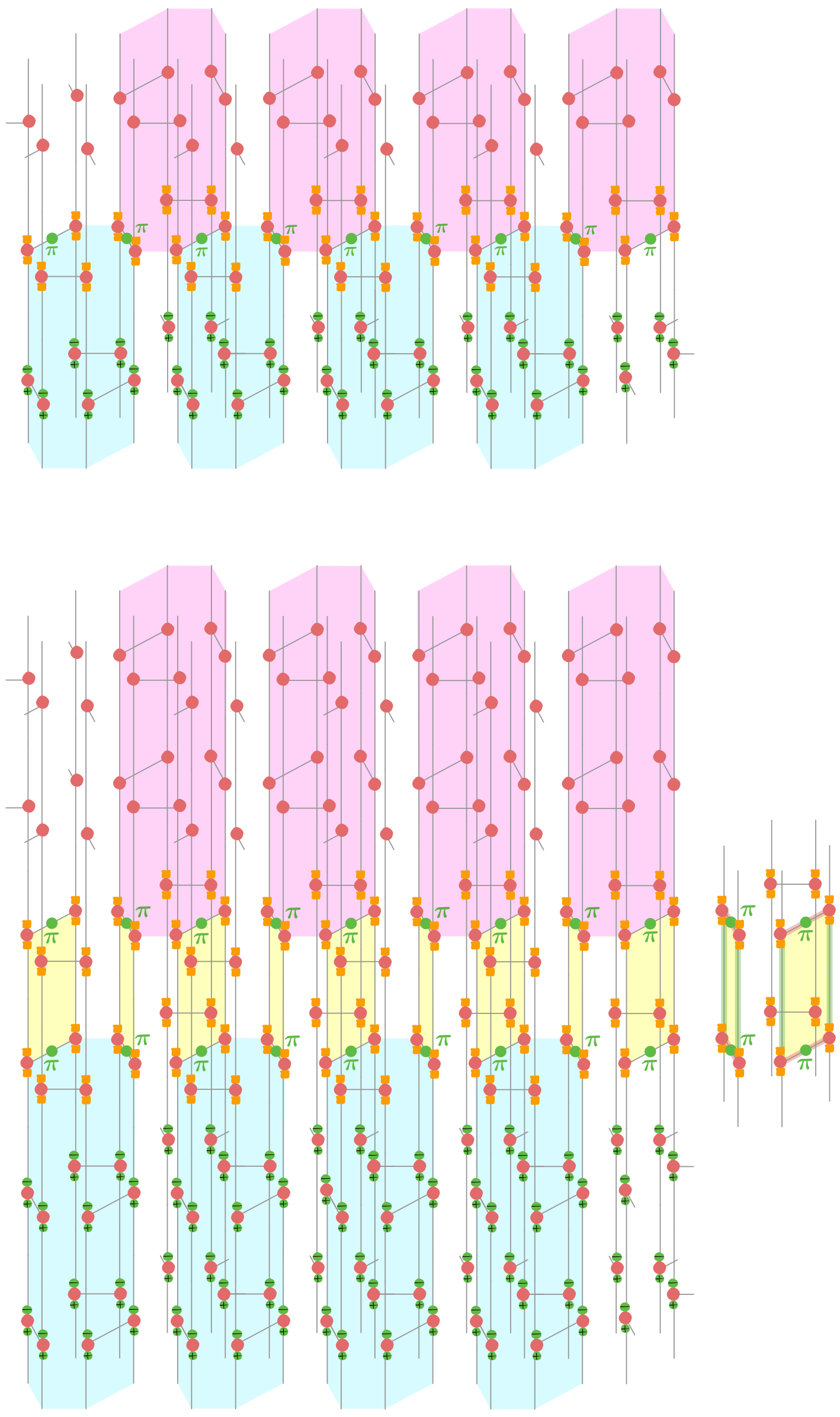}
        \caption{
            Top: a spacelike logical error in the $X^1 Y^1 Z^1$ code consisting entirely of measurement errors.
            Bottom: the analogous spacelike logical error in the $X^2 Y^2 Z^2$ code.
            This again consists entirely of measurement errors, but there are twice as many of them as before, because of the addition of the yellow edge detectors in the $X^2 Y^2 Z^2$ code.
            In both diagrams we only show one type ($E$ or $M$) of detectors.
            We draw the Pauli webs for two edge detectors to show that the logical error really doesn't violate any detectors (see \Cref{fig:XYZ_measurement_error} or \Cref{sec:zx-calculus-and-pauli-webs}),
            but otherwise omit the Pauli webs -- they can be found in \Cref{sec:detectors-cheatsheet}.
            There is a bijection between spacelike logical errors consisting entirely of measurement errors in the $X^1 Y^1 Z^1$ code and in any $X^a Y^b Z^c$ code.
            Hence repeating measurements in general increases the spacelike distance with respect to pure measurement noise (\Cref{tab:repeated_xyz_css_distances}).
            One might then expect this to improve the teraquop width $\tilde{n}_{E}$ and length $\tilde{n}_M$,
            especially in more measurement-biased noise models (\Cref{dif:repeated_spacelike_logicals}).
            But the simulations don't seem to back this up (\Cref{fig:X1Z1_X3Z3_comparison}).
        }
    \label{fig:spacelike_logical_from_measurement_errors_repeated_DCCC}
\end{figure}

\begin{figure}[htbp]
    \centering
    \includegraphics[height=0.75\textheight]{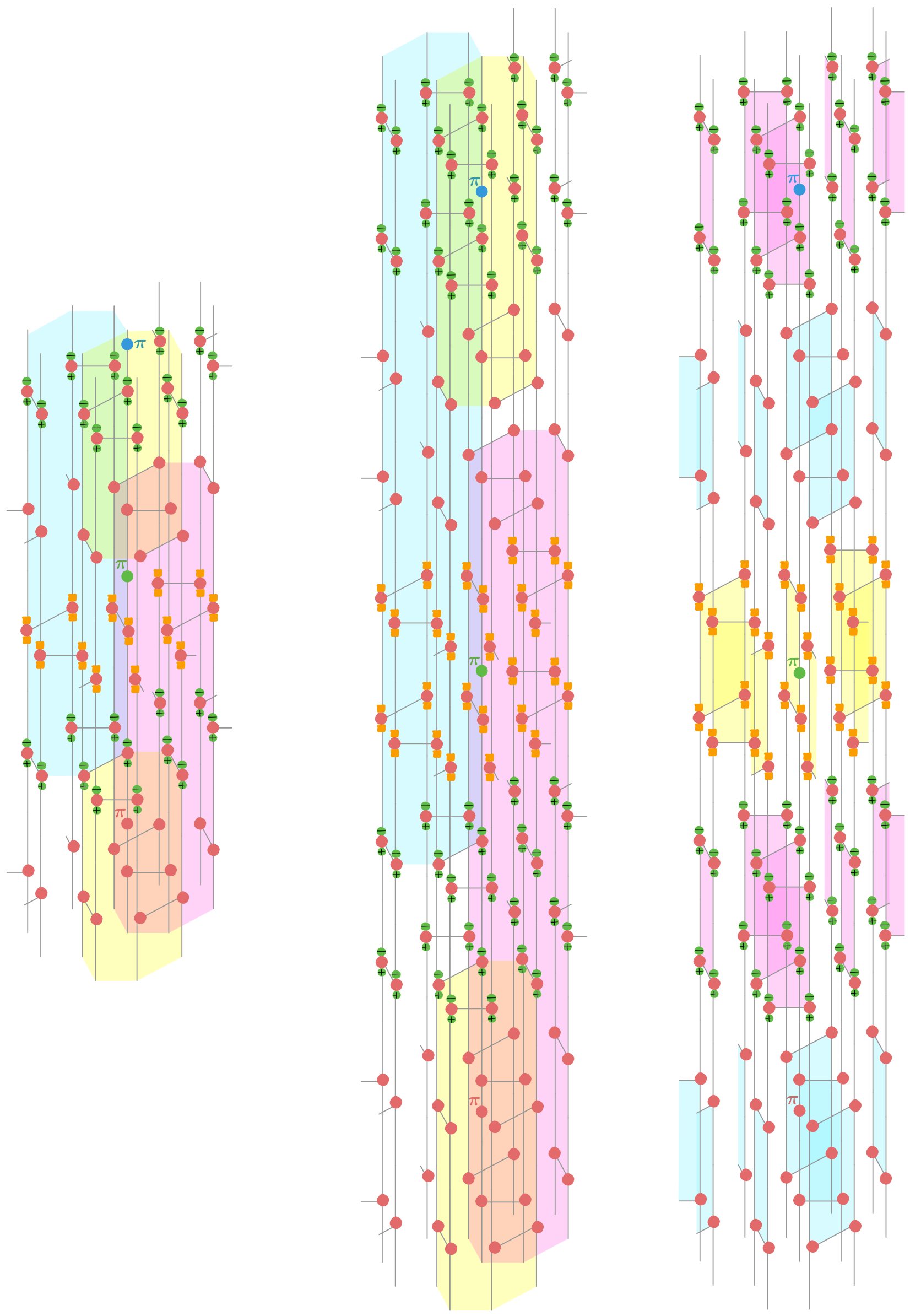}
    \caption{
        Left: a subset of a timelike logical error in the $X^1 Y^1 Z^1$ code consisting entirely of data qubit errors.
        Middle: the analogous timelike logical error in the $X^2 Y^2 Z^2$ code, with half ($E$ or $M$) of the face detectors indicated.
        Right: the same logical error but with the edge detectors indicated.
        In all diagrams, the Pauli webs are omitted -- they can be found in \Cref{sec:detectors-cheatsheet}.
        The logical error in the $X^2 Y^2 Z^2$ code in the middle and right subfigures again consists entirely of data qubit errors, but these data qubit errors now occur once every four rounds rather than once every two rounds.
        In other words, if these two codes were implemented for the same number of rounds, the logical in the $X^2 Y^2 Z^2$ code would have half the weight of the one in the $X^1 Y^1 Z^1$ code.
        Indeed, repeating measurements in general decreases the timelike distance with respect to pure data qubit noise (\Cref{tab:repeated_xyz_css_distances}).
        One would then expect this to worsen the teraquop heights $\tilde{h}_{E}$ and $\tilde{h}_M$,
        especially in more $Z$-biased noise models (\Cref{dif:repeated_timelike_logicals}).
        While the simulation results do show this to be the case (\Cref{fig:X1Z1_X3Z3_comparison}), the fact that this difference is greatest for unbiased noise rather than $Z$-biased noise suggests its cause is something more than the reason discussed here.
    }
    \label{fig:timelike_logical_from_data_qubit_errors_repeated_DCCC}
\end{figure}

The $(n_M, m_E, h_\mathbf{r})$-block definition from \Cref{subsec:estimating-teraquop-volume} still makes sense for any $\mathbf{r}$-repeated DCCC;
we just write $h_{\mathbf{r}}$ instead of $h$ to make it clear that we're considering a repeated DCCC.
%Likewise we can again consider $(s_X, s_Z, h_{\mathbf{r}})_X$- and $(s_X, s_Z, h_{\mathbf{r}})_Z$-blocks.
% as defined above, but with the measurement of edge operators $\dcccOp{E}{\kappa_t}{P_t}$ at timestep $t$ repeated $r_t$ times.
Repeating measurements can increase our code's resilience against spacelike logical errors consisting only of measurement errors.
Consider such a logical error in the unrepeated DCCC, consisting of measurement errors on a set of edge operators $\{\dcccOp{e}{\kappa}{P}\}$ all occurring at the timestep $t$.
For every error in this set,
consider an error on the measurement of $\dcccOp{e}{\kappa}{P}$ at \emph{all} $r_t$ repeated timesteps $R_t + u_t$ for $u_t \in \{0, \ldots, r_t - 1\}$.
The union of all such errors forms a spacelike logical error in the repeated DCCC.
The key here is that each new edge detector
$m_{R_t + u_t}(\dcccOp{e}{\kappa}{P}) m_{R_t + u_t - 1}(\dcccOp{e}{\kappa}{P})$
would be violated by a measurement error on $\dcccOp{e}{\kappa}{P}$ at timestep $R_t + u_t$ or $R_t + u_t - 1$.
So in order for no such detector to be violated, we need the measurements at both timesteps to be erroneous, thereby cancelling each other out.
All pure measurement noise spacelike errors in the repeated DCCC are formed in this way.
So the spacelike distance with respect to these types of logical errors specifically is $(\min_{t} r_t) \cdot d_s$.
An example is shown in \Cref{fig:spacelike_logical_from_measurement_errors_repeated_DCCC}.
% This provides more evidence that repeating measurements is a good idea, especially if the noise is biased towards measurement errors.

\begin{difference}\label{dif:repeated_spacelike_logicals}
    Repeating measurements increases a DCCC's resilience against spacelike logical errors consisting entirely of measurement errors.
    This is particularly relevant in measurement biased noise models.
\end{difference}

On the other hand, repeating measurements generally decreases a DCCC's resilience against timelike logical errors consisting only of data qubit errors.
Again, all such logical errors are analogues of those in the unrepeated DCCC.
Consider such an error in the unrepeated DCCC.
For every Pauli $P$ data qubit error on qubit $j$ immediately after timestep $t$ in the unrepeated DCCC,
the analogous logical error in the repeated DCCC consists of a Pauli $P$ error on qubit $j$ at \emph{any} timestep $R_t + u_t$ for $u_t \in \{0, \ldots, r_t - 1\}$.
Letting $\tau$ again be the set of timesteps of the measurement errors in the unrepeated DCCC,
where before the logical error consisted of $|\tau|$ errors across $h$ timesteps,
it's now $|\tau|$ across $h_{\mathbf{r}} = \sum_{0 < t < h} r_t$.
As soon as $r_t > 1$ for some $t$, this is a decrease.
% That is, in order to maintain the same timelike distance, we have to use block height $h_{\mathbf{r}} = \sum_{0 < t < h} r_t \geq h$.
%Reaching again for the example of an $(s_X, s_Z, h)_X$ block of CSS honeycomb code,
For example, in an $(n_M, n_E, h)$ block of the $X^1 Y^1 Z^1$ code,
in which there exist timelike logical errors consisting of a data qubit error every 2 rounds.
Now suppose we repeat all measurements some fixed number of times $r > 1$ to get an $(n_M, n_E, h_{\mathbf{r}})$ block of the $X^r Y^r Z^r$ code.
Under pure data qubit noise, the unrepeated code has a timelike distance of $\frac{h}{2}$,
whereas the repeated code has a (worse) timelike distance of $\frac{h_{\mathbf{r}}}{2r}$.
See \Cref{fig:timelike_logical_from_data_qubit_errors_repeated_DCCC}.

% Alternatively, as in the last subsection, we could have chosen to repeat all $XX$ measurements $r_X$ times,
% and all $ZZ$ measurements $r_Z$ times.
% Note that $X$ data qubit errors can only contribute to timelike $X$ logical errors, not $Z$ ones.
% Likewise $Z$ data qubit errors can only contribute to timelike $Z$ logical errors.

\begin{difference}\label{dif:repeated_timelike_logicals}
    Repeating measurements decreases the resilience of a DCCC to timelike logical errors made of data qubit errors.
\end{difference}

\begin{figure}[p]
    \centering
    \includegraphics[width=0.9\linewidth]{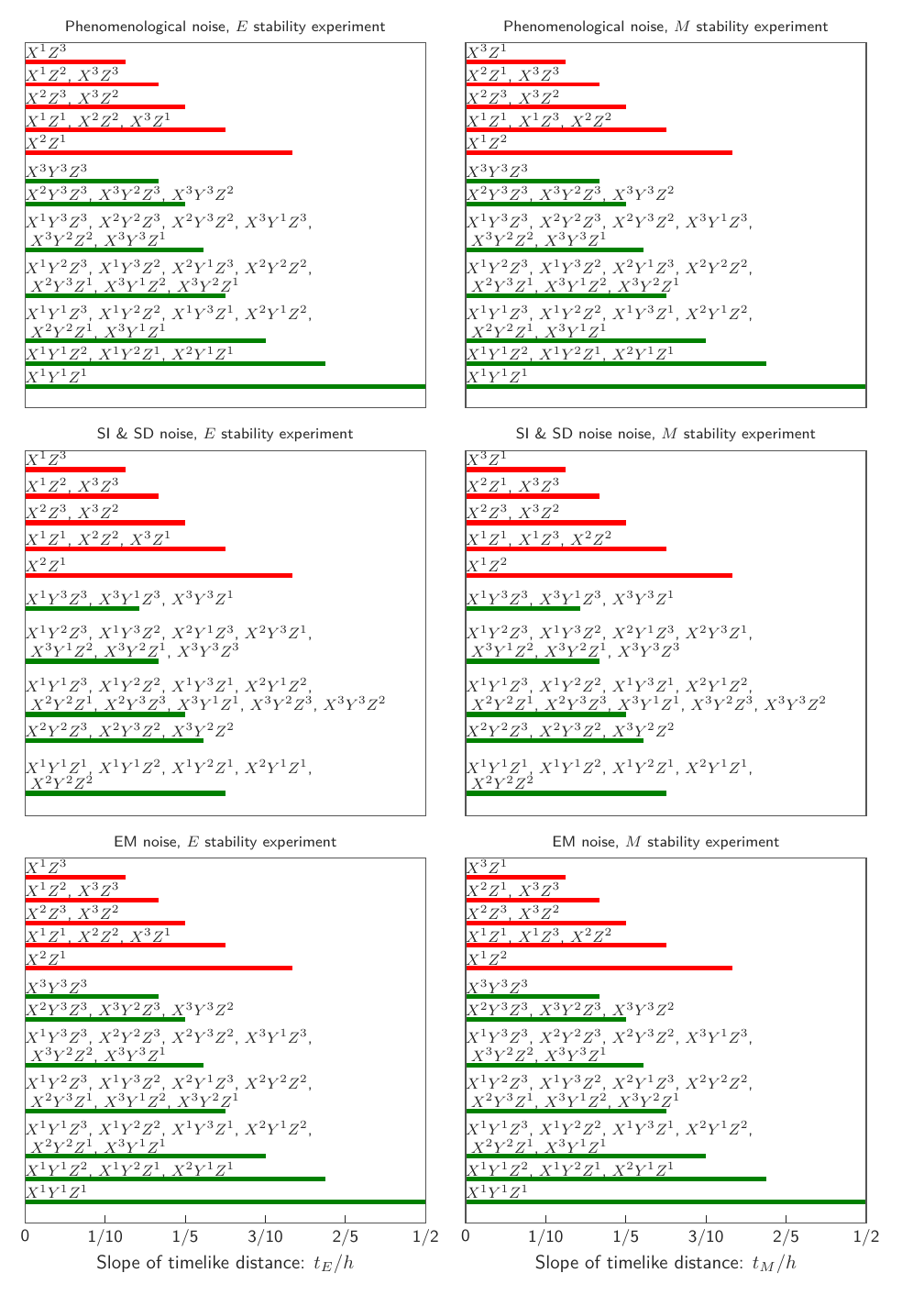}
    \caption{
        Plots showing the timelike distances of $(n_E, n_M, h)$-blocks of $X^a Y^b Z^c$ and $X^a Z^b$ codes under different noise models as a fraction of the block height $h$.
        Since standard depolarizing noise and superconducting-inspired noise differ only in the probabilites of the certain gates occurring, and not in which gates can occur, any notions of distance are identical across both models.
        So we bundle these models together into a `single-qubit measurements' noise model (middle row).
        One can see that the $X^a Z^b$ code timelike distances are unchanged across all three noise models.
        The timelike distances of $X^a Y^b Z^c$ codes are in general reduced under single-qubit measurements noise as compared to direct parity measurement noise,
        but preserved under entangling measurements noise.
        This feature doesn't seem to be significant enough to appear in the simulation results -- e.g.\ in \Cref{fig:beliefmatching_circuit_level_noise_volume_plot}, under the two single-qubit measurements noise models the $X^1 Y^1 Z^1$ code still has the best teraquop volume.
    }
    \label{fig:repeated_measurements_timelike_distances}
\end{figure}

The spacelike distance under pure Pauli noise and the timelike distance under pure measurement noise are unchanged from the unrepeated case.
So altogether we expect the two differences from this section to mean that a repeated DCCC's teraquop footprint $f(\tilde{n}_M \tilde{n}_E)$ is decreased, but that its teraquop height $\tilde{h}$ is increased.
In fact, though it gets hard to discuss analytically for asymmetric schedules, repeating measurements generally reduces not just the timelike distance for logical errors made entirely of data qubit errors, but the timelike distance in general.
We show this in \Cref{fig:repeated_measurements_timelike_distances} -- the topmost subfigure shows the direct parity measurement noise case.

\subsection{Decoding graph}\label{subsec:repeating_measurements_decoding_graph}

\begin{figure}[p]
    \centering
    \begin{subfigure}[c]{0.18\textwidth}
        \centering
        \includegraphics[height=200pt]{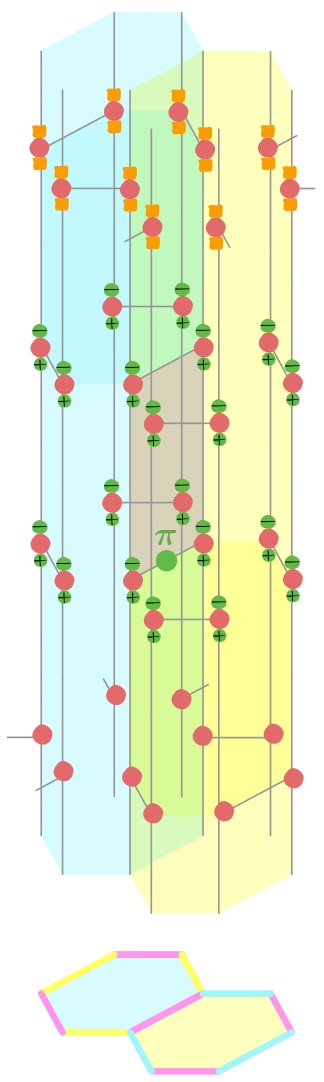}
        \caption{}
        \label{fig:repeated_DCCC_ordinary_edge_measurement_error_a}
    \end{subfigure}
    \begin{subfigure}[c]{0.18\textwidth}
        \centering
        \includegraphics[height=200pt]{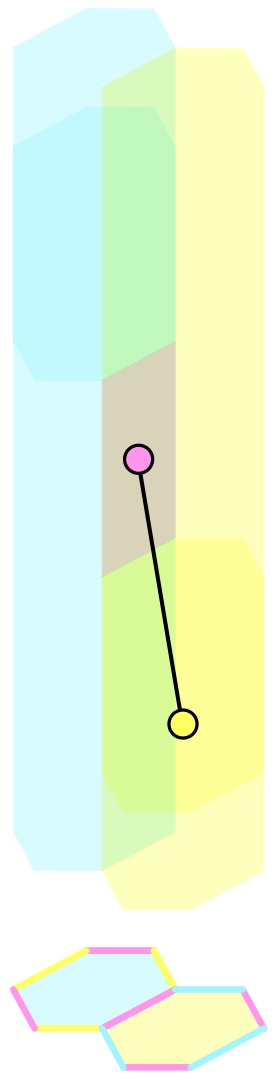}
        \caption{}
        \label{fig:repeated_DCCC_ordinary_edge_measurement_error_b}
    \end{subfigure}
    \begin{subfigure}[c]{0.14\textwidth}
        \centering
        \includegraphics[height=200pt]{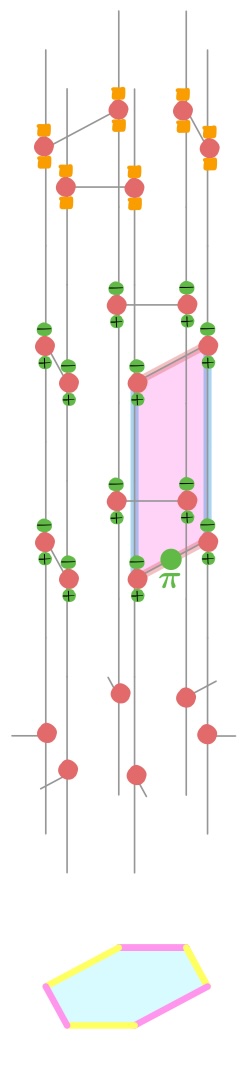}
        \caption{}
        \label{fig:repeated_DCCC_ordinary_edge_measurement_error_c}
    \end{subfigure}
    \begin{subfigure}[c]{0.14\textwidth}
        \centering
        \includegraphics[height=200pt]{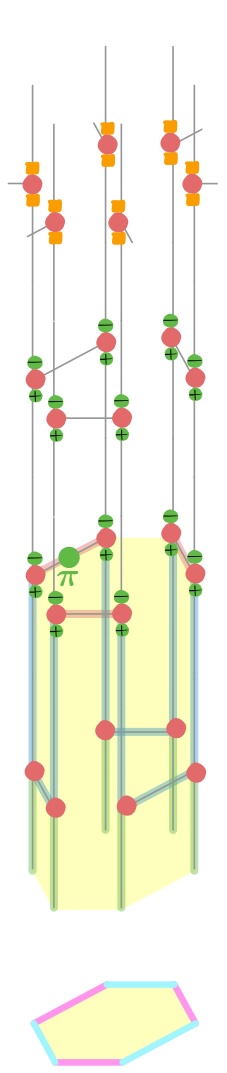}
        \caption{}
        \label{fig:repeated_DCCC_ordinary_edge_measurement_error_d}
    \end{subfigure}
    \begin{subfigure}[c]{0.14\textwidth}
        \centering
        \includegraphics[height=200pt]{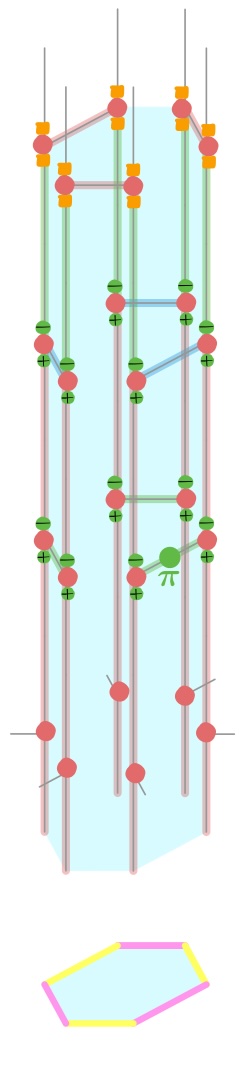}
        \caption{}
        \label{fig:repeated_DCCC_ordinary_edge_measurement_error_e}
    \end{subfigure}
    \begin{subfigure}[c]{0.14\textwidth}
        \centering
        \includegraphics[height=200pt]{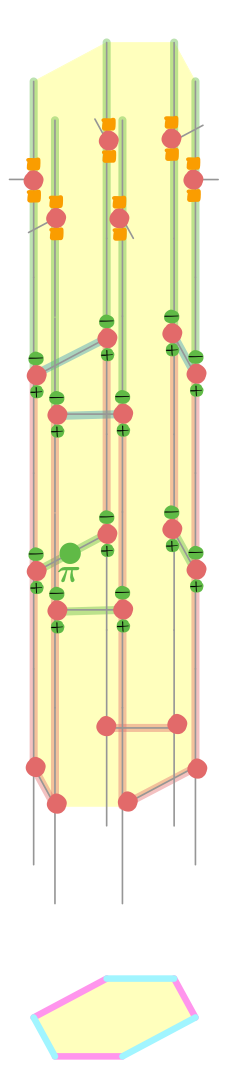}
        \caption{}
        \label{fig:repeated_DCCC_ordinary_edge_measurement_error_f}
    \end{subfigure}
    \caption{
        \textbf{(a)} A measurement error in the $X^2 Y^2 Z^2$ code.
        \textbf{(b)} The ordinary edge it corresponds to in the decoding graph.
        \textbf{(c, d)} The $E$ detectors the error violates, drawn as Pauli webs.
        \textbf{(e, f)} The $M$ detectors the error sits on but does not violate, drawn as Pauli webs.
        Compare this figure with \Cref{fig:XYZ_measurement_error}, where an analogous error in the $X^1 Y^1 Z^1$ code corresponded to a hyperedge in the decoding graph.
        This would still have been the case in the $X^2 Y^2 Z^2$ code had the measurement error occurred in the last of the $r=2$ repeated rounds of measurements.
        But for all other rounds such an error now corresponds to an ordinary edge.
        In general, repeating measurements reduces the number of measurement errors corresponding to hyperedges, which would suggest a plateau in the improvement from using belief matching over MWPM (\Cref{dif:repeated_meas_errors_hyperedges}), and indeed this seems to be the case numerically (\Cref{fig:XaYbZc_teraquop_volumes_measurement_bias}).
    }
    \label{fig:repeated_DCCC_ordinary_edge_measurement_error}

    \vspace{20pt}

    \begingroup
    \renewcommand*{\arraystretch}{1.2}
        \centering
        \begin{tabular}{|r|c|c|c|c|c|c|}
            \hline
            & \multicolumn{3}{c|}{$X^r Y^r Z^r$} & \multicolumn{3}{c|}{$X^r Z^r$} \\
            \hline
            & \makecell{Direct parity \\ meas.} & \makecell{Meas.\\noise} & \makecell{Z noise} & \makecell{Direct parity \\ meas.} & \makecell{Meas.\\noise} & \makecell{Z noise} \\
            \hline
            $d_{s, E}$ & $n_E$ & $r n_E \to \infty$ & $n_E$ & $n_E$ & $r n_E \to \infty$ & $n_E$ \\
            \hline
            $d_{s, M}$ & $\frac{4}{3}n_M$ & $r \frac{4}{3}n_M \to \infty$ & $\frac{4}{3}n_{M}$ & $\frac{4}{3}n_M$ & $r \frac{4}{3}n_M \to \infty$ & $\infty$ \\
            \hline
            $d_{t, E}$ & $\frac{h_\mathbf{r}}{2r} \to 0$ & $\frac{h_\mathbf{r}}{2}$ & $\frac{h_\mathbf{r}}{2r} \to 0$ & $\min\{\frac{h_\mathbf{r}}{2r}, \frac{h_\mathbf{r}}{4}\} \to 0$ & $\frac{h_\mathbf{r}}{4}$ & $\frac{h_\mathbf{r}}{2r} \to 0$ \\
            \hline
            $d_{t, M}$ & $\frac{h_\mathbf{r}}{2r} \to 0$ & $\frac{h_\mathbf{r}}{2}$ & $\frac{h_\mathbf{r}}{2r} \to 0$ & $\min\{\frac{h_\mathbf{r}}{2r}, \frac{h_\mathbf{r}}{4}\} \to 0$ & $\frac{h_\mathbf{r}}{4}$ & $\infty$ \\
            \hline
            $p_h$ & $\frac{2}{9} + \frac{1}{3r} \to \frac{2}{9}$& $\frac{1}{r} \to 0$ & $\frac{1}{3}$ & $\frac{2}{9}$ & $0$ & $0$ \\
            \hline
        \end{tabular}
        \captionof{table}{
            Comparison of the $X^r Y^r Z^r$ and $X^r Z^r$ codes, for $r > 1$.
            The first four rows show the spacelike $E$ and $M$ distances and timelike $E$ and $M$ distances respectively.
            The final row shows the hyperedge likelihood.
            The arrows denote the limit of an expression as $r$ tends to infinity.
            Repeating measurements increases the spacelike distances under pure measurement noise (\Cref{dif:repeated_spacelike_logicals}),
            but decreases the timelike distances under pure data qubit noise (\Cref{dif:repeated_timelike_logicals}).
            One might expect this to be reflected in the teraquop widths, lengths and heights, but the simulation results show this isn't the case (\Cref{fig:X1Z1_X3Z3_comparison}).
            Repeating measurements also decreases the frequency with which measurement errors correspond to hyperedges in the decoding graph.
            This explains why the improvement given by decoding with belief matching rather than MWPM plateaus the more we repeat measurements (\Cref{fig:XaYbZc_teraquop_volumes_measurement_bias}).
        }
        \label{tab:repeated_xyz_css_distances}
\endgroup
\end{figure}

Repeating measurements has an effect on the number of measurement errors that correspond to hyperedges in the decoding graph.
For example, in the unrepeated case, all measurement errors in the $X^1 Y^1 Z^1$ honeycomb code corresponded to hyperedges (\Cref{fig:XYZ_measurement_error}).
But repeating each measurement at time $t$ a number of times $r_t$, 
only one out of every $r_t$ measurement errors now corresponds to a hyperedge.
The remaining measurement errors correspond to ordinary edges -- such as the example in \Cref{fig:repeated_DCCC_ordinary_edge_measurement_error}.
(For the $X^1 Z^1$ code, in which all measurement errors already corresponded to ordinary edges in the hypergraph in the unrepeated case,
nothing changes in this regard when we repeat measurements.)

\begin{difference}\label{dif:repeated_meas_errors_hyperedges}
    The more we repeat measurements, the fewer measurement errors correspond to hyperedges in the decoding graph.
\end{difference}

So the more we repeat measurements, the less we expect the repeated $X^a Y^b Z^c$ code to benefit from using belief matching.
In particular, noise models biased towards measurement errors will see less benefit from using belief matching.

%\subsubsection{Decoding graph vertex degree}\label{subsec:repeating_measurements_vertex_degree}

\begin{figure}[p]
    \centering
    \begin{subfigure}[b]{0.24\textwidth}
        \centering
        \includegraphics[height=330pt]{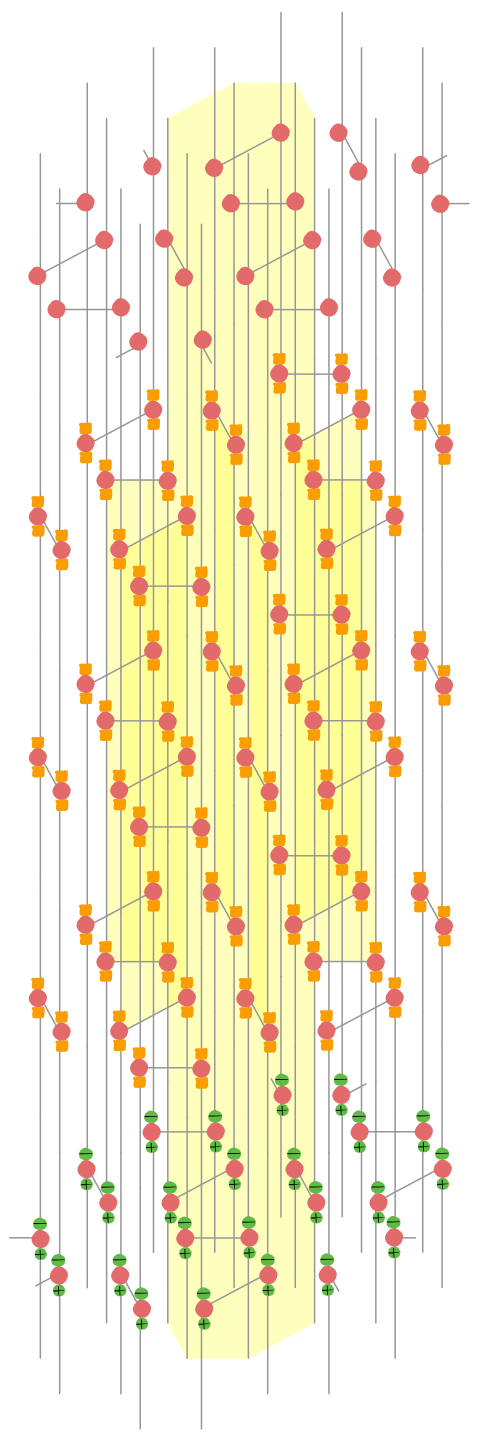}
        \caption{}
        \label{fig:ZX_plus_init}
    \end{subfigure}
    \begin{subfigure}[b]{0.24\textwidth}
        \centering
        \includegraphics[height=330pt]{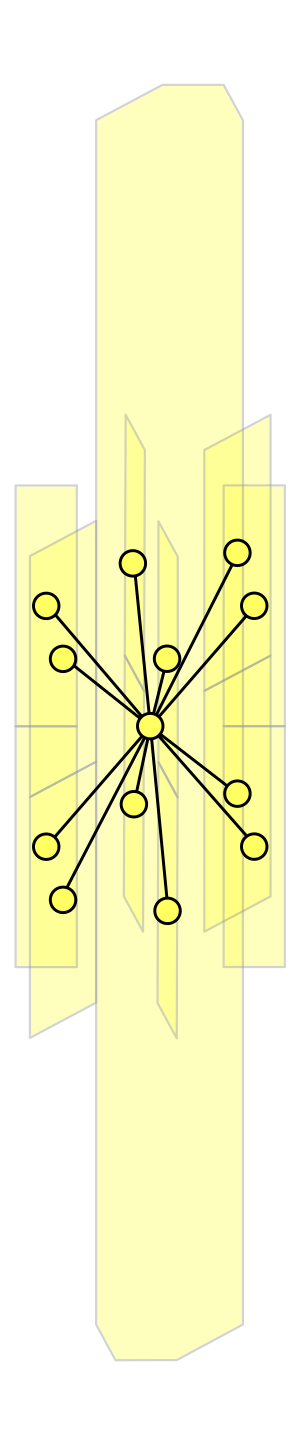}
        \caption{}
        \label{fig:ZX_X_meas}
    \end{subfigure}
    \begin{subfigure}[b]{0.24\textwidth}
        \centering
        \includegraphics[height=330pt]{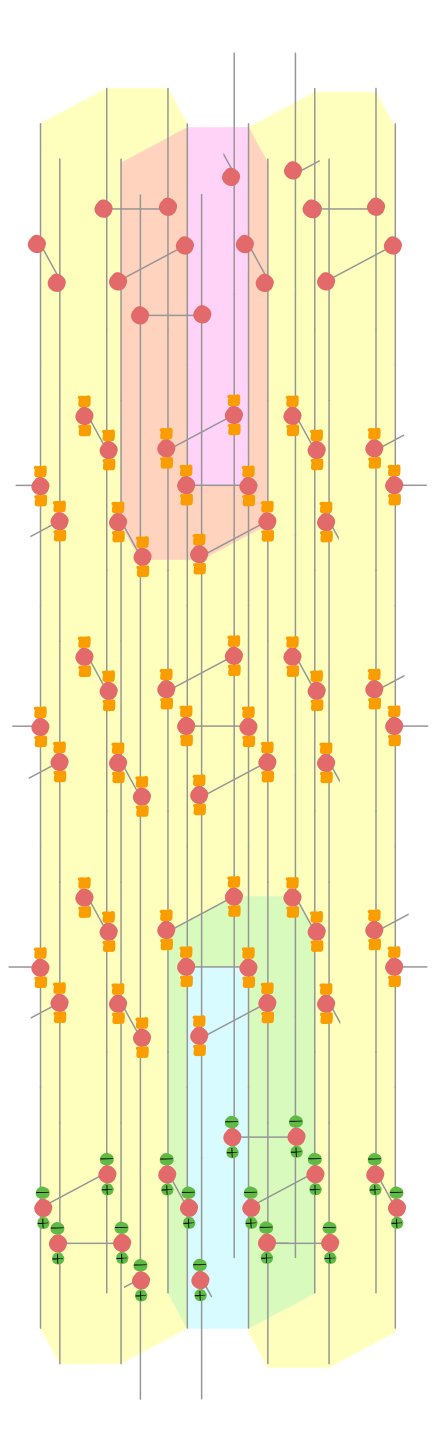}
        \caption{}
        \label{fig:ZX_CNOT}
    \end{subfigure}
    \begin{subfigure}[b]{0.24\textwidth}
        \centering
        \includegraphics[height=330pt]{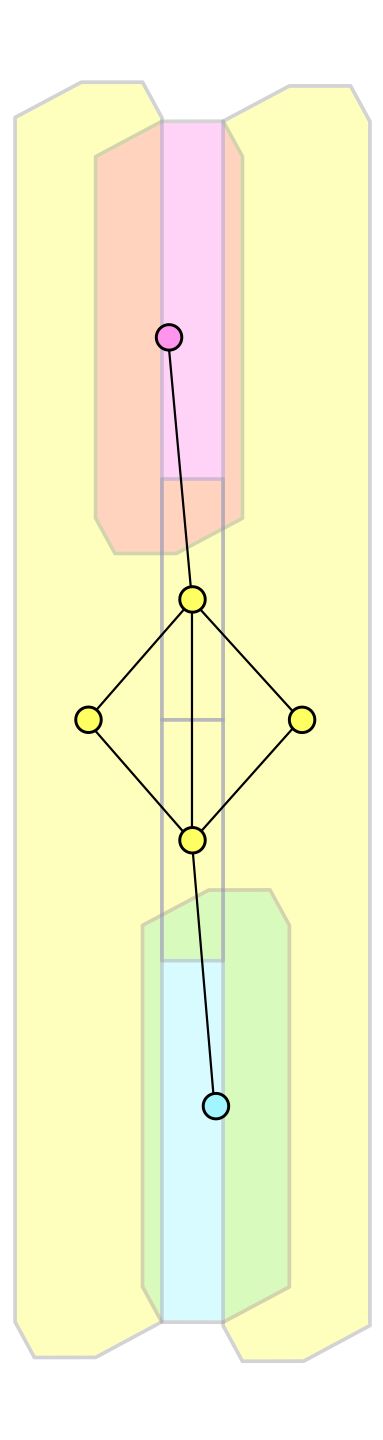}
        \caption{}
        \label{fig:ZX_XX_meas_circuit_YZ_error}
    \end{subfigure}
    \caption{
        \textbf{(a)} Five timesteps of the $X^3 Y^3 Z^3$ code, centered on a yellow face.
        The shaded yellow areas denote a face detector and 12 edge detectors.
        \textbf{(b)} The edges between the face detector and the 12 edge detectors in the decoding graph.
        The face detector has 12 further edges (unshown) to other detectors in the ordinary decoding graph, exactly analogous to the 12 edges it has in $\Cref{fig:decoding_graph_face_detector_neighbourhood}$ in the $X^1 Y^1 Z^1$ code.
        So altogether it now has 24 edges.
        \textbf{(c)} The same five timesteps of the $X^3 Y^3 Z^3$ code, but now centered on a yellow edge, with shaded areas indicating detectors.
        \textbf{(d)} The neighbourhoods of the two edge detectors in the ordinary decoding graph.
        Note each has degree 4.
        All of the above holds more generally, in that in the ordinary decoding graph for each $X^a Y^b Z^c$ and $X^a Z^b$ code, every edge detector will have degree 4, and every face detector will have degree $12 + 6(r-1)$, where $r$ is the number of repeated measurement rounds for the edges in consideration (here $r=3$).
        So repeating measurements generally increases the maximum vertex degree but reduces the average vertex degree in the decoding graph (\Cref{tab:repeated_xyz_css_averages}).
        The former suggests a general decrease in performance while the latter suggests an increase.
    }
    \label{fig:repeated_decoding_graph_neighbourhoods}
\end{figure}

The other key difference pertains to vertex degrees in a DCCC's decoding graph.
The degree of vertices in a code's decomposed decoding graph in general affects decoding performance; the lower the better~\cite{OscarSubsystemGaugeFixing}.
In any unrepeated DCCC, we said every vertex (i.e.\ detector) has degree 12 (\Cref{fig:decoding_graph_face_detector_neighbourhood}).
Repeating measurements introduces edge detectors, whose corresponding vertices always have degree 4 (\Cref{fig:repeated_decoding_graph_neighbourhoods}),
but it also increases the degree of the vertices corresponding to face detectors.
For the $X^a Y^b Z^c$ and $X^a Z^b$ codes this is overall a positive change,
in the sense that the average vertex degree in the decomposed decoding graph tends from 12 down to 6 as the number of repeated measurements increases.

\begin{difference}\label{dif:repeating_decoding_graph_degrees}
    Repeating measurements decreases the average vertex degree in the decoding graph of a DCCC.
\end{difference}

This difference would seem to predict a general performance increase for repeated DCCCs versus their unrepeated counterparts.

\subsection{Direct parity measurement noise results}\label{subsec:repeating_measurements_direct_parity_measurement_noise}

\Cref{fig:phenomenological_noise_volume_plot_pymatching,fig:phenomenological_noise_volume_plot_beliefmatching} show the teraquop volumes of the best $X^a Y^b Z^c$ and $X^a Z^b$ codes under direct parity measurement noise models of different biases, decoded with MWPM and belief matching respectively.

\begin{keypoint}
    Repeating measurements improves the teraquop volumes of both codes under both decoders when the noise is measurement-biased, but otherwise it has a negligible effect.
\end{keypoint}

One can see this most clearly by looking at the hearts ($\heartsuit$) subplot in \Cref{fig:phenomenological_noise_volume_plot_pymatching,fig:phenomenological_noise_volume_plot_beliefmatching};
under MWPM (\Cref{fig:phenomenological_noise_volume_plot_pymatching}) all but one of the ten best codes involve repeating measurements in some way,
and under belief matching (\Cref{fig:phenomenological_noise_volume_plot_beliefmatching}) it's all ten.
This chimes with the intuition that repeating measurements gives us more confidence about faulty measurements.
At the same time it seems to contradict the intuition that repeating measurements might improve performance in general (\Cref{dif:repeating_detectors,dif:repeating_decoding_graph_degrees}), and particularly in the case of $Z$-biased noise by spending more time measuring certain operators (\Cref{dif:repeating_asymmetric_schedule}). That said, we only tested a $Z$-noise bias of up to $\eta_Z = 16$, whereas in other work orders of magnitude higher values are considered -- e.g. bias of up to $\eta_Z = 10^6$ is considered in \Ccite{Chamberland_2022}.

\subsection{Auxiliary qubit circuit noise results}\label{subsec:repeating_measurements_auxiliary_qubit_circuit_noise}

\Cref{fig:pymatching_circuit_level_noise_volume_plot,fig:beliefmatching_circuit_level_noise_volume_plot} show the teraquop volumes of the best codes under the three auxiliary qubit circuit noise models, decoded with MWPM and belief matching respectively.
In all three cases, there is not much to be gained from repeating measurements.
Perhaps the only exception is that in the belief matching case under the EM3 noise model, the $X^2 Z^2$ code has a noticeably better teraquop volume than the $X^1 Z^1$ code.
But both are outperformed by the $X^1 Y^1 Z^1$ code, which has the best teraquop footprint out of all the codes in this regime.

%The bottom two subfigures in \Cref{fig:repeated_measurements_timelike_distances} show the timelike distances of the various $X^a Y^b Z^c$ and $X^a Z^b$ codes under circuit-level noise.
%While there's a lot going on, perhaps the main thing to notice is that the timelike distances of the $X^a Z^b$ codes are unchanged regardless of the noise model used, whereas the timelike distances of the $X^a Y^b Z^c$ codes are in general reduced when moving to either circuit-level noise model.

\begin{figure}[t]
    \centering
    \includegraphics[width=\linewidth]{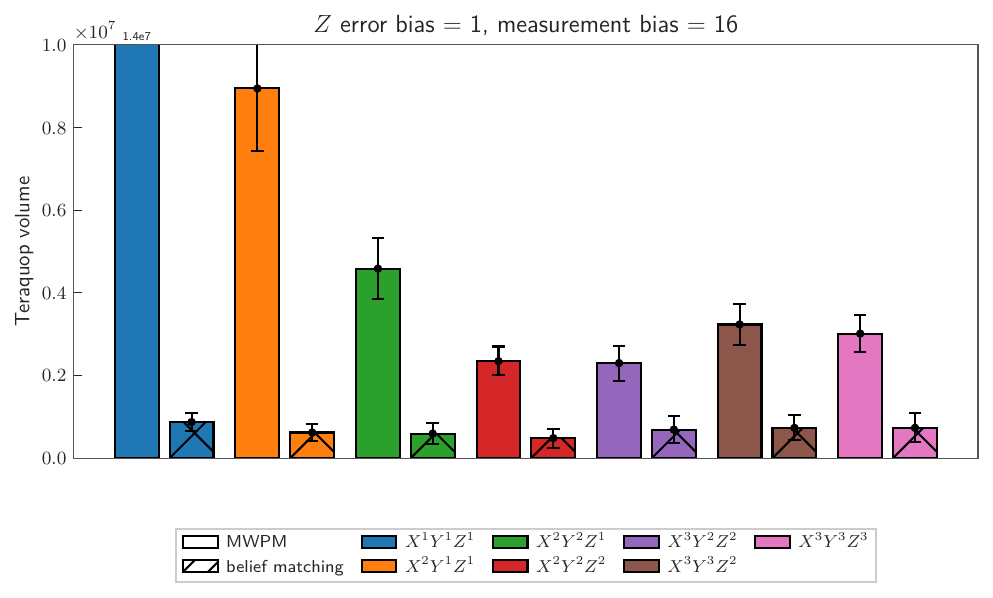}
    \caption{
        Teraquop volumes of the $X^aY^bZ^c$ code for direct parity measurement noise with measurement bias 16,
        and an increasing number $a + b + c$ of repeated rounds of measurements.
        As predicted by \Cref{dif:repeated_meas_errors_hyperedges}, using belief matching rather than MWPM initially makes a huge difference to the volume, but this plateaus the more we repeat measurements.
    }
    \label{fig:XaYbZc_teraquop_volumes_measurement_bias}
\end{figure}

\begin{figure}[p]
    \centering
    \includegraphics[width=\linewidth]{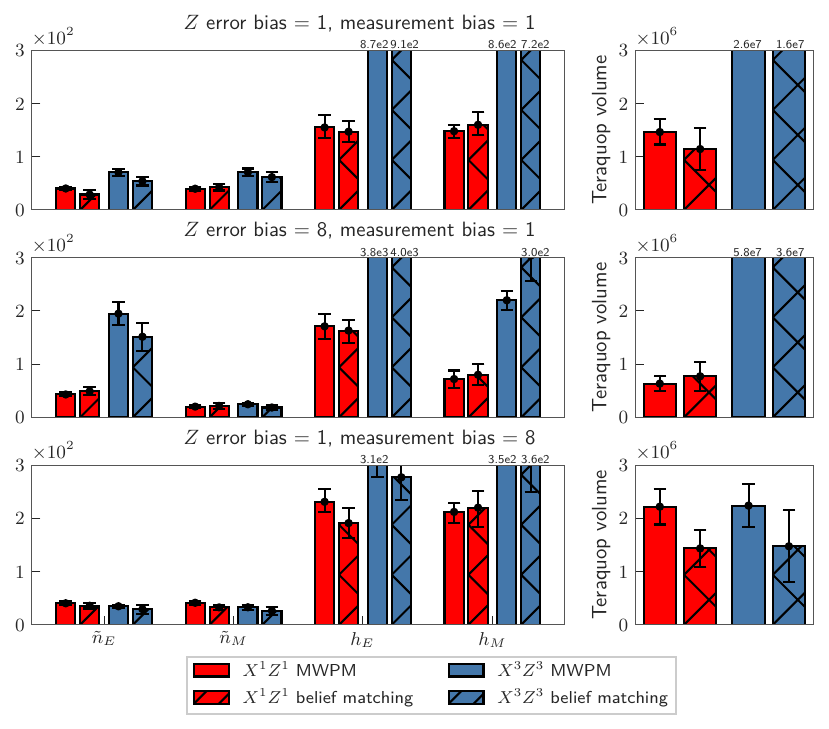}
    \caption{
        Comparison of $\tilde{n}_E, \tilde{n}_M, \tilde{h}_E, \tilde{h}_M,$ and teraquop volume of the $X^1Z^1$ and $X^3Z^3$ code decoded using MWPM and belief matching for unbiased, $Z$-biased, and measurement-biased direct parity measurement noise.
        Based on \Cref{dif:repeated_spacelike_logicals}, we expected that repeating measurements would reduce the teraquop width and length $\tilde{n}_E$ and $\tilde{n}_M$, especially in the case of measurement-biased noise.
        Similarly, based on \Cref{dif:repeated_timelike_logicals}, we expected that repeating measurements would increase the teraquop heights $\tilde{h}_E$ and $\tilde{h}_M$, especially in the case of $Z$-biased noise.
        Neither of these predictions are borne out by the simulation results here.
        In no cases does repeating measurements significantly reduce the teraquop width or length,
        and while it's true that repeating measurements increases the teraquop heights, the fact that this increase is greatest for unbiased noise suggests the cause is something more than \Cref{dif:repeated_timelike_logicals}.
    }
    \label{fig:X1Z1_X3Z3_comparison}
\end{figure}

     \section{Discussion and outlook}
\label{sec:summary_and_outlook}

DCCCs can be used to implement a surface code block, the basic unit of a lattice surgery quantum computer.
In this work, we have optimized the overhead of a surface code block by tailoring the measurement schedule of DCCCs to direct parity measurement noise and auxiliary qubit circuit noise models.
This was done by repeating measurements and measuring in different bases.

The metric we have introduced and optimized here is the teraquop volume, the spacetime overhead required to reach spacelike and timelike logical error rate $10^{-12}$.
We have shown that the optimal measurement schedule depends on the noise model and the type of decoder used. 
We hope to see experimental demonstrations of DCCCs with tailored measurement schedules.
Additionally, we hope that the teraquop volume metric gains wider adoption, as we consider it a crucial measure of overhead in quantum error correction and a tool to make reasonable design choices when setting up specific schemes for quantum error correction.

It would be interesting to see if tailoring measurement schedules can be combined with other techniques to further reduce the teraquop volume.
Taking inspiration from \Ccite{berthusen2025adaptivesyndromeextraction}, one could envision an adaptive measurement schedule.
Furthermore, for strongly biased noise we expect a gain from combining repeated measurements with Clifford conjugated DCCCs as in \Ccite{setiawan2024tailoring}.
Time vortex defects~\cite{kishony2025increasingdistancetopologicalcodes} may further reduce teraquop volumes. It is also the hope that ideas such as those presented here will assist in taking steps for designing fault tolerant schemes for quantum computing beyond quantum memories.
    \section{Author contributions}\label{sec:contributions}

Peter-Jan Derks and Alex Townsend-Teague led this project.
Alex wrote most of the code to generate the circuits used in the numerical experiments; Peter-Jan wrote the rest of the code and produced the numerical results.
Oscar Higgott wrote the implementation of the belief matching decoder used in this work.
All authors contributed to the design of numerical experiments and the analysis of the results.
Peter-Jan and Alex drafted the manuscript with input from all of the authors.
    \section{Acknowledgements}\label{sec:acknowledgements}

We thank Andreas Wallraff for discussions. The Berlin team
 acknowledges funding from BMBF (RealistiQ, QSolid, MuniQC-Atoms),
 the Munich Quantum Valley (K-8), the Quantum Flagship
 (Millenion, PasQuans2), Berlin Quantum, 
 the Einstein Foundation (Einstein Research Unit on 
 quantum devices),
 the European Research Council (DebuQC),
 and the DFG (CRC 183).
O.H.~acknowledges support from the Engineering and Physical
Sciences Research Council (Grant No. EP/L015242/1) and
a Google PhD fellowship.

    \bibliographystyle{quantum}
    \bibliography{base}

@article{hastings2021dynamically,
  title = {Dynamically generated logical qubits},
  author = {Hastings, Matthew B. and Haah, Jeongwan},
  journal = {Quantum},
  volume = {5},
  pages = {564},
  year = {2021},
  publisher = {Verein zur F{\"o}rderung des Open Access Publizierens in den Quantenwissenschaften},
  doi = {10.22331/q-2021-10-19-564},
  eprint = {2107.02194},
  archivePrefix = {arXiv}
}

@article{Srivastava2022xyzhexagonal,
  doi = {10.22331/q-2022-04-27-698},
  title = {The {XYZ}{$^2$} hexagonal stabilizer code},
  author = {Srivastava, Basudha and Frisk Kockum, Anton and Granath, Mats},
  journal = {{Quantum}},
  issn = {2521-327X},
  publisher = {{Verein zur F{\"{o}}rderung des Open Access Publizierens in den Quantenwissenschaften}},
  volume = {6},
  pages = {698},
  month = {apr},
  year = {2022},
  eprint = {2112.06036},
  archivePrefix = {arXiv}
}

@article{Miguel2023cellularautomaton,
  doi = {10.22331/q-2023-03-09-940},
  title = {A cellular automaton decoder for a noise-bias tailored color code},
  author = {Miguel, Jonathan F. San and Williamson, Dominic J. and Brown, Benjamin J.},
  journal = {{Quantum}},
  issn = {2521-327X},
  publisher = {{Verein zur F{\"{o}}rderung des Open Access Publizierens in den Quantenwissenschaften}},
  volume = {7},
  pages = {940},
  month = {mar},
  year = {2023},
  eprint = {2203.16534},
  archivePrefix = {arXiv}
}

@article{Tuckett2019,
  title = {Tailoring Surface Codes for Highly Biased Noise},
  author = {Tuckett, David K. and Darmawan, Andrew S. and Chubb, Christopher T. and Bravyi, Sergey and Bartlett, Stephen D. and Flammia, Steven T.},
  journal = {Phys. Rev. X},
  volume = {9},
  issue = {4},
  pages = {041031},
  numpages = {22},
  year = {2019},
  month = {Nov},
  publisher = {American Physical Society},
  doi = {10.1103/PhysRevX.9.041031},
  eprint = {1812.08186},
  archivePrefix = {arXiv}
}

@article{Tuckett2018,
  title = {Ultrahigh Error Threshold for Surface Codes with Biased Noise},
  author = {Tuckett, David K. and Bartlett, Stephen D. and Flammia, Steven T.},
  journal = {Phys. Rev. Lett.},
  volume = {120},
  issue = {5},
  pages = {050505},
  numpages = {5},
  year = {2018},
  month = {Jan},
  publisher = {American Physical Society},
  doi = {10.1103/PhysRevLett.120.050505},
  eprint = {1708.08474},
  archivePrefix = {arXiv}
}

@article{Tuckett2020,
  title = {Fault-Tolerant Thresholds for the Surface Code in Excess of 5\% under Biased Noise},
  author = {Tuckett, David K. and Bartlett, Stephen D. and Flammia, Steven T. and Brown, Benjamin J.},
  journal = {Phys. Rev. Lett.},
  volume = {124},
  issue = {13},
  pages = {130501},
  numpages = {6},
  year = {2020},
  month = {Mar},
  publisher = {American Physical Society},
  doi = {10.1103/PhysRevLett.124.130501},
  eprint = {1907.02554},
  archivePrefix = {arXiv}
}

@article{harper2025characterising,
  title = {Characterising the failure mechanisms of error-corrected quantum logic gates},
  author = {Robin Harper and Constance Lainé and Evan Hockings and Campbell McLauchlan and Georgia M. Nixon and Benjamin J. Brown and Stephen D. Bartlett},
  year = {2026},
  eprint = {2504.07258},
  archivePrefix = {arXiv},
  primaryClass = {quant-ph},
  journal = {Nature Communications},
  volume = {17},
  pages = {5039},
  doi = {10.1038/s41467-026-71773-6}
}

@article{gidney2022stability,
  title = {Stability experiments: The overlooked dual of memory experiments},
  author = {Gidney, Craig},
  journal = {Quantum},
  volume = {6},
  pages = {786},
  year = {2022},
  publisher = {Verein zur F{\"o}rderung des Open Access Publizierens in den Quantenwissenschaften},
  doi = {10.22331/q-2022-08-24-786},
  eprint = {2204.13834},
  archivePrefix = {arXiv}
}

@article{MWPM,
author={Jack Edmonds},title={Paths, trees, and flowers}, 
doi={10.4153/CJM-1965-045-4},
journal={Canad. J. Math.}, volume=17,pages={449-467}, year=1965}

@article{kesselring2022anyon,
  title = {Anyon condensation and the color code},
  author = {Kesselring, Markus S. and de la Fuente, Julio C. Magdalena and Thomsen, Felix and Eisert, Jens and Bartlett, Stephen D. and Brown, Benjamin J.},
  journal = {PRX Quantum},
  volume = {5},
  pages = {010342},
  year = {2024},
  DOI = {10.1103/PRXQuantum.5.010342},
  eprint = {2212.00042},
  archivePrefix = {arXiv}
}

@article{MWPMColor,
  author = {Aleksander Kubica and Nicolas Delfosse},
  title = {Efficient color code decoders in $d \geq 2$ dimensions from toric code decoders},
  journal = {Quantum},
  year = {2023},
  eprint = {1905.07393},
  archivePrefix = {arXiv},
  volume = {7},
  pages = {929},
  doi = {10.22331/q-2023-02-21-929}
}

@article{UnionFind,
  author = {Nicolas Delfosse and Naomi H. Nickerson},
  title = {Almost-linear time decoding algorithm for topological codes},
  journal = {Quantum},
  volume = {5},
  pages = {595},
  year = {2021},
  doi = {10.22331/q-2021-12-02-595},
  eprint = {1709.06218},
  archivePrefix = {arXiv}
}

@article{PhysRevResearch.2.033042,
  title = {Linear-time maximum likelihood decoding of surface codes over the quantum erasure channel},
  author = {Delfosse, Nicolas and Z\'emor, Gilles},
  journal = {Phys. Rev. Res.},
  volume = {2},
  issue = {3},
  pages = {033042},
  numpages = {5},
  year = {2020},
  optmonth = {Jul},
  publisher = {American Physical Society},
  doi = {10.1103/PhysRevResearch.2.033042},
  eprint = {1703.01517},
  archivePrefix = {arXiv}
}

@article{gidney2021fault,
  title = {A fault-tolerant honeycomb memory},
  author = {Gidney, Craig and Newman, Michael and Fowler, Austin G. and Broughton, Michael},
  journal = {Quantum},
  volume = {5},
  pages = {605},
  year = {2021},
  publisher = {Verein zur F{\"o}rderung des Open Access Publizierens in den Quantenwissenschaften},
  doi = {10.22331/q-2021-12-20-605},
  eprint = {2108.10457},
  archivePrefix = {arXiv}
}

@article{gidney2022benchmarking,
  title = {Benchmarking the planar honeycomb code},
  author = {Gidney, Craig and Newman, Michael and McEwen, Matt},
  journal = {Quantum},
  volume = {6},
  pages = {813},
  year = {2022},
  publisher = {Verein zur F{\"o}rderung des Open Access Publizierens in den Quantenwissenschaften},
  doi = {10.22331/q-2022-09-21-813},
  eprint = {2202.11845},
  archivePrefix = {arXiv}
}

@article{davydova2022floquet,
  title = {Floquet codes without parent subsystem codes},
  author = {Davydova, Margarita and Tantivasadakarn, Nathanan and Balasubramanian, Shankar},
  journal = {PRX Quantum},
  year = {2023},
  volume = {4},
  pages = {020341},
  doi = {10.1103/PRXQuantum.4.020341},
  eprint = {2210.02468},
  archivePrefix = {arXiv}
}

@article{OscarSubsystemGaugeFixing,
  title = {Subsystem Codes with High Thresholds by Gauge Fixing and Reduced Qubit Overhead},
  author = {Higgott, Oscar and Breuckmann, Nikolas P.},
  journal = {Phys. Rev. X},
  volume = {11},
  issue = {3},
  pages = {031039},
  numpages = {30},
  year = {2021},
  optmonth = {Aug},
  publisher = {American Physical Society},
  doi = {10.1103/PhysRevX.11.031039},
  eprint = {2010.09626},
  archivePrefix = {arXiv}
}

@article{higgott2022fragile,
  title = {Improved decoding of circuit noise and fragile boundaries of tailored surface codes},
  author = {Higgott, Oscar and Bohdanowicz, Thomas C. and Kubica, Aleksander and Flammia, Steven T. and Campbell, Earl T.},
  journal = {Phys. Rev. X},
  volume = {13},
  optnumber = {3},
  pages = {031007},
  year = {2023},
  publisher = {APS},
  doi = {10.1103/PhysRevX.13.031007},
  eprint = {2203.04948},
  archivePrefix = {arXiv}
}

@article{Gidney2021stimfaststabilizer,
  doi = {10.22331/q-2021-07-06-497},
  title = {Stim: a fast stabilizer circuit simulator},
  author = {Gidney, Craig},
  journal = {{Quantum}},
  issn = {2521-327X},
  publisher = {{Verein zur F{\"{o}}rderung des Open Access Publizierens in den Quantenwissenschaften}},
  volume = {5},
  pages = {497},
  optmonth = {jul},
  year = {2021},
  eprint = {2103.02202},
  archivePrefix = {arXiv}
}

@article{higgott2025sparse,
  title = {Sparse blossom: correcting a million errors per core second with minimum-weight matching},
  author = {Higgott, Oscar and Gidney, Craig},
  journal = {Quantum},
  volume = {9},
  pages = {1600},
  year = {2025},
  publisher = {Verein zur F{\"o}rderung des Open Access Publizierens in den Quantenwissenschaften},
  doi = {10.22331/q-2025-01-20-1600},
  eprint = {2303.15933},
  archivePrefix = {arXiv}
}

@misc{GottesmanQECBook2024,
  title = {Surviving as a Quantum Computer in a Classical World - 2024 Draft},
  author = {Gottesman, Daniel},
  year = {2024},
  url = {https://www.cs.umd.edu/~dgottesm/QECCbook-2024.pdf},
  howpublished = {\url{https://www.cs.umd.edu/~dgottesm/QECCbook-2024.pdf}}
}

@article{Litinski2019gameofsurfacecodes,
  doi = {10.22331/q-2019-03-05-128},
  title = {A {g}ame of {s}urface {c}odes: {L}arge-{s}cale {q}uantum {c}omputing with {l}attice {s}urgery},
  author = {Litinski, Daniel},
  journal = {{Quantum}},
  issn = {2521-327X},
  publisher = {{Verein zur F{\"{o}}rderung des Open Access Publizierens in den Quantenwissenschaften}},
  volume = {3},
  pages = {128},
  optmonth = {mar},
  year = {2019},
  eprint = {1808.02892},
  archivePrefix = {arXiv}
}

@article{acharya2024quantum,
  author = {Acharya, Rajeev and Abanin, Dmitry A. and Aghababaie-Beni, Laleh and Aleiner, Igor and Andersen, Trond I. and Ansmann, Markus and Arute, Frank and Arya, Kunal and Asfaw, Abraham and Astrakhantsev, Nikita and Atalaya, Juan and others},
  title = {Quantum error correction below the surface code threshold},
  journal = {Nature},
  year = {2025},
  doi = {10.1038/s41586-024-08449-y},
  volume = {638},
  pages = {920--926},
  eprint = {2408.13687},
  archivePrefix = {arXiv}
}

@article{bombin2024unifying,
  title = {{Unifying flavors of fault tolerance with the ZX calculus}},
  author = {Bombin, Hector and Litinski, Daniel and Nickerson, Naomi and Pastawski, Fernando and Roberts, Sam},
  journal = {Quantum},
  volume = {8},
  pages = {1379},
  year = {2024},
  publisher = {Verein zur F{\"o}rderung des Open Access Publizierens in den Quantenwissenschaften},
  doi = {10.22331/q-2024-06-18-1379},
  eprint = {2303.08829},
  archivePrefix = {arXiv}
}

@article{bonilla2021xzzx,
  title = {{The XZZX surface code}},
  author = {Bonilla Ataides, J. Pablo and Tuckett, David K. and Bartlett, Stephen D. and Flammia, Steven T. and Brown, Benjamin J.},
  journal = {Nature Comm.},
  volume = {12},
  optnumber = {1},
  pages = {2172},
  year = {2021},
  publisher = {Nature Publishing Group UK London},
  doi = {10.1038/s41467-021-22274-1},
  eprint = {2009.07851},
  archivePrefix = {arXiv}
}

@article{darmawan2021practical,
  title = {{Practical quantum error correction with the XZZX code and Kerr-cat qubits}},
  author = {Darmawan, Andrew S. and Brown, Benjamin J. and Grimsmo, Arne L. and Tuckett, David K. and Puri, Shruti},
  journal = {PRX Quantum},
  volume = {2},
  optnumber = {3},
  pages = {030345},
  year = {2021},
  publisher = {APS},
  doi = {10.1103/PRXQuantum.2.030345},
  eprint = {2104.09539},
  archivePrefix = {arXiv}
}

@article{lee2021even,
  title = {Even more efficient quantum computations of chemistry through tensor hypercontraction},
  author = {Lee, Joonho and Berry, Dominic W. and Gidney, Craig and Huggins, William J. and McClean, Jarrod R. and Wiebe, Nathan and Babbush, Ryan},
  journal = {PRX Quantum},
  volume = {2},
  optnumber = {3},
  pages = {030305},
  year = {2021},
  publisher = {APS},
  doi = {10.1103/PRXQuantum.2.030305},
  eprint = {2011.03494},
  archivePrefix = {arXiv}
}

@article{gidney2021factor,
  title = {How to factor 2048 bit {RSA} integers in 8 hours using 20 million noisy qubits},
  author = {Gidney, Craig and Eker{\aa}, Martin},
  journal = {Quantum},
  volume = {5},
  pages = {433},
  year = {2021},
  publisher = {Verein zur F{\"o}rderung des Open Access Publizierens in den Quantenwissenschaften},
  doi = {10.22331/q-2021-04-15-433},
  eprint = {1905.09749},
  archivePrefix = {arXiv}
}

@inproceedings{haner2020improved,
  title = {Improved quantum circuits for elliptic curve discrete logarithms},
  author = {H{\"a}ner, Thomas and Jaques, Samuel and Naehrig, Michael and Roetteler, Martin and Soeken, Mathias},
  booktitle = {Post-Quantum Cryptography: 11th International Conference, PQCrypto 2020, Paris, France, April 15--17, 2020, Proceedings 11},
  pages = {425--444},
  year = {2020},
  organization = {Springer},
  doi = {10.1007/978-3-030-44223-1_23},
  eprint = {2001.09580},
  archivePrefix = {arXiv}
}

@article{caune2024demonstrating,
  title = {Demonstrating real-time and low-latency quantum error correction with superconducting qubits},
  author = {Caune, Laura and Skoric, Luka and Blunt, Nick S. and Ruban, Archibald and McDaniel, Jimmy and Valery, Joseph A. and Patterson, Andrew D. and Gramolin, Alexander V. and Majaniemi, Joonas and Barnes, Kenton M. and others},
  journal = {Nature Communications},
  year = {2026},
  doi = {10.1038/s41467-026-73331-6},
  eprint = {2410.05202},
  archivePrefix = {arXiv}
}

@book{NielsenChuang,
  title = {Quantum computation and quantum information},
  Author = {Nielsen, Michael A. and Chuang, I. L.},
  Publisher = {Cambridge University Press},
  Year = {2010},
  doi = {10.1017/CBO9780511976667},
  ISBN = {9781107002173},
  edition = {10th Anniversary}
}

@article{wootton2022measurements,
  title = {{Measurements of Floquet code plaquette stabilizers}},
  author = {Wootton, James R.},
  journal = {arXiv preprint arXiv:2210.13154},
  year = {2022},
  eprint = {2210.13154},
  archivePrefix = {arXiv}
}

@article{PhysRevResearch.2.043423,
  title = {Decoding across the quantum low-density parity-check code landscape},
  author = {Roffe, Joschka and White, David R. and Burton, Simon and Campbell, Earl T.},
  journal = {Phys. Rev. Res.},
  volume = {2},
  issue = {4},
  pages = {043423},
  numpages = {13},
  year = {2020},
  optmonth = {Dec},
  publisher = {American Physical Society},
  doi = {10.1103/PhysRevResearch.2.043423},
  eprint = {2005.07016},
  archivePrefix = {arXiv}
}

@article{McKay,
author={David J. C. MacKay}, title={Good error-correcting codes based on very sparse matrices}, journal={IEEE Trans. Inf. Th.}, volume=45, pages={399-431}, 
doi={10.1109/18.748992},year=1999}

@article{Leifer,
  author = {Matt Leifer and David Poulin},
  title = {Quantum graphical models and belief propagation},
  journal = {Ann. Phys.},
  volume = {323},
  pages = {1899-1946},
  year = {2008},
  doi = {10.1016/j.aop.2007.10.001},
  eprint = {0708.1337},
  archivePrefix = {arXiv}
}

@article{NonIID,
  title = {Correcting non-independent and non-identically distributed errors with surface codes},
  author = {Konstantin Tiurev and Peter-Jan H. S. Derks and Joschka Roffe and Jens Eisert and Jan-Michael Reiner},
  journal = {Quantum},
  volume = {7},
  pages = {1123},
  doi = {10.22331/q-2023-09-26-1123},
  year = {2023},
  eprint = {2208.02191},
  archivePrefix = {arXiv}
}

@article{PhysRevLett.133.110601,
  title = {Domain wall color code},
  author = {Tiurev, Konstantin and Pesah, Arthur and Derks, Peter-Jan H. S. and Roffe, Joschka and Eisert, Jens and Kesselring, Markus S. and Reiner, Jan-Michael},
  journal = {Phys. Rev. Lett.},
  volume = {133},
  issue = {11},
  pages = {110601},
  numpages = {6},
  year = {2024},
  optmonth = {Sep},
  publisher = {American Physical Society},
  doi = {10.1103/PhysRevLett.133.110601},
  eprint = {2307.00054},
  archivePrefix = {arXiv}
}

@article{derks2024,
  journal = {Quantum},
  title = {Designing fault-tolerant circuits using detector error models},
  author = {Peter-Jan H. S. Derks and Alex Townsend-Teague and Ansgar G. Burchards and Jens Eisert},
  year = {2025},
  eprint = {2407.13826},
  archivePrefix = {arXiv},
  primaryClass = {quant-ph},
  volume = {9},
  pages = {1905},
  doi = {10.22331/q-2025-11-06-1905}
}

@article{dennis2002topological,
  author = {Dennis, Eric and Kitaev, Alexei and Landahl, Andrew and Preskill, John},
  title = {Topological quantum memory},
  journal = {J. Math. Phys.},
  volume = {43},
  optnumber = {9},
  pages = {4452-4505},
  year = {2002},
  optmonth = {09},
  issn = {0022-2488},
  doi = {10.1063/1.1499754},
  eprint = {quant-ph/0110143},
  archivePrefix = {arXiv}
}

@article{Horsman_2012_lattice_surgery,
  doi = {10.1088/1367-2630/14/12/123011},
  year = {2012},
  optmonth = {dec},
  publisher = {IOP Publishing},
  volume = {14},
  optnumber = {12},
  pages = {123011},
  author = {Dominic Horsman and Austin G. Fowler and Simon Devitt and Rodney Van Meter},
  title = {Surface code quantum computing by lattice surgery},
  journal = {New J. Phys.},
  eprint = {1111.4022},
  archivePrefix = {arXiv}
}

@article{eickbusch2024demonstrating,
  title = {Demonstration of dynamic surface codes},
  author = {Eickbusch, Alec and McEwen, Matt and Sivak, Volodymyr and Bourassa, Alexandre and Atalaya, Juan and Claes, Jahan and Kafri, Dvir and Gidney, Craig and Warren, Christopher W. and Gross, Jonathan and others},
  journal = {Nature Physics},
  year = {2025},
  volume = {21},
  pages = {1994--2001},
  doi = {10.1038/s41567-025-03070-w},
  eprint = {2412.14360},
  archivePrefix = {arXiv}
}

@article{ali2024reducing,
  title = {Reducing the error rate of a superconducting logical qubit using analog readout information},
  author = {Ali, Hany and Marques, Jorge and Crawford, Ophelia and Majaniemi, Joonas and Serra-Peralta, Marc and Byfield, David and Varbanov, Boris and Terhal, Barbara M. and DiCarlo, Leonardo and Campbell, Earl T.},
  journal = {Phys. Rev. Appl.},
  volume = {22},
  optnumber = {4},
  pages = {044031},
  year = {2024},
  publisher = {APS},
  doi = {10.1103/PhysRevApplied.22.044031},
  eprint = {2403.00706},
  archivePrefix = {arXiv}
}

@article{krinner2022realizing,
  title = {Realizing repeated quantum error correction in a distance-three surface code},
  author = {Krinner, Sebastian and Lacroix, Nathan and Remm, Ants and Di Paolo, Agustin and Genois, Elie and Leroux, Catherine and Hellings, Christoph and Lazar, Stefania and Swiadek, Francois and Herrmann, Johannes and others},
  journal = {Nature},
  volume = {605},
  optnumber = {7911},
  pages = {669--674},
  year = {2022},
  publisher = {Nature Publishing Group UK London},
  doi = {10.1038/s41586-022-04566-8},
  eprint = {2112.03708},
  archivePrefix = {arXiv}
}

@article{gidney2023pair,
  title = {A pair measurement surface code on pentagons},
  author = {Gidney, Craig},
  journal = {Quantum},
  volume = {7},
  pages = {1156},
  year = {2023},
  publisher = {Verein zur F{\"o}rderung des Open Access Publizierens in den Quantenwissenschaften},
  doi = {10.22331/q-2023-10-25-1156},
  eprint = {2206.12780},
  archivePrefix = {arXiv}
}

@article{fu2024errorcorrectiondynamicalcodes,
  title = {Error Correction in Dynamical Codes},
  journal = {Quantum},
  author = {Esther Xiaozhen Fu and Daniel Gottesman},
  year = {2025},
  eprint = {2403.04163},
  archivePrefix = {arXiv},
  primaryClass = {quant-ph},
  volume = {9},
  pages = {1886},
  doi = {10.22331/q-2025-10-20-1886}
}

@article{Townsend_Teague_2023,
  title = {Floquetifying the Colour Code},
  volume = {384},
  ISSN = {2075-2180},
  DOI = {10.4204/eptcs.384.14},
  journal = {Electronic Proc. Th. Compu. Sc.},
  publisher = {Open Publishing Association},
  author = {Townsend-Teague, Alex and Magdalena de la Fuente, Julio and Kesselring, Markus},
  year = {2023},
  optmonth = {aug},
  pages = {265–303},
  eprint = {2307.11136},
  archivePrefix = {arXiv}
}

@misc{gidney2022stim_error_model,
  author = {Gidney, Craig},
  title = {{Where do lines like `error(0.00033) D0 D9 L0 \textasciicircum{} D7 \textasciicircum{} D8' come from in a Stim detector error model?}},
  year = {2022},
  url = {https://quantumcomputing.stackexchange.com/q/23768},
  note = {Accessed: 2025-01-21},
  howpublished = {\url{https://quantumcomputing.stackexchange.com/q/23768}}
}

@misc{stim_command_line_doc,
  title = {Stim: Command Line Usage Documentation},
  author = {Gidney, Craig},
  note = {Accessed: 2025-01-21},
  url = {https://github.com/quantumlib/Stim/blob/main/doc/usage_command_line.md},
  howpublished = {\url{https://github.com/quantumlib/Stim/blob/main/doc/usage_command_line.md}}
}

@article{setiawan2024tailoring,
  title = {{Tailoring dynamical codes for biased noise: The $X ^3 Z^3$ Floquet code}},
  author = {Setiawan, Fnu and McLauchlan, Campbell},
  journal = {npj Quantum Information},
  year = {2025},
  volume = {11},
  pages = {149},
  doi = {10.1038/s41534-025-01074-1},
  eprint = {2411.04974},
  archivePrefix = {arXiv}
}

@article{higgott2024constructions,
  title = {{Constructions and performance of hyperbolic and semi-hyperbolic Floquet codes}},
  author = {Higgott, Oscar and Breuckmann, Nikolas P.},
  journal = {PRX Quantum},
  volume = {5},
  pages = {040327},
  year = {2024},
  publisher = {APS},
  doi = {10.1103/PRXQuantum.5.040327},
  eprint = {2308.03750},
  archivePrefix = {arXiv}
}

@inproceedings{sutcliffe2025distributedquantumerrorcorrection,
  title = {Distributed quantum error correction based on hyperbolic {Floquet} codes},
  author = {Evan Sutcliffe and Bhargavi Jonnadula and Claire Le Gall and Alexandra E. Moylett and Coral M. Westoby},
  year = {2025},
  eprint = {2501.14029},
  archivePrefix = {arXiv},
  primaryClass = {quant-ph},
  booktitle = {2025 IEEE International Conference on Quantum Computing and Engineering (QCE)},
  pages = {649--657},
  publisher = {IEEE},
  doi = {10.1109/QCE65121.2025.00076}
}

@article{Hetnyi2024,
  title = {Creating Entangled Logical Qubits in the Heavy-Hex Lattice with Topological Codes},
  volume = {5},
  ISSN = {2691-3399},
  DOI = {10.1103/prxquantum.5.040334},
  number = {4},
  pages = {040334},
  journal = {PRX Quantum},
  publisher = {American Physical Society (APS)},
  author = {Hetényi, Bence and Wootton, James R.},
  year = {2024},
  optmonth = {dec},
  eprint = {2404.15989},
  archivePrefix = {arXiv}
}

@article{hesner2024using,
  title = {Using detector likelihood for benchmarking quantum error correction},
  author = {Hesner, Ian and Het{\'e}nyi, Bence and Wootton, James R},
  journal = {Phys. Rev. A},
  year = {2025},
  volume = {111},
  pages = {052452},
  doi = {10.1103/PhysRevA.111.052452},
  eprint = {2408.02082},
  archivePrefix = {arXiv}
}

@misc{crumble,
  title = {{Crumble}},
  author = {Gidney, Craig},
  year = {2023},
  note = {In-development tool for exploring and inspecting 2D stabilizer circuits},
  url = {https://algassert.com/crumble},
  howpublished = {\url{https://algassert.com/crumble}}
}

@article{Coecke_2011,
  title = {Interacting quantum observables: categorical algebra and diagrammatics},
  volume = {13},
  ISSN = {1367-2630},
  DOI = {10.1088/1367-2630/13/4/043016},
  journal = {New J. Phys.},
  publisher = {IOP Publishing},
  author = {Coecke, Bob and Duncan, Ross},
  year = {2011},
  optmonth = {apr},
  pages = {043016},
  eprint = {0906.4725},
  archivePrefix = {arXiv}
}

@article{vandewetering2020zxcalculusworkingquantumcomputer,
journal={arXiv preprint arXiv:2012.13966},
    title={{ZX-calculus for the working quantum computer scientist}},
    author={John van de Wetering},
    year={2020},
    eprint={2012.13966},
    archivePrefix={arXiv},
    primaryClass={quant-ph},
    url={https://arxiv.org/abs/2012.13966},
}

@book{KissingerWetering2024Book,
  author = {Kissinger, Aleks and van de Wetering, John},
  title = {{Picturing Quantum Software: An Introduction to the ZX-Calculus and Quantum Compilation}},
  year = {2024},
  publisher = {Preprint},
  url = {https://zxcalc.github.io/book/}
}

@article{rodatz2024floquetifyingstabilisercodesdistancepreserving,
    journal={arXiv preprint arXiv:2410.17240},title={Floquetifying stabiliser codes with distance-preserving rewrites},
    author={Benjamin Rodatz and Boldizsár Poór and Aleks Kissinger},
    year={2024},
    eprint={2410.17240},
    archivePrefix={arXiv},
    primaryClass={quant-ph},
    url={https://arxiv.org/abs/2410.17240},
}

@article{wan2025pauliwebyranglestate,
journal={arXiv preprint arXiv:2501.15566},
    title={Pauli web of the {$|Y\rangle$} state surface code injection},
    author={Kwok Ho Wan and Zhenghao Zhong},
    year={2025},
    eprint={2501.15566},
    archivePrefix={arXiv},
    primaryClass={quant-ph},
    url={https://arxiv.org/abs/2501.15566},
}

@article{McEwen_2023,
  title = {Relaxing Hardware Requirements for Surface Code Circuits using Time-dynamics},
  volume = {7},
  ISSN = {2521-327X},
  DOI = {10.22331/q-2023-11-07-1172},
  journal = {Quantum},
  publisher = {Verein zur Forderung des Open Access Publizierens in den Quantenwissenschaften},
  author = {McEwen, Matt and Bacon, Dave and Gidney, Craig},
  year = {2023},
  optmonth = {nov},
  pages = {1172},
  eprint = {2302.02192},
  archivePrefix = {arXiv}
}

@article{delafuente2024xyzrubycodemaking,
  title = {The {XYZ} ruby code: Making a case for a three-colored graphical calculus for quantum error correction in spacetime},
  author = {Julio C. Magdalena de la Fuente and Josias Old and Alex Townsend-Teague and Manuel Rispler and Jens Eisert and Markus Müller},
  year = {2025},
  journal = {PRX Quantum},
  volume = {6},
  pages = {010360},
  doi = {10.1103/PRXQuantum.6.010360},
  eprint = {2407.08566},
  archivePrefix = {arXiv}
}

@article{gidney2023baconthreshold,
    title={Less {Bacon} More Threshold},
journal={arXiv preprint arXiv:2305.12046},
    author={Craig Gidney and Dave Bacon},
    year={2023},
    eprint={2305.12046},
    archivePrefix={arXiv},
    primaryClass={quant-ph},
    url={https://arxiv.org/abs/2305.12046},
}

@article{berthusen2025adaptivesyndromeextraction,
  journal = {PRX Quantum},
  title = {Adaptive syndrome extraction},
  author = {Noah Berthusen and Shi J. S. Tan and Eric Huang and Daniel Gottesman},
  year = {2025},
  eprint = {2502.14835},
  archivePrefix = {arXiv},
  primaryClass = {quant-ph},
  volume = {6},
  pages = {030307},
  doi = {10.1103/ps3r-wf84}
}

@article{kishony2025increasingdistancetopologicalcodes,
  journal = {Quantum},
  title = {Increasing the distance of topological codes with time vortex defects},
  author = {Gilad Kishony and Erez Berg},
  year = {2026},
  eprint = {2502.12236},
  archivePrefix = {arXiv},
  primaryClass = {quant-ph},
  volume = {10},
  pages = {2006},
  doi = {10.22331/q-2026-02-23-2006}
}

@misc{AG47_2024,
  author = {AG47},
  title = {Uncertainty of estimates computed by stim/sinter},
  year = {2024},
  note = {Accessed: 2025-03-13},
  url = {https://quantumcomputing.stackexchange.com/q/37267},
  howpublished = {\url{https://quantumcomputing.stackexchange.com/q/37267}}
}

@misc{higgott_beliefmatching_2023,
  author = {Higgott, Oscar},
  title = {{BeliefMatching}},
  year = {2023},
  publisher = {GitHub},
  journal = {GitHub repository},
  url = {https://github.com/oscarhiggott/BeliefMatching},
  howpublished = {\url{https://github.com/oscarhiggott/BeliefMatching}}
}

@article{paetznick2023performance,
  title = {Performance of Planar {Floquet} Codes with {Majorana}-Based Qubits},
  author = {Paetznick, Adam and Knapp, Christina and Delfosse, Nicolas and Bauer, Bela and Haah, Jeongwan and Hastings, Matthew B. and da Silva, Marcus P.},
  journal = {PRX Quantum},
  volume = {4},
  issue = {1},
  pages = {010310},
  numpages = {15},
  year = {2023},
  month = {Jan},
  publisher = {American Physical Society},
  doi = {10.1103/PRXQuantum.4.010310},
  eprint = {2202.11829},
  archivePrefix = {arXiv}
}

@misc{rigetti_qpus,
  author = {Rigetti Computing},
  title = {Quantum Processing Units},
  note = {Accessed: 2025-05-09},
  url = {https://qcs.rigetti.com/qpus},
  howpublished = {\url{https://qcs.rigetti.com/qpus}}
}

@misc{iqm_radiance,
  author       = {{IQM Quantum Computers}},
  title        = {{IQM Radiance – High-Performance Quantum Computing Platform}},
  year         = {2025},
  url          = {https://meetiqm.com/products/iqm-radiance/},
  note         = {Accessed: 2025-05-09},
  howpublished = {\url{https://meetiqm.com/products/iqm-radiance/}}
}

@article{Chamberland_2022,
  title = {Building a Fault-Tolerant Quantum Computer Using Concatenated Cat Codes},
  volume = {3},
  ISSN = {2691-3399},
  DOI = {10.1103/prxquantum.3.010329},
  number = {1},
  pages = {010329},
  journal = {PRX Quantum},
  publisher = {American Physical Society (APS)},
  author = {Chamberland, Christopher and Noh, Kyungjoo and Arrangoiz-Arriola, Patricio and Campbell, Earl T. and Hann, Connor T. and Iverson, Joseph and Putterman, Harald and Bohdanowicz, Thomas C. and Flammia, Steven T. and Keller, Andrew and Refael, Gil and Preskill, John and Jiang, Liang and Safavi-Naeini, Amir H. and Painter, Oskar and Brandão, Fernando G.S.L.},
  year = {2022},
  month = {feb},
  eprint = {2012.04108},
  archivePrefix = {arXiv}
}

@misc{dessertaine2025,
      title={Enhanced Fault-tolerance in Photonic Quantum Computing: Comparing the Honeycomb {Floquet} Code and the Surface Code in Tailored Architecture}, 
      author={Théo Dessertaine and Boris Bourdoncle and Aurélie Denys and Grégoire de Gliniasty and Pierre Colonna d'Istria and Gerard Valentí-Rojas and Shane Mansfield and Paul Hilaire},
      year={2025},
      eprint={2410.07065},
      archivePrefix={arXiv},
      primaryClass={quant-ph},
      url={https://arxiv.org/abs/2410.07065}, 
}

@article{chan2025tailoring,
  title = {Tailoring fusion-based photonic quantum computing schemes to quantum emitters},
  author = {Chan, Ming Lai and Bell, Thomas J and Pettersson, Love A and Chen, Susan X and Yard, Patrick and S{\o}rensen, Anders S and Paesani, Stefano},
  journal = {PRX Quantum},
  volume = {6},
  number = {2},
  pages = {020304},
  year = {2025},
  publisher = {APS},
  doi = {10.1103/PRXQuantum.6.020304},
  eprint = {2410.06784},
  archivePrefix = {arXiv}
}

@article{SohaibAlam_2025,
  title = {Dynamical logical qubits in the {Bacon-Shor} code},
  author = {Alam, M. Sohaib and Rieffel, Eleanor},
  journal = {Phys. Rev. A},
  volume = {112},
  issue = {2},
  pages = {022436},
  numpages = {22},
  year = {2025},
  month = {Aug},
  publisher = {American Physical Society},
  doi = {10.1103/nfxv-3dp7},
  eprint = {2403.03291},
  archivePrefix = {arXiv}
}

@article{bourdoncle2026two,
  title = {Two Layers, No Swaps: Biplanar {SPOQC} Architecture Improves Runtime of {Fermi-Hubbard} Simulation},
  author = {Bourdoncle, Boris and Derks, Peter-Jan and Dessertaine, Th{\'e}o and Frank, Johannes},
  journal = {arXiv preprint arXiv:2605.05315},
  year = {2026},
  eprint = {2605.05315},
  archivePrefix = {arXiv}
}

@article{gidney2025factor,
  title = {How to factor 2048 bit {RSA} integers with less than a million noisy qubits},
  author = {Gidney, Craig},
  journal = {arXiv preprint arXiv:2505.15917},
  year = {2025},
  eprint = {2505.15917},
  archivePrefix = {arXiv}
}

    \appendix

    \section{Methodology}
\label{sec:methodology}

This section provides details on the numerics, though not exhaustively. For full reproducibility,  \href{https://github.com/peter-janderks/floquet_colour_codes_numerics}{the source code} includes instructions for generating circuits, running simulations, and using Jupyter notebooks to produce all plots.

We built circuits that implement the memory and stability experiments using Stim \cite{Gidney2021stimfaststabilizer}. 
For convenience, we simulate the DCCCs on a torus, but we check the error rate of 1 logical $X$ operator and 1 logical $Z$ operator.
For phenomenological, SD, and SI noise we simulate distances $4,8,12,$ and $16$.
For EM noise we simulate distances $2,4,6,$ and $8$.
MWPM decoding is performed using the PyMatching package \cite{higgott2025sparse} and belief matching via the 
BeliefMatching package \cite{higgott2022fragile,higgott_beliefmatching_2023}. 
We collect shots using Sinter.
When decoding each circuit with MWPM, we collect $10^6$ shots or as many shots required until $100$ logical errors occur, whichever comes first.
With belief matching, we collect $10^4$ shots  or as many shots required until $100$ logical errors occur, whichever comes first.

The highlighted regions in \Cref{fig:volume_verification,fig:X1Y1Z1_X1Z1_teraquop_linefits} are computed with Sinter and show hypothesis probabilities within $1000 \times$ of 
the max likelihood hypothesis \cite{AG47_2024}.
The error in $\tilde{n}_E, \tilde{n}_M, \tilde{h}_E, \tilde{h}_M$ shown in \Cref{fig:X1Y1Z1_X1Z1_comparison_cln,fig:X1Y1Z1_X1Z1_comparison,fig:X1Z1_X3Z3_comparison} is determined by fitting two lines with a maximum additional squared error set to 1 to the data using Sinter.
The error is then given by the value at which these two lines intersect the horizontal line $y=10^{-12}$. 
More info on the error in the line fits is given in \Ccite[Appendix A]{gidney2023pair}.
We denote the maximum/minimum errors as $\tilde{n}_{E,\text{max/min}}, \, \tilde{n}_{M,\text{max/min}}, \, \tilde{h}_{E,\text{max/min}}, $ and $\tilde{h}_{M,\text{max/min}}$.
The error $\tilde{h}_{\text{max/min}}$ is given by $\frac{\tilde{h}_{E,\text{max/min}} + \tilde{h}_{M,\text{max/min}}}{2}$.
We denote the relative error of a variable $x$ as 
\begin{equation}\Delta x =  \frac{x_\text{high} - x_\text{best}}{x_\text{best}}.
\end{equation}
The maximum and minimum error in the teraquop volume is given as
\begin{equation}
\epsilon_{\text{max/min}} = \sqrt{ \Delta \tilde{n}_{E,\text{max/min}}^2 + \Delta\tilde{n}_{M,\text{max/min}}^2 + \Delta \tilde{h}_{\text{max/min}}^2}.
\end{equation}

    \section{ZX-calculus and Pauli webs}\label{sec:zx-calculus-and-pauli-webs}

Throughout this work we've used the \textit{ZX-calculus} and \textit{Pauli webs} in the figures.
Here we give a very quick explainer of some of our notation, and give references for the reader interested in learning more.
The ZX-calculus is a graphical language for reasoning about quantum mechanics.
Two great introductions are \Ccite{vandewetering2020zxcalculusworkingquantumcomputer,KissingerWetering2024Book}.
A ZX-diagram consists of green and red \textit{spiders} with input and output legs, which one can think of as just very good notation for families of linear maps
\begin{equation}\label{eq:spiders_defn}
    \begin{aligned}
        \includegraphics[height=50pt,valign=c]{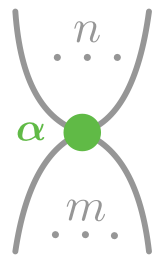}
        &\coloneqq \ket{0}^{\otimes n}\bra{0}^{\otimes m} + e^{i\alpha}\ket{1}^{\otimes n}\bra{1}^{\otimes m}, \\
        \includegraphics[height=50pt,valign=c]{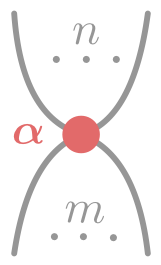}
        &\coloneqq \ket{+}^{\otimes n}\bra{+}^{\otimes m} + e^{i\alpha}\ket{-}^{\otimes n}\bra{-}^{\otimes m}.
    \end{aligned}
\end{equation}
These spider legs can be connected together, corresponding to composition of linear maps, and spiders can be placed alongside each other, corresponding to taking tensor products.
In this work, we additionally used the shorthands
\begin{equation}\label{eq:zx_shorthands}
    \includegraphics[height=55pt, valign=c]{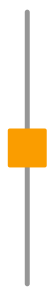} \coloneqq
    \includegraphics[height=55pt, valign=c]{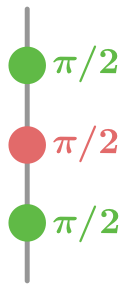} ,\qqquad
    \includegraphics[height=40pt, valign=c]{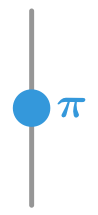} \coloneqq
    \includegraphics[height=40pt, valign=c]{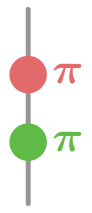} ,\qqquad
    \includegraphics[height=40pt, valign=c]{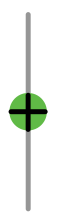} \coloneqq
    \includegraphics[height=40pt, valign=c]{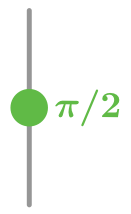} ,\qqquad
    \includegraphics[height=40pt, valign=c]{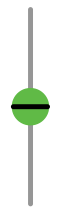} \coloneqq
    \includegraphics[height=40pt, valign=c]{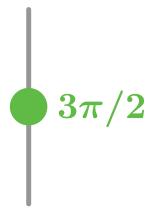}.
\end{equation}
We also got a bit cheeky and drew some spiders/shorthands right on top of each other, hiding the connecting wire
\begin{equation}\label{eq:compressed_ZZ_measurement}
    \includegraphics[height=40pt, valign=c]{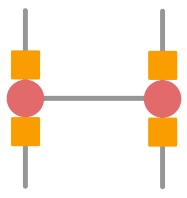} \quad = \quad
    \includegraphics[height=40pt, valign=c]{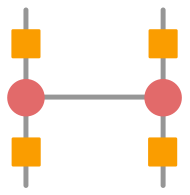}.
\end{equation}
We continued to be cheeky when we said that certain diagrams correspond to measurements of $X \otimes X$, $Y \otimes Y$ and $Z \otimes Z$ in \Cref{fig:schedules_XYZ_and_CSS}.
In reality, the diagrams there are equal (up to positive scalar) to the projections onto the $+1$ eigenspaces of $X \otimes X$, $Y \otimes Y$ and $Z \otimes Z$ respectively.
In other words, it's as if we're assuming throughout that we always get measurement outcome $+1$, despite many of these measurements being totally random.
It turns out that for reasoning about QEC protocols, we can get away with this, as explained (very briefly!) in \Ccite[Appendix G]{Townsend_Teague_2023}.

ZX-diagrams can additionally be annotated with \textit{Pauli webs}.
One can think of these as describing how Pauli gates commute with individual spiders in a ZX-diagram.
A nice introduction to them can be found in \Ccite{rodatz2024floquetifyingstabilisercodesdistancepreserving}.
The important part with respect to making sense of the figures in this work is that `closed' Pauli webs (ones in which no overall input or output leg of the ZX-diagram is highlighted) correspond to detectors.
Whether a set of Pauli errors violates a detector can then be seen graphically -- this happens if and only if there are an odd number of edges that are highlighted in a colour other than the colour of the Pauli error that sits on it.
For example, the following sets of Pauli errors on consecutive $Z \otimes Z$ measurements violate the corresponding detector
\begin{equation}\label{eq:ZZ_detector_violated}
    \includegraphics[height=70pt, valign=c]{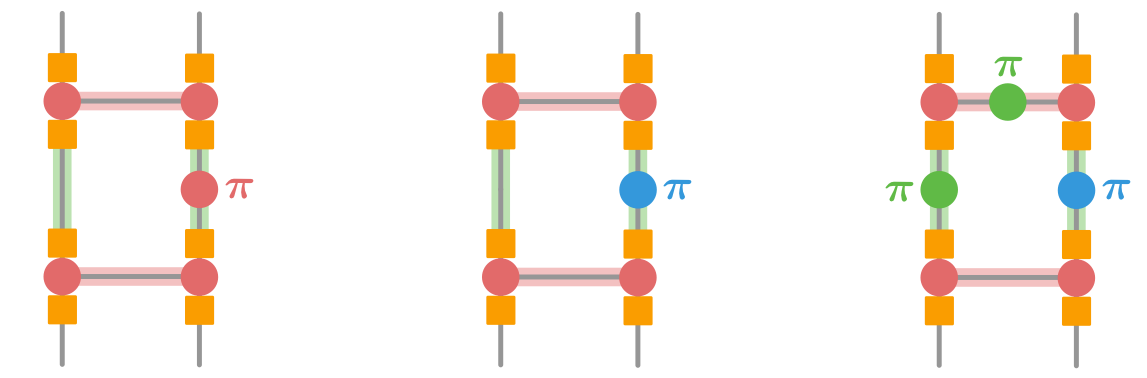}.
\end{equation}
On the other hand, these sets of Pauli errors don't violate it
\begin{equation}\label{eq:ZZ_detector_not_violated}
    \includegraphics[height=70pt, valign=c]{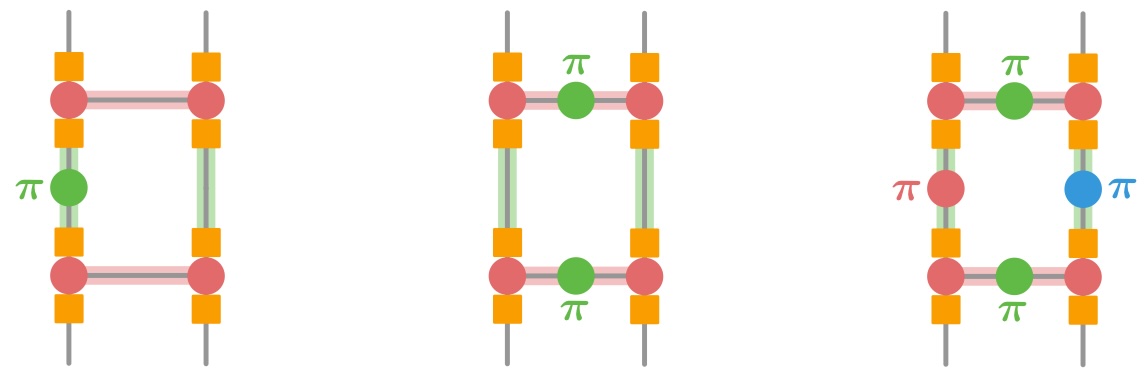}.
\end{equation}
So given a closed Pauli web, one can read off which measurements make up the corresponding detector, and which types of Pauli errors this detector can detect at different time steps.
Namely, since measurement errors correspond to 
green two-legged 
$\pi$-spiders 
on `horizontal' 
wires (\Cref{fig:phenom_error_examples}), the corresponding measurement is only included in a detector 
if this wire is 
highlighted red or blue.
For example, in \Cref{fig:detectors_XYZ_and_CSS} we can now read off that the detector in the $X^1 Y^1 Z^1$ honeycomb code consists of twelve detectors, while the detector in the $X^1 Z^1$ honeycomb code consists of only six.
For more details, see \Ccite{rodatz2024floquetifyingstabilisercodesdistancepreserving}, or the handful of other works that make use of Pauli webs~\cite{bombin2024unifying,McEwen_2023,wan2025pauliwebyranglestate,delafuente2024xyzrubycodemaking, Townsend_Teague_2023}.

    \section{Detectors cheatsheet}\label{sec:detectors-cheatsheet}

In the following figures, we show Pauli webs of all detectors in the unrepeated $X^1 Y^1 Z^1$ and $X^1 Z^1$ codes
and the repeated $X^2 Y^2 Z^2$ and $X^2 Z^2$ codes.

\newpage

\begin{figure}[p]
    \centering
    \includegraphics[width=\linewidth]{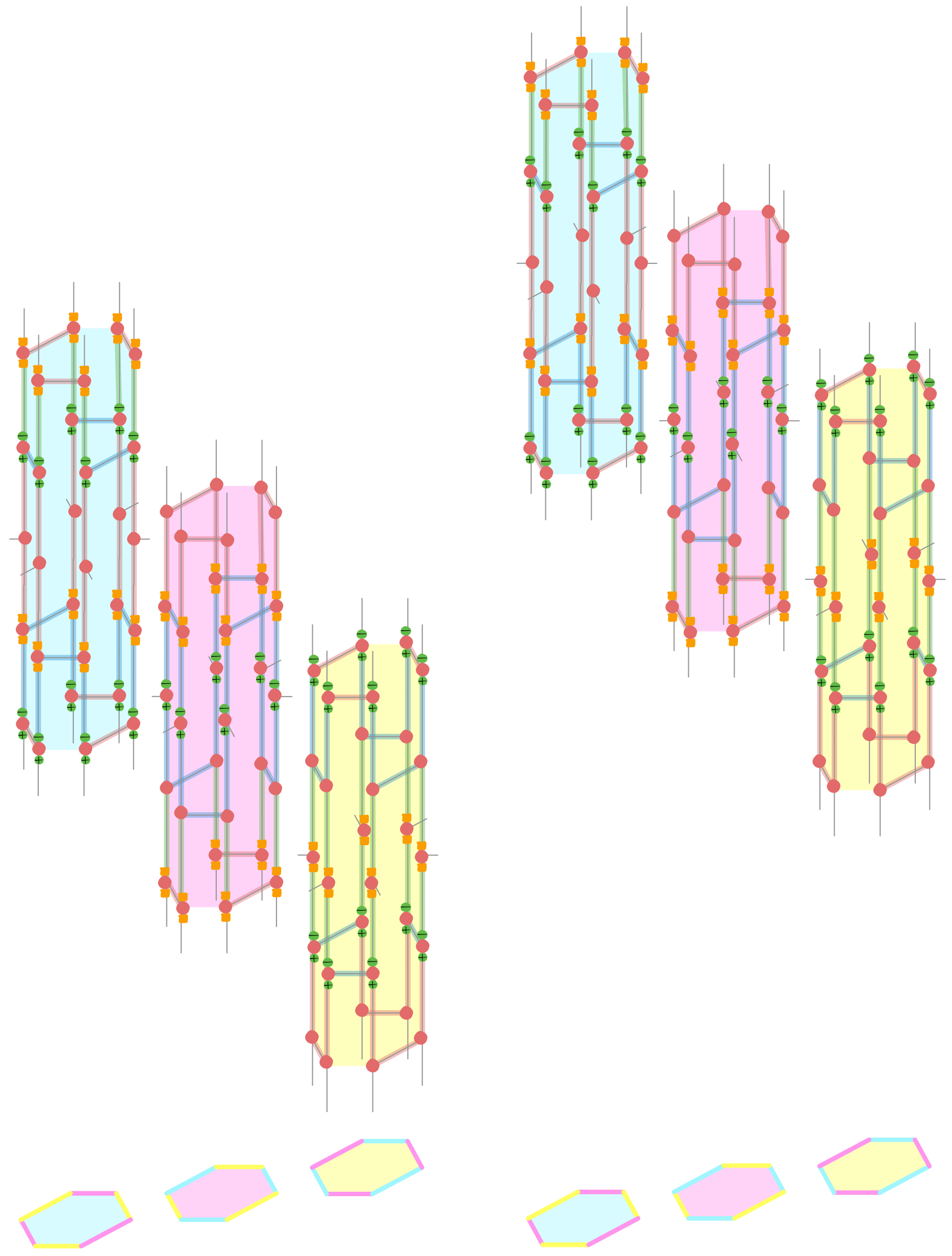}
    \caption{$E$ (left) and $M$ (right) detectors of the $X^1 Y^1 Z^1$ code.}
    \label{fig:XYZ_all_detector_webs}
\end{figure}

\begin{figure}[p]
    \centering
    \includegraphics[width=\linewidth]{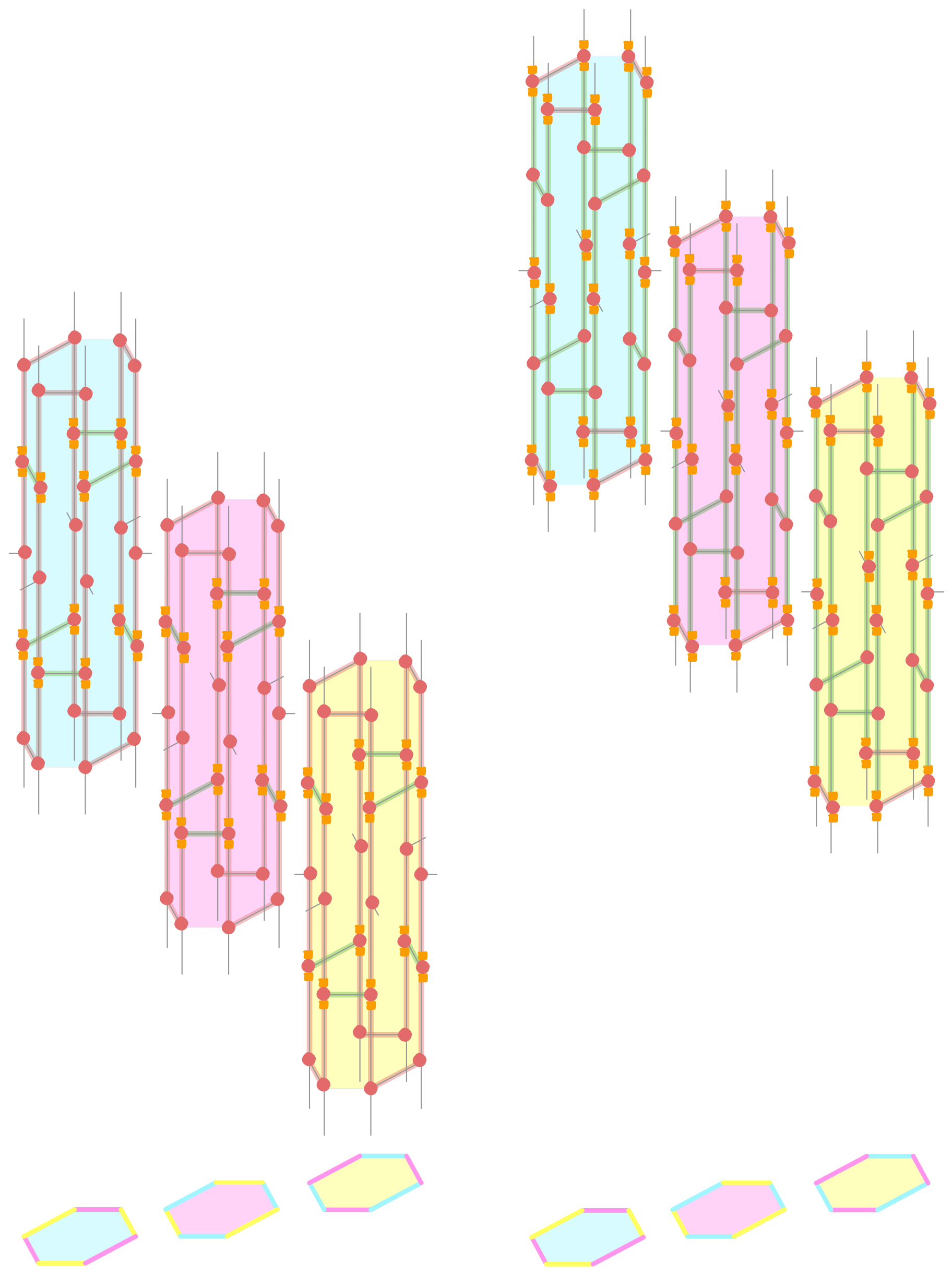}
    \caption{$E$ (left) and $M$ (right) detectors of the $X^1 Z^1$ code.}
    \label{fig:CSS_all_detector_webs}
\end{figure}

\begin{figure}[p]
    \centering
    \includegraphics[height=0.95\textheight]{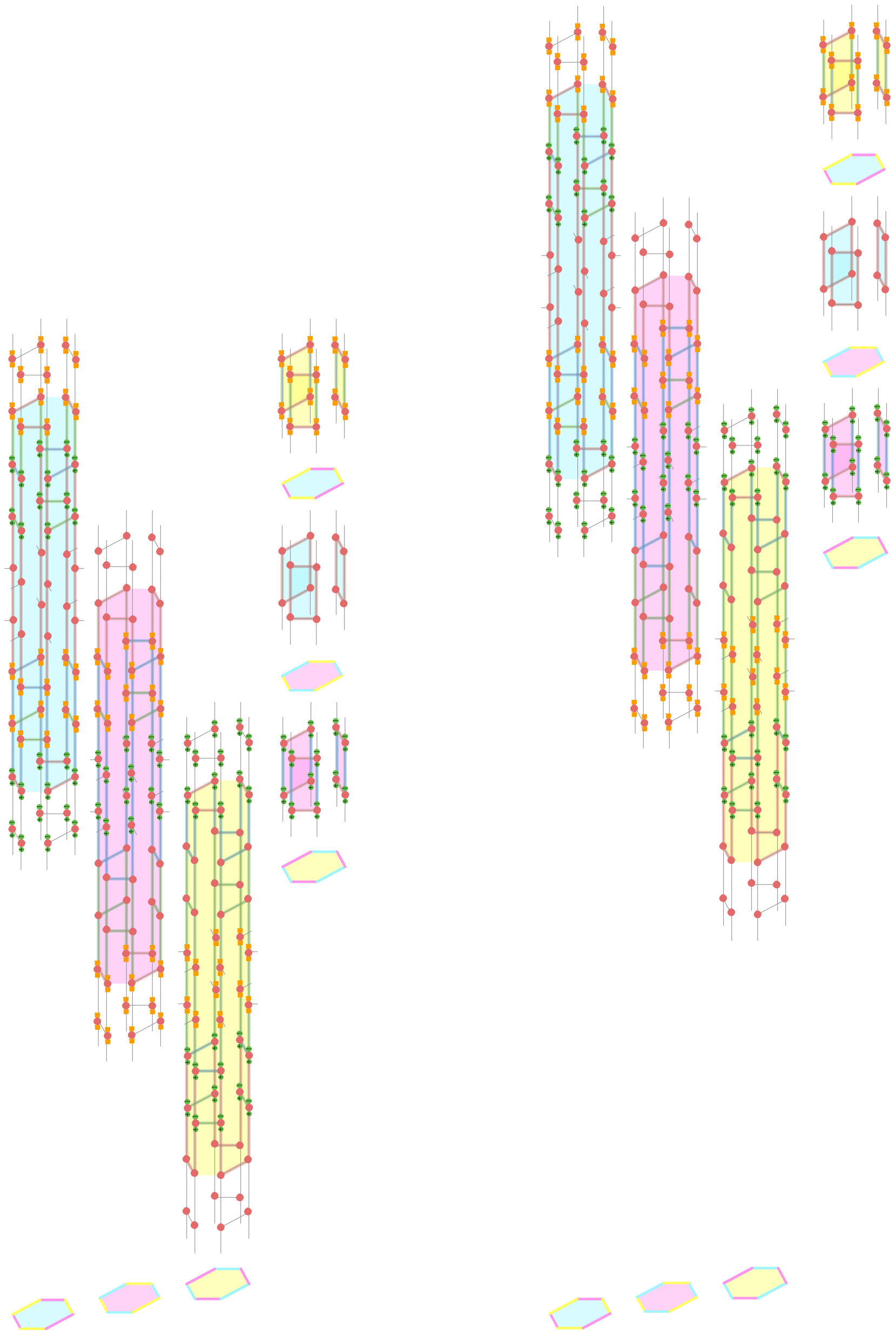}
    \caption{$E$ (left) and $M$ (right) detectors of the $X^2 Y^2 Z^2$ code.}
    \label{fig:repeated_XYZ_all_detector_webs}
\end{figure}

\begin{figure}[p]
    \centering
    \includegraphics[height=0.95\textheight]{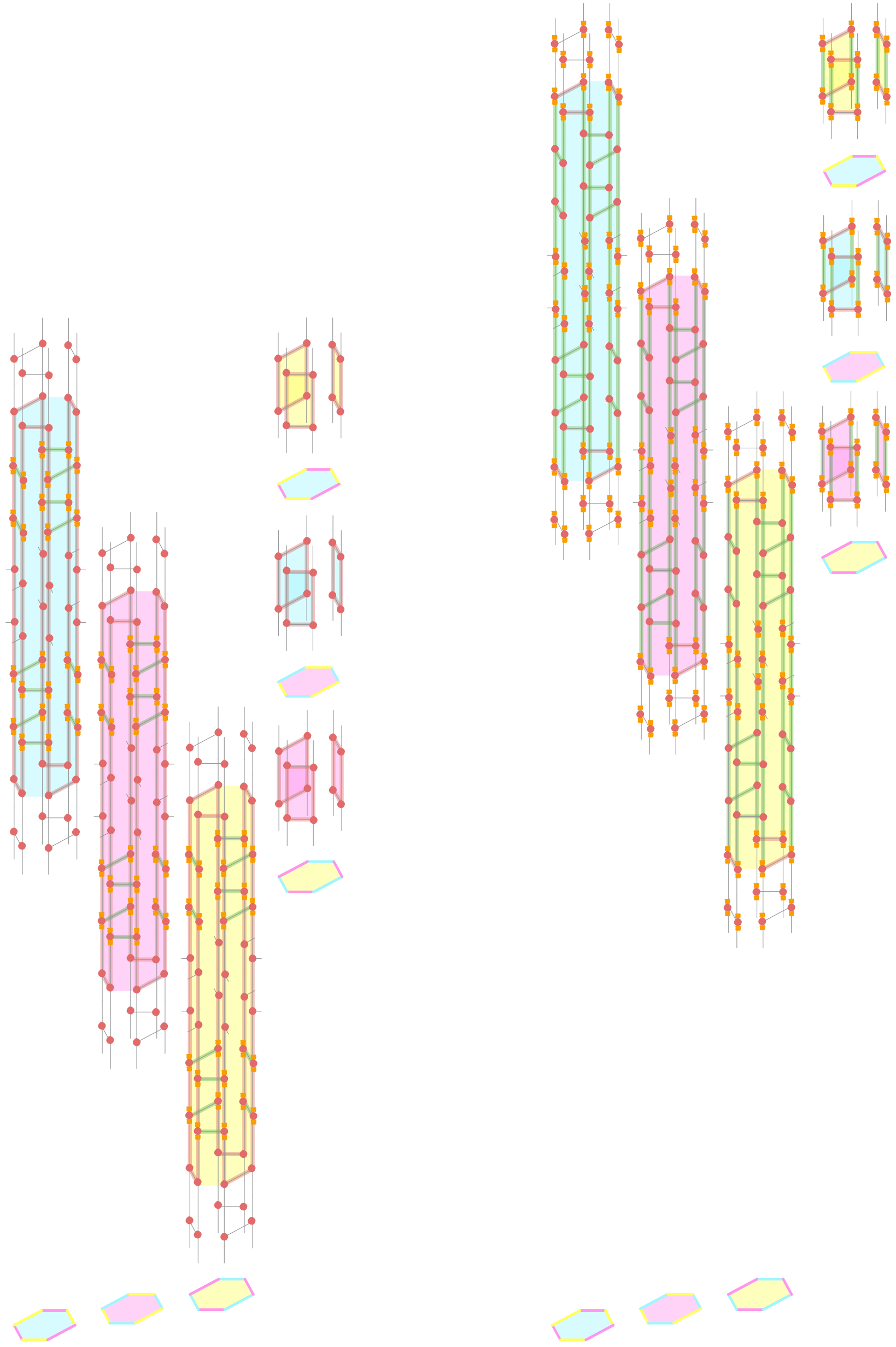}
    \caption{$E$ (left) and $M$ (right) detectors of the $X^2 Z^2$ code.}
    \label{fig:repeated_CSS_all_detector_webs}
\end{figure}

\end{document}